%% file: main.tex
\documentclass[journal]{IEEEtran}
\ifCLASSINFOpdf
\else
\fi
\input{preamble}

\begin{document}

\title{Generative AI Meets 6G and Beyond:\\Diffusion Models for Semantic Communications}

\author{Hai-Long~Qin,~\IEEEmembership{Graduate Student Member,~IEEE},
	    Jincheng~Dai,~\IEEEmembership{Member,~IEEE},
	    Guo~Lu,~\IEEEmembership{Member,~IEEE},\\
	    Shuo~Shao,~\IEEEmembership{Member,~IEEE},
	    Sixian~Wang,~\IEEEmembership{Member,~IEEE},
	    Tongda~Xu,~\IEEEmembership{Graduate Student Member,~IEEE},\\
	    Wenjun~Zhang,~\IEEEmembership{Fellow,~IEEE},
	    Ping~Zhang,~\IEEEmembership{Fellow,~IEEE},
	    and Khaled~B.~Letaief,~\IEEEmembership{Fellow,~IEEE}

\thanks{This work was supported in part by the National Key Research and Development Program of China under Grant 2024YFF0509700; in part by the National Natural Science Foundation of China under Grants 62371063, 62293481, 62471290, 62471294, 62231022, and 62321001; in part by the Beijing Municipal Natural Science Foundation under Grant L232047; in part by the Postdoctoral Fellowship Program of China Postdoctoral Science Foundation (CPSF) under Grant GZB20250810; in part by the Beijing Nova Program; in part by the Hong Kong Research Grant Council under Grant 16215624; and in part by the Hong Kong RGC Areas of Excellence (AoE) Scheme under Grant AoE/E-601/22-R. \textit{(Corresponding author: Jincheng Dai.)}}

\thanks{Hai-Long Qin, Jincheng Dai, and Ping Zhang are with Beijing University of Posts and Telecommunications. Guo Lu, Sixian Wang, and Wenjun Zhang are with Shanghai Jiao Tong University. Shuo Shao is with East China Normal University. Tongda Xu is with Tsinghua University. Khaled B. Letaief is with Hong Kong University of Science and Technology.}

\thanks{A public GitHub repository on diffusion models for semantic communications is available at: \url{https://github.com/qin-jingyun/Awesome-DiffComm}.}

\vspace{-1em}
}

\maketitle
\input{sec/0_abstract}
\input{sec/1_introduction}
\input{sec/2_preliminary}
\input{sec/3_diffusion-model}
\input{sec/4_inverse-problem}
\input{sec/5_application}
\input{sec/6_open-issue}
\input{sec/7_conclusion}
\bibliographystyle{IEEEbib}
\bibliography{ref}

\end{document}

%% file: preamble.tex
\usepackage{microtype} 
\usepackage{graphicx} 

\usepackage[table,xcdraw,dvipsnames]{xcolor}
\usepackage{url}
\usepackage{cite}
\definecolor{bupt-blue}{rgb}{0,0.2,0.6}
\definecolor{cvpr-blue}{rgb}{0.21,0.49,0.74}
\definecolor{dark-red}{rgb}{0.6,0,0}
\definecolor{thu-purple}{rgb}{0.4,0.031,0.455}
\definecolor{nyu-purple}{rgb}{0.333,0.204,0.533}
\definecolor{harvard-crimson}{rgb}{0.647,0,0.118}
\usepackage[colorlinks=true, linkcolor=dark-red, citecolor=bupt-blue]{hyperref}

\usepackage{enumitem}
\usepackage{wrapfig}
\usepackage{siunitx}

\usepackage{times}
\usepackage{helvet}
\usepackage{courier}

\usepackage{amsmath,amssymb,amsfonts}
\usepackage{bm}
\usepackage{ntheorem}

\newtheorem*{*thm}{Theorem}

\newtheorem*{*lemma}{Lemma}

\def\st{{\mathrm{s.t.}}}
\def\etal{{\em et al.}}
\def\ie{\mbox{\textit{i.e.}, }}
\def\eg{\mbox{\textit{e.g.}, }}

\def\aka{\mbox{\textit{a.k.a.}, }}
\def\vs{\mbox{\textit{vs.}}}

\usepackage{graphicx}
\usepackage{wrapfig}

\usepackage{caption}
\usepackage{booktabs}
\usepackage{array}
\usepackage{tikz-cd}
\usepackage{multirow}
\usepackage{multicol}
\usepackage{colortbl}
\usepackage{makecell}
\usepackage{tabularx}
\usepackage{adjustbox}

\usepackage{pifont}

\definecolor{ratinggood}{RGB}{34,139,34}
\definecolor{ratingmedium}{RGB}{240,170,0}
\definecolor{ratingpoor}{RGB}{192,50,50}
\newcommand{\good}{{\color{ratinggood}$\bullet\bullet\bullet$}}
\newcommand{\medium}{{\color{ratingmedium}$\bullet\bullet\circ$}}
\newcommand{\poor}{{\color{ratingpoor}$\bullet\circ\circ$}}

\usepackage[most]{tcolorbox}
\usepackage{etoolbox}

\newcounter{remarkcount}
\newtcolorbox{remark}[1][]{
	enhanced,
	breakable,
	colback=black!1!white,
	colbacktitle=gray!10,
	coltitle=black,
	colframe=gray!10,
	fonttitle=\small\bfseries,
	title={\refstepcounter{remarkcount}Remark~\Roman{remarkcount}\ifstrempty{#1}{}{~(#1)}},
	boxrule=1.0pt,
	arc=3pt,
	left=6pt, right=6pt, top=4pt, bottom=4pt,
	toptitle=2pt, bottomtitle=2pt,
	fontupper=\small,
}

\definecolor{pink}{rgb}{1, 0.753, 0.796}
\definecolor{prim}{rgb}{0.965, 0.914, 0.945}
\definecolor{aquaisland}{rgb}{0.6352, 0.851, 0.8078}
\definecolor{whitelilac}{rgb}{0.945, 0.914, 0.965}
\definecolor{ecruwhite}{rgb}{0.945, 0.965, 0.914}
\definecolor{teal}{rgb}{0, 0.502, 0.502}
\definecolor{sidecar}{rgb}{0.9686, 0.9059, 0.8078}
\definecolor{peachcream}{rgb}{1, 0.9451, 0.8784}
\definecolor{catskillwhite}{rgb}{0.9529, 0.9765, 0.9765}
\definecolor{botticelli}{rgb}{0.8235, 0.9098, 0.9098}
\definecolor{junglemist}{rgb}{0.6824, 0.8314, 0.8314}
\definecolor{neptune}{rgb}{0.5412, 0.7530, 0.7530}
\definecolor{azalea1}{rgb}{0.9686, 0.7922, 0.7883}
\definecolor{azalea2}{rgb}{0.9804, 0.8588, 0.8471}
\definecolor{prelude}{rgb}{0.8235, 0.7059, 0.8706}
\definecolor{lightred}{RGB}{255, 248, 248}
\definecolor{lightblue}{RGB}{245, 250, 255}

\colorlet{colorFst}{Green!25}
\colorlet{colorSnd}{SpringGreen!45}
\colorlet{colorTrd}{Yellow!30}
\colorlet{colorLow}{darkgray!30}
\colorlet{colorDeg}{Orange!30}

\newcolumntype{C}[1]{>{\centering\arraybackslash}p{#1}}
\newcolumntype{L}{>{\raggedright\arraybackslash}X}
\newcolumntype{R}{>{\raggedleft\arraybackslash}X}

\usepackage{algorithm,algpseudocode}

\usepackage{stackengine}
\usepackage{threeparttable}

\newcommand{\redcross}{\vcenter{\hbox{\includegraphics[height=0.5em]{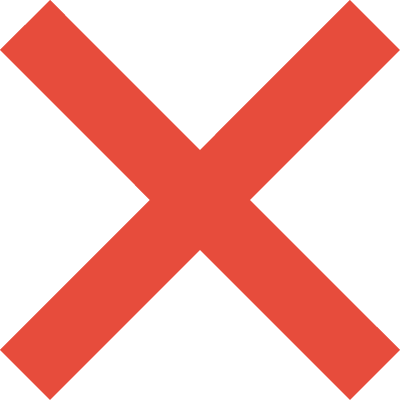}}}}
\newcommand{\bluedot}{\vcenter{\hbox{\includegraphics[height=0.5em]{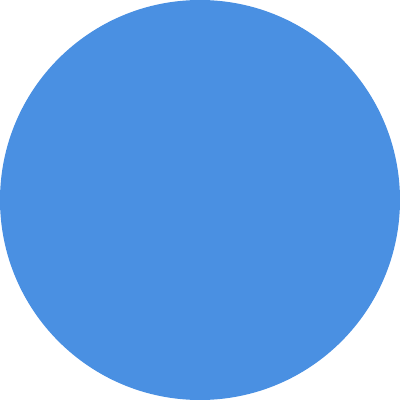}}}}

\newcommand{\github}{\adjustbox{valign=c}{\includegraphics[height=1em]{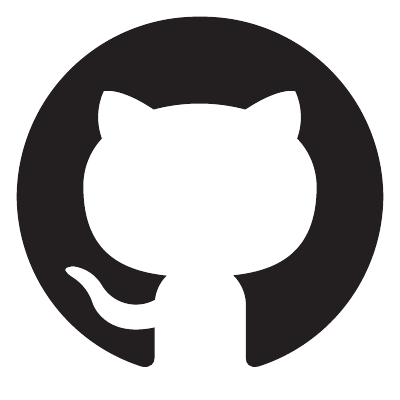}}}
\newcommand{\huggingface}{\adjustbox{valign=c}{\includegraphics[height=1em]{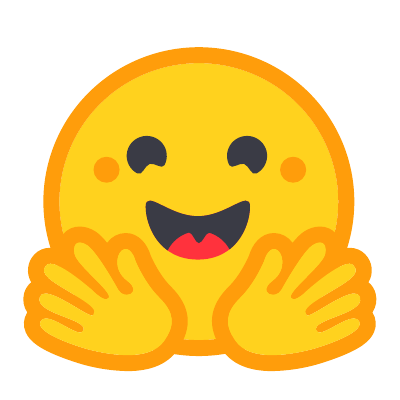}}}
\newcommand{\link}{\adjustbox{valign=c}{\includegraphics[height=1em]{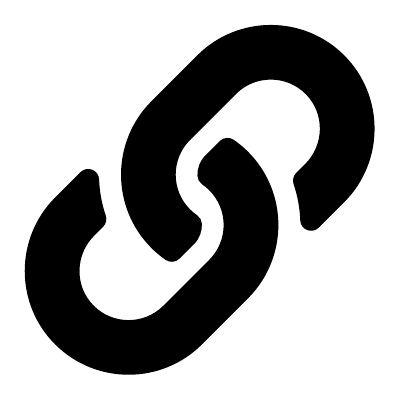}}}

%% file: sec/0_abstract.tex
\begin{abstract}
	Semantic communications mark a paradigm shift from bit-accurate transmission toward meaning-centric communication, essential as wireless systems approach theoretical capacity limits. The emergence of generative AI has catalyzed generative semantic communications, where receivers reconstruct content from minimal semantic cues by leveraging learned priors. Among generative approaches, diffusion models stand out for their superior generation quality, stable training dynamics, and rigorous theoretical foundations. However, the field currently lacks systematic guidance connecting diffusion techniques to communication system design, forcing researchers to navigate disparate literatures. This article provides the first comprehensive tutorial on diffusion models for generative semantic communications. We present score-based diffusion foundations and systematically review three technical pillars: conditional diffusion for controllable generation, efficient diffusion for accelerated inference, and generalized diffusion for cross-domain adaptation. In addition, we introduce an inverse problem perspective that reformulates semantic decoding as posterior inference, bridging semantic communications with computational imaging. Through analysis of human-centric, machine-centric, and agent-centric scenarios, we illustrate how diffusion models enable extreme compression while maintaining semantic fidelity and robustness. By bridging generative AI innovations with communication system design, this article aims to establish diffusion models as foundational components of next-generation wireless networks and beyond.
\end{abstract}

\begin{IEEEkeywords}
	6G, semantic communications, generative AI, diffusion models, tutorial.
\end{IEEEkeywords}

%% file: sec/1_introduction.tex
\section{Introduction}

\subsection{Background}

\IEEEPARstart{W}{ireless} communication systems have evolved remarkably from first-generation (1G) to fifth-generation (5G) technologies. However, emerging applications in beyond-5G (B5G) and sixth-generation (6G) networks, including Industry 4.0, the metaverse, brain-computer interfaces, and digital twins, which demand ultra-low latency and ultra-high data rates that far exceed current 5G capabilities. Compounding this challenge, existing physical-layer technologies are approaching their practical performance limits, necessitating innovative paradigms for future wireless communications.

\begin{figure}[t]
	\centering
	\includegraphics[width=\columnwidth]{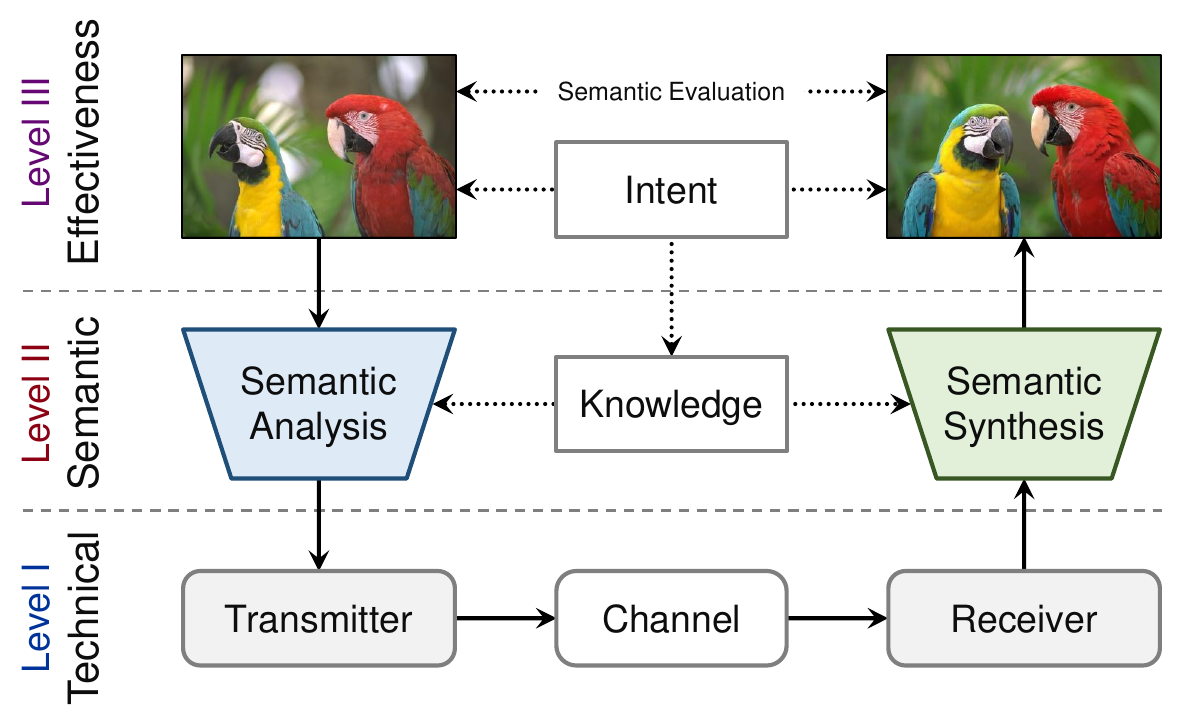}
	\caption{Schematic diagram of Weaver's three-level communication model. \textcolor[HTML]{003399}{Level I} focuses on accurate bit transmission, \textcolor[HTML]{8B0012}{Level II} on conveyance of meaning, and \textcolor[HTML]{660874}{Level III} on achieving desired outcomes.}
	\label{fig:level}
\end{figure}

These limitations compel us to reconsider communication beyond Shannon's information-theoretic framework. In 1949, Weaver articulated a prescient three-level communication model~\cite{weaver-semantics}: the technical level (accurate symbol transmission), the semantic level (conveying intended meaning), and the effectiveness level (achieving desired outcomes), illustrated in Fig.~\ref{fig:level}. While Shannon's theory elegantly addresses technical-level challenges, it treats all bits equally despite their varying significance to communication objectives. This insight redirects our focus from bit-perfect transmission toward effectively conveying meaning and achieving communication intent at higher levels.

These practical and theoretical imperatives have catalyzed exploration of semantic communications as a transformative paradigm~\cite{guo2024survey, nguyen2025contemporary}. Semantic communications enable truly AI-native systems, forming the foundation for next-generation intellicise wireless networks~\cite{zhang2024intellicise}. By integrating deep learning (DL) for end-to-end optimization, semantic communications achieve substantial efficiency gains over conventional approaches. Crucially, semantic communications are inherently \emph{knowledge-driven}~\cite{chaccour2024less}: semantics constitute context-dependent abstractions requiring supporting knowledge for interpretation. This knowledge, which is whether derived from data patterns, task-specific databases, or domain expertise~\cite{von2021informed}, guides semantic feature extraction at transmitters and enables desired reconstruction from compressed representations at receivers.

The emergence of powerful generative models has catalyzed a paradigm shift in semantic communications~\cite{grassucci2024generative, ren2024generative}. Pre-trained on massive datasets, these models capture rich semantic distributions serving as \emph{prior knowledge} for communication systems. Receivers can now reconstruct complete content from minimal ``semantic cues'', exploiting the inherent robustness of semantic features to channel impairments. This synergy between generative AI and semantic communications enables unprecedented compression ratios while maintaining quality, heralding the era of \emph{generative semantic communications}~\cite{dai2024deep}.

\subsection{Motivation}

\begin{figure}[t]
	\centering
	\includegraphics[width=\columnwidth]{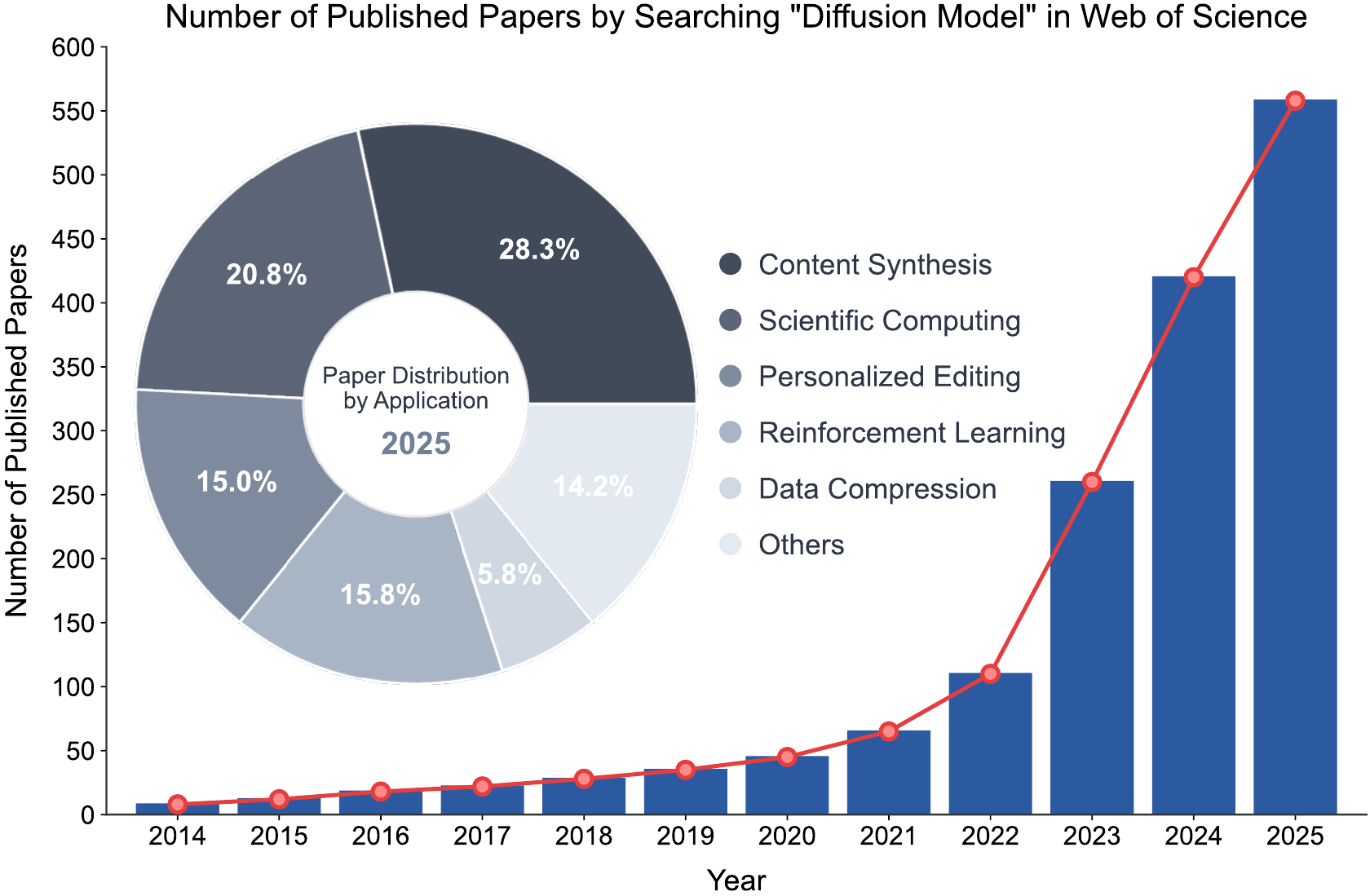}
	\caption{Statistical results manifesting the rapid development of diffusion models in recent years. The number of published papers by searching ``diffusion models'' in Web of Science is accessed by December 31, 2025. For 2025, the paper distribution indicates the wide application potential of diffusion models across science and engineering, particularly in content synthesis and scientific computing.}
	\label{fig:develop}
\end{figure}

The success of large language models (LLMs), particularly the GPT family~\cite{radford2018improving, radford2019language, brown2020language}, has sparked renewed interest in generative modeling across diverse domains. Among generative approaches, diffusion models~\cite{sohl2015deep} have emerged as particularly compelling for semantic communications, experiencing rapid development in recent years (Fig.~\ref{fig:develop}). Their capacity to generate high-quality visual content, coupled with stable training dynamics, flexible conditioning mechanisms, and rigorous theoretical foundations, positions them ideally for next-generation immersive communication systems.

Early investigations have validated this potential. Researchers have successfully applied diffusion models to semantic source coding for superior compression~\cite{yang2024diffusion}, channel denoising with enhanced robustness~\cite{wu2024cddm}, and end-to-end system optimization yielding remarkable quality improvements~\cite{guo2024diffusion}. These pioneering efforts demonstrate that diffusion models can fundamentally transform how semantic communication systems encode, transmit, and reconstruct information.

Despite these promising advances, the field lacks a comprehensive tutorial bridging diffusion model techniques with practical semantic communication applications. While existing surveys and perspective articles offer valuable overviews~\cite{du2024enhancing, fan2025generative, liang2025diffsg}, they typically lack the technical depth and implementation details essential for researchers entering this interdisciplinary domain. The absence of such a resource forces newcomers to navigate disparate literatures across machine learning (ML) and communications, creating unnecessary barriers that impede progress. A systematic tutorial paper addressing this gap would accelerate research by providing clear foundations, practical guidance, and a unified framework for understanding this rapidly evolving field.

\subsection{Contribution}

This article makes three key contributions to the semantic communications research community:

\begin{itemize}
	\item \textbf{Comprehensive Technical Foundation}: We present the first systematic tutorial on diffusion models specifically tailored for semantic communication researchers. Beginning from fundamental principles, we develop three critical technical threads (\ie conditional diffusion for controllable generation, efficient diffusion for accelerated sampling, and generalized diffusion for task adaptation), each essential for practical semantic communication system design.
	
	\item \textbf{Unified Theoretical Framework}: We introduce a novel perspective by formulating semantic decoding as an inverse problem, establishing connections between semantic communications and mature fields such as computational imaging. This framework unlocks powerful mathematical tools from inverse problem theory while providing fresh insights into the fundamental nature of semantic reconstruction under uncertainty and imperfect channel conditions.
	
	\item \textbf{Practical Implementation Resources}: We provide extensive practical resources including curated open-source implementations, educational materials, and deployment guidelines. Through detailed analysis of human-centric, machine-centric, and agent-centric semantic communication scenarios, we identify critical research challenges and chart promising directions for future investigation, bridging the gap between theoretical advances and practical applications.
\end{itemize}

\subsection{Organization}
The remainder of this article is structured as follows. Section~\ref{sec:pre} introduces generative modeling preliminaries essential for understanding score-based diffusion models. Section~\ref{sec:diffusion} systematically presents diffusion model fundamentals, including conditioning mechanisms, acceleration techniques, and generalization strategies, accompanied by comprehensive literature review. Section~\ref{sec:inverse} formulates semantic decoding as an inverse problem and explores diffusion-based solution methodologies. Section~\ref{sec:app} examines applications of diffusion models across human-centric, machine-centric, and agent-centric semantic communication scenarios. Section~\ref{sec:open} identifies theoretical limitations, technical challenges, and promising future research directions. Finally, Section~\ref{sec:conclusion} provides concluding remarks.

\begin{figure*}[t]
	\centering
	\includegraphics[width=\textwidth]{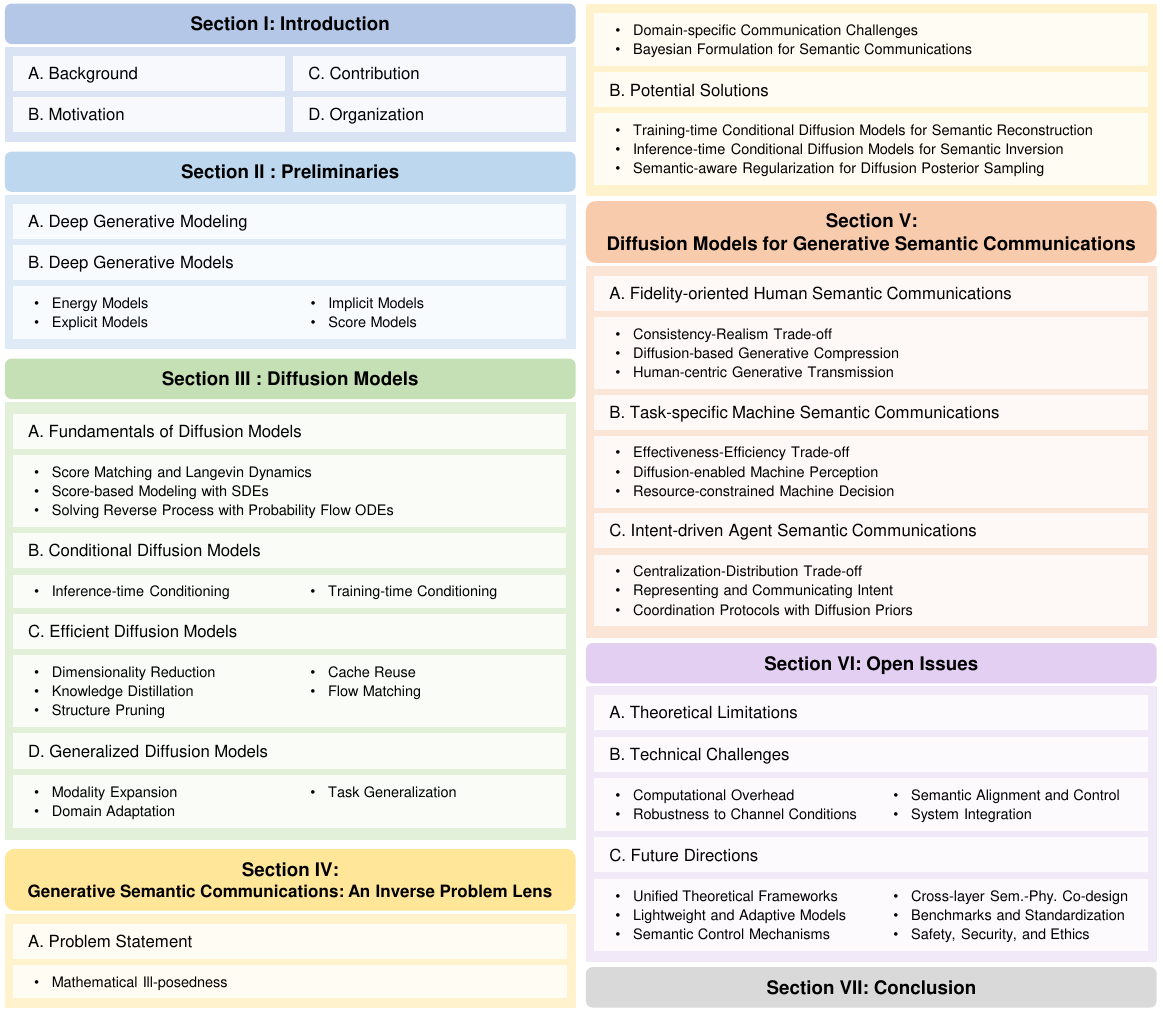}
	\caption{Overview of the article organization. Each colored box represents a major section of the article.}
	\label{fig:outline}
\end{figure*}

Fig.~\ref{fig:outline} presents a visual overview of the article structure, while Table~\ref{tab:terminology-notation} summarizes the primary academic terminology and mathematical notation employed throughout.

\begin{table*}[t]
	\centering
	\caption{Academic terminology and mathematical notation used in this article. All vectors, matrices, and tensors are set in boldface; non-bold symbols denote scalars unless otherwise specified. Symbols not explicitly listed are defined in context at first use.}
	\label{tab:terminology-notation}
	\renewcommand{\arraystretch}{1.5}
	\begin{tabularx}{\textwidth}{>{\raggedright\arraybackslash}p{0.07\textwidth}>{\raggedright\arraybackslash}p{0.35\textwidth}|>{\centering\arraybackslash}p{0.1\textwidth}>{\raggedright\arraybackslash}X}
		\toprule
		\multicolumn{2}{c|}{\textbf{Academic Terminology}} & \multicolumn{2}{c}{\textbf{Mathematical Notation}} \\
		\midrule \midrule
		\textbf{Acronym} & \textbf{Definition} & \textbf{Symbol} & \textbf{Meaning} \\
		\midrule
		\rowcolor{black!3!white}
		DNN & deep neural network & $\mathcal{X}$ & data space for source and reconstructions \\
		AIGC & AI-generated content & $\mathcal{Y}$ & measurement space at the receiver \\
		\rowcolor{black!3!white}
		ARM & autoregressive model & $\mathcal{Z}$ & latent space for semantic representation \\
		VAE & variational autoencoder & $\mathbf{x}\!\in\!\mathcal{X}$ & raw data variable \\
		\rowcolor{black!3!white}
		GAN & generative adversarial network & $\mathbf{y}\!\in\!\mathcal{Y}$ & degraded measurement variable \\
		LDM & latent diffusion model & $\mathbf{z}\!\in\!\mathcal{Z}$ & low-dimensional latent variable \\
		\rowcolor{black!3!white}
		MCMC & Markov chain Monte Carlo & $\hat{\mathbf{x}}_{0|t}$ & posterior mean of clean data at time step $t$ \\
		DSM & denoising score matching & $\boldsymbol{\theta},\,\boldsymbol{\phi},\,\boldsymbol{\psi},\,\boldsymbol{\vartheta}$ & model parameters \\
		\rowcolor{black!3!white}
		SMLD & score matching with Langevin dynamics & $p(\mathbf{x})$ & data distribution or density \\
		DDPM & denoising diffusion probabilistic model & $\boldsymbol{s}_{\boldsymbol{\theta}}(\cdot,t)$ & learned score $\nabla_{\mathbf{x}}\log p_t(\mathbf{x})$ driving reverse sampling \\
		\rowcolor{black!3!white}
		VE SDE & variance exploding stochastic differential equation & $\boldsymbol{\epsilon}_{\boldsymbol{\theta}}(\cdot,t)$ & denoising network predicting noise; $\boldsymbol{\epsilon}_{\boldsymbol{\theta}}=-\sqrt{1-\bar{\alpha}_t}\,\boldsymbol{s}_{\boldsymbol{\theta}}$ \\
		VP SDE & variance preserving stochastic differential equation & $\boldsymbol{v}(\cdot,t)$ & velocity field governing deterministic trajectory evolution \\
		\rowcolor{black!3!white}
		PF ODE & probability flow ordinary differential equation & $\boldsymbol{\psi}(\cdot,t)$ & flow map transporting prior to data distribution \\
		CG & classifier guidance & $\boldsymbol{f}(\cdot,t)$ & vector-valued drift coefficient \\
		\rowcolor{black!3!white}
		DPS & diffusion posterior sampling & $g(t)$ & scalar-valued diffusion coefficient \\
		CFG & classifier-free guidance & $\mathcal{A}(\cdot)$ & general forward operator for inverse problems \\
		\bottomrule
	\end{tabularx}
\end{table*}

%% file: sec/2_preliminary.tex
\section{Preliminaries} \label{sec:pre}

\subsection{Deep Generative Modeling} \label{sec:dgm}
\begin{figure}[t]
	\centering
	\includegraphics[width=\columnwidth]{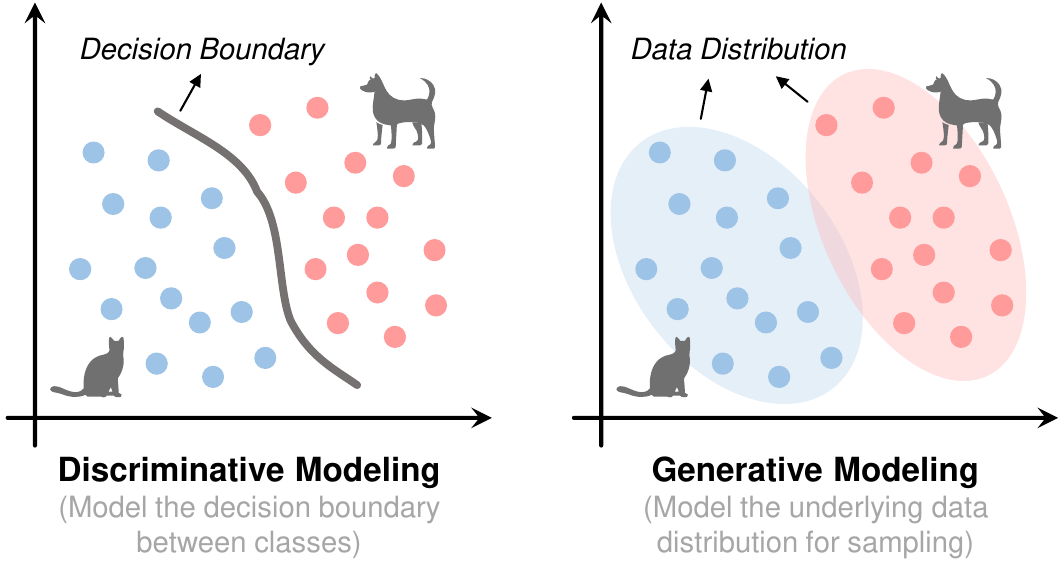}
	\caption{Comparison between discriminative and generative modeling in machine learning. Discriminative models directly learn the mapping from inputs to outputs, while generative models learn the underlying data distribution enabling synthesis.}
	\label{fig:dis-gen}
	\vskip -1.4em
\end{figure}

A major division in machine learning (ML) lies between discriminative and generative modeling (Fig.~\ref{fig:dis-gen}). Discriminative models learn decision boundaries between classes from observations, while generative models learn joint distributions over all variables, fitting the underlying data distribution. Generative models simulate real-world data generation processes and offer two key advantages. First, they enable AI-generated content (AIGC) applications and unsupervised representation learning that extracts disentangled, semantically meaningful, and statistically independent factors of variation~\cite{kingma-introduction}. Second, they can incorporate physical laws and constraints while treating unknown details as noise, making them intuitive and interpretable for validating theories about real-world mechanisms through prediction-observation comparison.

Let $\mathcal{X} \subset \mathbb{R}^D$ denote the data space with dimensionality $D \in \mathbb{N}^+$. The true data distribution $p_{\mathrm{data}}(\mathbf{x}): \mathbb{R}^D \to \mathbb{R}_{\geqslant 0}$ satisfies $\int_{\mathbb{R}^D} p_{\mathrm{data}}(\mathbf{x}) \,\mathrm{d}\mathbf{x} = 1$, where $\mathbf{x} = (x_1, \ldots, x_D)^{\top} \in \mathbb{R}^D$ is a data point. Generative modeling aims to estimate $p_{\mathrm{data}}(\mathbf{x})$ from a dataset $\{ \mathbf{x}_i \}_{i=1}^{N}$ to enable sampling and probability evaluation. A parametric model $p_{\boldsymbol{\theta}}(\mathbf{x}): \mathbb{R}^D \to \mathbb{R}_{\geqslant 0}$ 
with $P \in \mathbb{N}^+$ parameters $\boldsymbol{\theta} \in \Theta \subset \mathbb{R}^P$ 
serves as a proxy for $p_{\mathrm{data}}(\mathbf{x})$. The objective is to find optimal parameters $\boldsymbol{\theta}^{\star}$ such that $p_{\boldsymbol{\theta}^{\star}}(\mathbf{x}) \approx p_{\mathrm{data}}(\mathbf{x})$. In deep learning (DL), such models are parameterized by deep neural networks (DNNs), forming \emph{deep generative models}.

Valid probability distributions require $p_{\boldsymbol{\theta}}(\mathbf{x})$ to satisfy:
\begin{enumerate}[label=\arabic*.]
	\item (Non-negativity) $\forall \mathbf{x} \in \mathbb{R}^D : p_{\boldsymbol{\theta}}(\mathbf{x}) \geqslant 0$.
	\item (Normalization) $\int_{\mathbb{R}^D} p_{\boldsymbol{\theta}}(\mathbf{x}) \,\mathrm{d}\mathbf{x} = 1$.
\end{enumerate}

While non-negativity is straightforward, normalization is challenging~\cite{hyvarinen2005estimation}: it requires integrating over the entire high-dimensional data space, which is typically intractable for complex models, motivating specialized strategies in modern deep generative models to address the normalization challenge.

\subsection{Deep Generative Models} \label{sec:generative-models}
\subsubsection{Energy Models}
When exact normalization is intractable, \emph{approximation} becomes necessary. Energy models~\cite{lecun2006tutorial} parameterize distributions using Boltzmann machines~\cite{ackley1985learning}, inspired by Boltzmann distributions in statistical physics:
\begin{equation}
	p_{\boldsymbol{\theta}}(\mathbf{x}) = \frac{\exp \left( -\beta E_{\boldsymbol{\theta}}(\mathbf{x}) \right)}{Z_{\boldsymbol{\theta}}},
\end{equation}
where $E_{\boldsymbol{\theta}}(\mathbf{x})$ is an energy function with parameter $\boldsymbol{\theta}$. For example, it can be interpreted as the potential energy $-\log p_{\boldsymbol{\theta}}(\mathbf{x})$. $\beta$ is an arbitrary positive constant akin to an inverse temperature, and $Z_{\boldsymbol{\theta}} = \int_{\mathbb{R}^D} \exp \left( -\beta E_{\boldsymbol{\theta}}(\mathbf{x}) \right) \mathrm{d} \mathbf{x}$ is the partition function ensuring the normalization of $p_{\boldsymbol{\theta}}(\mathbf{x})$.

The energy function can be flexibly parameterized by DNNs without normalization constraints, but evaluating $Z_{\boldsymbol{\theta}}$ requires high-dimensional integration, making it intractable. Fortunately, Markov chain Monte Carlo (MCMC)~\cite{van2018simple} enables approximation without computing $Z_{\boldsymbol{\theta}}$ explicitly. Since $\nabla_{\boldsymbol{\theta}} \log Z_{\boldsymbol{\theta}} = -\mathbb{E}_{p_{\boldsymbol{\theta}}(\mathbf{x})} \left[ \nabla_{\boldsymbol{\theta}} E_{\boldsymbol{\theta}}(\mathbf{x}) \right]$, one can sample $\hat{\mathbf{x}} \sim p_{\boldsymbol{\theta}}(\mathbf{x})$ via MCMC and approximate $\nabla_{\boldsymbol{\theta}} \log Z_{\boldsymbol{\theta}} \approx -\nabla_{\boldsymbol{\theta}} E_{\boldsymbol{\theta}}(\hat{\mathbf{x}})$, making training feasible.

While MCMC sampling avoids explicit $Z_{\boldsymbol{\theta}}$ computation, probability evaluation still requires estimating $Z_{\boldsymbol{\theta}}$. Methods like annealed importance sampling~\cite{neal2001annealed} exist but yield inevitable estimation errors, limiting accurate probability computation.

\subsubsection{Explicit Models}
In generative modeling, besides using specific sampling methods for approximation to address the normalization challenge, one can also directly employ an explicit formulation to achieve this approximation. Two representative examples of explicit deep generative models are autoregressive models~\cite{xiong2024autoregressive} and variational autoencoders~\cite{kingma2013auto}, which illustrate different formulation ways of enforcing normalization.

\emph{Autoregressive Models (ARMs):} ARMs leverage the probability \emph{chain rule} to factorize high-dimensional distributions into products of univariate conditionals, which are easier to parameterize and normalize. Specifically, ARMs define:
\begin{equation}
	p_{\boldsymbol{\theta}}(\mathbf{x}) = \prod_{i=1}^{D} p_{\boldsymbol{\theta}}(x_i | \mathbf{x}_{<i}), \label{eq:autoregressive}
\end{equation}
where $D$ is the dimensionality of $\mathbf{x}$, $x_i$ is the $i$-th element, and $\mathbf{x}_{<i} = \{x_1, \ldots, x_{i-1}\}$ with $\mathbf{x}_{<1} := \varnothing$. This factorization guarantees exact normalization of $p_{\boldsymbol{\theta}}(\mathbf{x})$ when each conditional $p_{\boldsymbol{\theta}}(x_i | \mathbf{x}_{<i})$ is normalized.

However, autoregressive factorization requires a sequential ordering of data dimensions. While natural for sequential data (\eg text, audio), many domains lack inherent ordering. For instance, pixel arrangements in images. This constraint limits architectural flexibility, often requiring specialized constructs such as masked convolutions~\cite{van2016pixel}. Consequently, ARMs excel at sequential data generation but face challenges with ultra-high-resolution images or videos.

\emph{Variational Autoencoders (VAEs):} Similar to ARMs, VAEs achieve exact normalization by imposing architectural constraints on the probability distribution. Building on autoencoders (AEs) for data compression, VAEs introduce an auxiliary latent variable $\mathbf{z} \sim p(\mathbf{z})$ to model the data distribution $p(\mathbf{x})$. The framework comprises three components: a tractable prior $p(\mathbf{z})$, an encoder $q_{\boldsymbol{\phi}}(\mathbf{z} | \mathbf{x})$ with parameters $\boldsymbol{\phi} \in \Phi \subset \mathbb{R}^P$, and a decoder $p_{\boldsymbol{\theta}}(\mathbf{x} | \mathbf{z})$ with parameters $\boldsymbol{\theta}$. Together, they define the marginal distribution:
\begin{equation}
	p_{\boldsymbol{\theta}}(\mathbf{x}) = \int_{\mathbb{R}^d} p(\mathbf{z}) p_{\boldsymbol{\theta}}(\mathbf{x} | \mathbf{z}) \mathrm{d} \mathbf{z}.
\end{equation}

This can be interpreted as an infinite mixture model~\cite{kingma-introduction}, where $p(\mathbf{z})$ provides mixture coefficients and $p_{\boldsymbol{\theta}}(\mathbf{x} | \mathbf{z})$ defines mixture components. Normalization of $p_{\boldsymbol{\theta}}(\mathbf{x})$ is guaranteed when both $p(\mathbf{z})$ and $p_{\boldsymbol{\theta}}(\mathbf{x} | \mathbf{z})$ are normalized. VAEs parameterize the encoder and decoder as specific distributions (\eg exponential families) with neural network backbones. The encoder approximates the intractable posterior $p(\mathbf{z} | \mathbf{x})$ via \emph{variational inference}, parameterizing a distribution over latent codes rather than producing deterministic embeddings. Training maximizes the Evidence Lower BOund (ELBO) on the log-likelihood~\cite{kingma2013auto}. Sampling follows ancestral generation: draw $\hat{\mathbf{z}} \sim p(\mathbf{z})$, then sample $\hat{\mathbf{x}} \sim p_{\boldsymbol{\theta}}(\mathbf{x} | \hat{\mathbf{z}})$. While VAEs construct exactly normalized distributions, they require architectural constraints and yield only approximate probability values.

\subsubsection{Implicit Models}
The challenge of normalization originates from modeling probability density or mass functions, thus may be avoided if we take alternative ways to represent the probability distribution implicitly. Below we introduce generative adversarial networks (GANs)~\cite{goodfellow2014generative}, a representative family of implicit deep generative models that directly model the sampling process of a probability distribution, thereby completely sidestepping the difficulty of normalization.

GANs model data generation as a two-step process: first sample $\mathbf{z} \sim p(\mathbf{z})$ from a simple prior (\eg standard Gaussian), then transform it via a deterministic generator $G_{\boldsymbol{\theta}}$ to obtain $\mathbf{x} = G_{\boldsymbol{\theta}}(\mathbf{z})$. The distribution $p_{\boldsymbol{\theta}}(\mathbf{x})$ is implicitly defined without direct parameterization, circumventing normalization concerns.

Training employs an auxiliary discriminator $D_{\boldsymbol{\phi}}: \mathbb{R}^D \to [0, 1]$ that distinguishes real data from generator samples, while the generator aims to fool the discriminator. This adversarial competition forms a two-player game. Unlike explicit models, GANs require no special architectural constraints and can leverage flexible DNNs. However, they cannot provide probability values, limiting applications requiring likelihood evaluation. Additionally, adversarial training often suffers from instability and mode collapse~\cite{salimans2016improved}, where the generator focuses on limited data modes that easily deceive the discriminator, failing to capture full data diversity.

\subsubsection{Score Models} \label{sec:score}
Score models have emerged as a state-of-the-art paradigm that circumvents key limitations of previous deep generative models. Rather than modeling normalized distributions or using adversarial training, score models learn the score function $\boldsymbol{s}(\mathbf{x}) := \nabla_{\mathbf{x}} \log p(\mathbf{x})$~\cite{stein1972bound}, the gradient of the log-density. A neural network estimates this quantity iteratively, yielding $\boldsymbol{s}_{\boldsymbol{\theta}}(\mathbf{x})$ with parameters $\boldsymbol{\theta}$.

The score model is actually a conservative vector field\footnote{In physics, the score $\nabla_{\mathbf{x}}\log p(\mathbf{x})$ corresponds to the negative gradient of potential energy $-\log p(\mathbf{x})$, acting as a ``force'' driving samples toward high-probability regions.} and must be the gradient of a scalar function. This is enforced by parameterizing an energy function $E_{\boldsymbol{\theta}}(\mathbf{x})$ with a neural network, then constructing the score model with  $\boldsymbol{s}_{\boldsymbol{\theta}}(\mathbf{x}) = -\nabla_{\mathbf{x}} E_{\boldsymbol{\theta}}(\mathbf{x})$~\cite{song2021train}. Computing the gradient of a DNN is often efficient with backpropagation or automatic differentiation~\cite{baydin2018automatic}. Therefore, a score model constructed in this way is fast to evaluate.

Crucially, score functions are oblivious to normalization. For unnormalized density $\tilde{p}(\mathbf{x})$ with $\int_{\mathbb{R}^D} \tilde{p}(\mathbf{x}) \,\mathrm{d}\mathbf{x} = Z \ne 1$:
\begin{equation*}
	\boldsymbol{s}(\mathbf{x}) = \nabla_{\mathbf{x}} \log \frac{\tilde{p}(\mathbf{x})}{Z} = \nabla_{\mathbf{x}} \log \tilde{p}(\mathbf{x}) - \underbrace{\nabla_{\mathbf{x}} \log Z}_{=0} = \nabla_{\mathbf{x}} \log \tilde{p}(\mathbf{x}).
\end{equation*}

One of the most representative score-based deep generative models is diffusion models~\cite{sohl2015deep}, which is the main focus of this article.

%% file: sec/3_diffusion-model.tex
\section{Diffusion Models} \label{sec:diffusion}
Diffusion models take their name and motivation from the thermodynamic phenomenon of diffusion~\cite{sohl2015deep}, such as how a drop of ink spreads in water. In physics, diffusion is a gradual, stochastic process that destroys local structure and drives the system toward maximum entropy. Diffusion models mimic this by defining a \emph{forward process} that progressively corrupts data with small increments of Gaussian noise, until all semantic structure is lost and the distribution approaches pure randomness. Crucially, they learn a \emph{reverse process} that inverts this evolution: starting from nearly isotropic noise, the model incrementally ``denoises'' in small steps, guided by estimates of the gradient of the log probability density, thereby reconstructing fine details just as carefully as they were erased.

Before delving into the technical details, it is instructive to position diffusion models among other deep generative models that can serve as semantic decoders at the communication receiver. Energy models~\cite{lecun2006tutorial} face significant practical challenges due to their intractable partition functions and heavy reliance on sophisticated sampling techniques. ARMs~\cite{van2016pixel} excel at discrete token generation with exact likelihoods, but their sequential sampling is computationally expensive for high-dimensional continuous data. VAEs~\cite{kingma2013auto} offer fast sampling through amortized inference and construct tractable approximate posteriors, yet they often produce blurred reconstructions due to the limitations of the variational bound. GANs~\cite{goodfellow2014generative} deliver sharp perceptual quality and rapid single-pass inference, but suffer from mode collapse and training instability that can compromise distribution coverage. In contrast, diffusion models circumvent these limitations by trading higher inference cost for superior distribution coverage, stable training via the denoising score matching objective, flexible conditional sampling through guidance mechanisms, and a natural fit for solving inverse problems in communication scenarios, as will be detailed further. Table~\ref{tbl:dgm-comparison} summarizes this comparison. This tutorial focuses on diffusion models exactly owing to their unique combination of stochastic posterior sampling, conditioning flexibility, and robustness to noisy measurements, all of which are directly aligned with the requirements of generative semantic communications~\cite{dai2024deep}.

\begin{table}[t]
	\centering
	\renewcommand{\arraystretch}{1.5}
	\caption{Qualitative comparison of representative deep generative models from seven aspects. Here, energy models (EMs) and diffusion models (DMs) are abbreviated accordingly.}
	\label{tbl:dgm-comparison}
	\newsavebox{\dgmtablebox}
	\savebox{\dgmtablebox}{
		\begin{tabular}{r | c c c c c c c}
			\toprule
			\textbf{Model} & \textbf{TS} & \textbf{CT} & \textbf{GE} & \textbf{PQ} & \textbf{MD} & \textbf{IP} & \textbf{LE} \\
			\midrule\midrule
			\rowcolor{black!3!white}
			EMs~\cite{lecun2006tutorial}    & \poor   & \poor   & \poor   & \poor   & \medium & \good   & \poor   \\
			\midrule
			ARMs~\cite{van2016pixel}        & \good   & \medium & \poor   & \medium & \good   & \good   & \good   \\
			\rowcolor{black!3!white}
			VAEs~\cite{kingma2013auto}      & \good   & \medium & \good   & \poor   & \medium & \good   & \medium \\
			\midrule
			GANs~\cite{goodfellow2014generative} & \poor & \medium & \good & \good & \poor & \poor & \poor \\
			\midrule
			\rowcolor{black!3!white}
			DMs~\cite{sohl2015deep}         & \good   & \good   & \poor   & \good   & \good   & \good   & \medium \\
			\bottomrule
		\end{tabular}
	}
	\usebox{\dgmtablebox}\\[6pt]
	\parbox{\wd\dgmtablebox}{\scriptsize
		\emph{Abbreviations:} \textbf{TS}: Training stability $\cdot$
		\textbf{CT}: Controllability $\cdot$
		\textbf{GE}: Generation efficiency $\cdot$
		\textbf{PQ}: Perceptual quality $\cdot$
		\textbf{MD}: Mode diversity $\cdot$
		\textbf{IP}: Interpretability $\cdot$
		\textbf{LE}: Likelihood evaluability.
		\emph{Ratings:} \good\;Good $\cdot$ \medium\;Medium $\cdot$ \poor\;Poor.
	}
\end{table}

This section examines diffusion models through four perspectives: Section \ref{sec:fundamentals} presents the mathematical foundations; Section \ref{sec:conditional} discusses conditioning mechanisms; Section \ref{sec:efficient} surveys acceleration techniques; and Section \ref{sec:generalized} explores task generalizations of diffusion models.

\subsection{Fundamentals of Diffusion Models} \label{sec:fundamentals}

\subsubsection{Score Matching and Langevin Dynamics}
As discussed in Section \ref{sec:pre}, learning unnormalized generative models proves challenging due to the intractable partition function $Z_{\boldsymbol{\theta}}$. This raises a natural question: how can we effectively train flexible score-based diffusion models from high-dimensional data? The answer lies in score matching~\cite{hyvarinen2005estimation, lai2025principles}, a well-established method for estimating unnormalized statistical models, and more generally, score models.

Score matching minimizes the distance between the scores of the data and model distributions. Without requiring evaluating the partition functions, score matching operates directly on the scores that remain oblivious to the intractable partition functions. From a statistical perspective, this approach corresponds to minimizing the Fisher divergence between $p_{\mathrm{data}}(\mathbf{x})$ and $p_{\boldsymbol{\theta}}(\mathbf{x})$, which is expressed in terms of scores as follows\footnote{The factor $\frac{1}{2}$ is conventional rather than essential: it simplifies gradient expressions by removing an extra factor of two, and it is consistent with the original score matching formulation~\cite{hyvarinen2005estimation}. Omitting it does not affect the minimizer of the objective, as it only rescales the loss by a constant.}:
\begin{equation}
	D_{F}(p_{\mathrm{data}} \parallel p_{\boldsymbol{\theta}}) :=  \mathbb{E}_{p_{\mathrm{data}}(\mathbf{x})} \left[ \frac{1}{2} \left\| \boldsymbol{s}(\mathbf{x}) - \boldsymbol{s}_{\boldsymbol{\theta}}(\mathbf{x}) \right\|_2^2  \right].  \label{eq:fisher}
\end{equation}

Since $\boldsymbol{s}_{\boldsymbol{\theta}}(\mathbf{x})$ does not involve the unnormalized $Z_{\boldsymbol{\theta}}$, the Fisher divergence is independent of this intractable term. However, directly computing the Fisher divergence remains infeasible as it requires access to the unknown ground-truth data score $\boldsymbol{s}(\mathbf{x})$. To circumvent this issue, the Fisher divergence can be reformulated through integration by parts~\cite{hyvarinen2005estimation}, yielding an equivalent objective that eliminates dependence on the ground-truth score. Specifically, the divergence can be decomposed as $D_{F}(p_{\mathrm{data}} \parallel p_{\boldsymbol{\theta}}) = \mathcal{L}(\boldsymbol{\theta}) + C$, where $C$ remains constant with respect to $\boldsymbol{\theta}$, and the tractable loss function is:
\begin{equation}
	\mathcal{L}(\boldsymbol{\theta}) := \mathbb{E}_{p_{\mathrm{data}}(\mathbf{x})}\left[ \frac{1}{2} \left\| \boldsymbol{s}_{\boldsymbol{\theta}}(\mathbf{x}) \right\|_2^2 + \mathrm{tr}(\nabla_{\mathbf{x}} \boldsymbol{s}_{\boldsymbol{\theta}}(\mathbf{x})) \right],  \label{eq:fisher-loss}
\end{equation}
where $\mathrm{tr}(\cdot)$ denotes the trace of a matrix. While the tractable loss function in Eq. \eqref{eq:fisher-loss} enables score matching without the ground-truth score, its practical implementation requires computing the trace of the Hessian matrix $\mathrm{tr}(\nabla_{\mathbf{x}} \boldsymbol{s}_{\boldsymbol{\theta}}(\mathbf{x}))$, which involves second-order derivatives that are computationally expensive in high-dimensional spaces. To circumvent this computational bottleneck, Denoising Score Matching (DSM)~\cite{vincent2011connection} has emerged as the standard approach for training score-based diffusion models.

DSM reformulates the score matching objective by introducing \emph{controlled noise corruption} to the data. Instead of matching scores on the original data distribution, DSM operates on a noise-corrupted distribution $q(\tilde{\mathbf{x}}) = \int_{\mathbb{R}^D} p_{\mathrm{data}}(\mathbf{x})q(\tilde{\mathbf{x}} | \mathbf{x})\mathrm{d}\mathbf{x}$, where $q(\tilde{\mathbf{x}} | \mathbf{x})$ represents the noise corruption process. The resulting DSM objective becomes:
\begin{equation}
	\mathcal{J}(\boldsymbol{\theta}) := \mathbb{E}_{p_{\mathrm{data}}(\mathbf{x}) q(\tilde{\mathbf{x}} | \mathbf{x})} \left[ \frac{1}{2} \left\| \nabla_{\tilde{\mathbf{x}}} \log q(\tilde{\mathbf{x}} | \mathbf{x}) - \boldsymbol{s}_{\boldsymbol{\theta}}(\tilde{\mathbf{x}}) \right\|_2^2 \right],  \label{eq:dsm}
\end{equation}
which can be evaluated without computing any Hessian matrices. It should be noted that DSM learns the score of the noise-corrupted distribution rather than the original data distribution, with the noise corruption process $q(\tilde{\mathbf{x}} | \mathbf{x})$ typically implemented as additive Gaussian noise for its analytical tractability and desirable theoretical properties.

Once $\boldsymbol{s}_{\boldsymbol{\theta}}(\mathbf{x})$ has been trained, samples can be generated through Langevin dynamics~\cite{welling2011bayesian}, an iterative sampling procedure rooted in statistical physics, originally formulated to describe the Brownian motion of particles in a fluid.

Mathematically, Langevin dynamics implements a discrete MCMC procedure that initializes from an arbitrary prior distribution $\mathbf{x}_0 \sim \pi(\mathbf{x})$ and iteratively updates the sample for $i = 1, 2, \ldots, N$ according to:
\begin{equation}
	\mathbf{x}_{i} = \mathbf{x}_{i-1} + \zeta  \boldsymbol{s}_{\boldsymbol{\theta}}(\mathbf{x}_{i-1} ) + \sqrt{2\zeta} \boldsymbol{\epsilon},
\end{equation}
where $\zeta$ is the step size, $\boldsymbol{\epsilon} \sim \mathcal{N}(\mathbf{0}, \mathbf{I})$ denotes standard Gaussian noise, and $\mathbf{I}$ is the identity matrix. The first term on the right-hand side is the current position $\mathbf{x}_{i-1}$, the second term provides a deterministic drift toward higher probability regions guided by the learned score $\boldsymbol{s}_{\boldsymbol{\theta}}(\mathbf{x}_{i-1} )$, and the third term introduces stochastic perturbations $\boldsymbol{\epsilon}$ that prevent the dynamics from getting trapped in local modes. When $\zeta \to 0$ and $N \to \infty$, the sampling endpoint $\mathbf{x}_N$ converges exactly to the target distribution $p_{\mathrm{data}}(\mathbf{x})$ under mild regularity conditions.

\begin{figure}[t]
	\centering
	\includegraphics[width=\columnwidth]{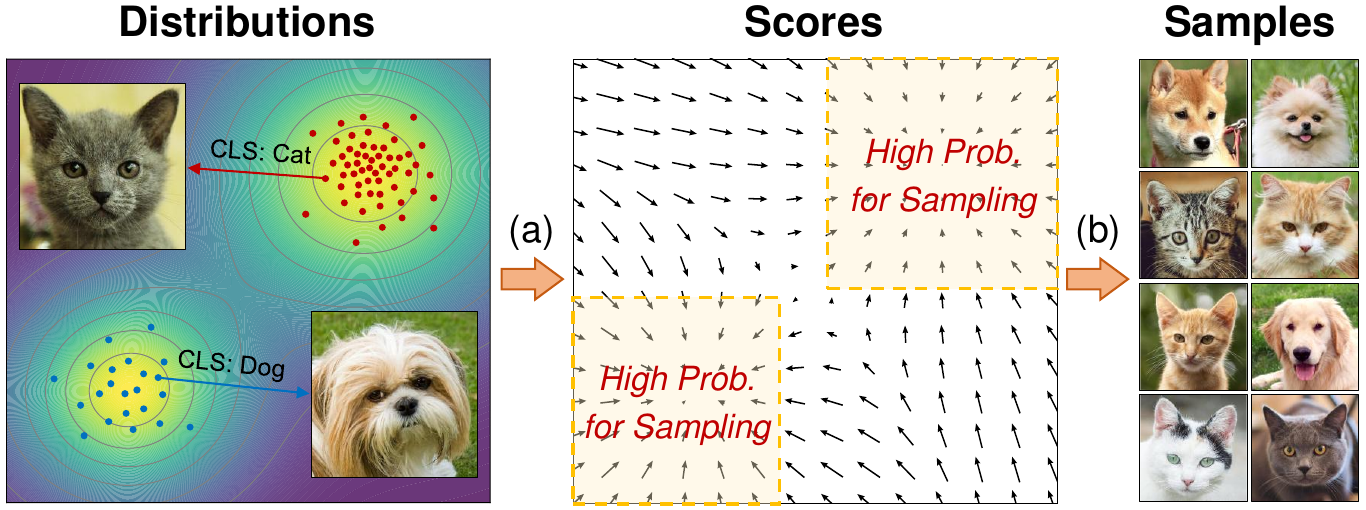}
	\caption{Score-based modeling pipeline for diffusion models. (a) Score matching: The model learns to approximate the score (gradient of log-density, indicated by arrows) of the data distribution through techniques such as denoising score matching. (b) Stochastic sampling: (Annealing) Langevin dynamics generates samples by following the learned score function with stochastic perturbations, where annealing refers to gradually decreasing noise levels during the sampling process.}
	\label{fig:score-pipeline}
\end{figure}

Fig. \ref{fig:score-pipeline} illustrates the overall pipeline of score-based modeling in diffusion models, demonstrating how data distributions are approximated through score matching and subsequently sampled via Langevin dynamics.

\subsubsection{Score-based Modeling with SDEs}
Building upon the standard score-based modeling pipeline for diffusion models, Song \etal~\cite{song2020score} introduced a unified framework that generalizes score matching and sampling procedures through the lens of stochastic differential equations (SDEs)~\cite{anderson1982reverse}. Instead of perturbing data with a finite number of noise distributions, this framework considers a continuum of intermediate distributions evolving over continuous time. Such evolution follows a prescribed SDE, independent of the data and containing no trainable parameters. The reverse-time SDE can then be derived and approximated by training a time-dependent neural network to estimate the score function, thereby enabling sample generation via numerical SDE solvers.

\paragraph{Perturbating Data with Forward SDEs} To establish the connection between discrete recursions and continuous-time SDEs, consider first the discrete gradient descent update for minimizing a differentiable convex function $f(\cdot)$:
\begin{equation}
	\mathbf{x}_{i+1} = \mathbf{x}_i - \beta_i \nabla f(\mathbf{x}_i),
\end{equation}
where $\beta_i$ represents the step size at iteration $i$. Taking the continuous-time limit as the step size approaches zero yields the ordinary differential equation (ODE):
\begin{equation}
	\frac{\mathrm{d}\mathbf{x}(t)}{\mathrm{d}t} = -\beta(t) \nabla f(\mathbf{x}(t)),
\end{equation}
where $\beta(t)$ represents the continuous-time analogue of the discrete step size. From the perspective of generative modeling, deterministic dynamics alone cannot capture the stochastic nature of data distributions~\cite{kingma-introduction}. Incorporating random perturbations into the ODE yields:
\begin{equation}
	\frac{\mathrm{d}\mathbf{x}(t)}{\mathrm{d}t} = -\beta(t) \nabla f(\mathbf{x}(t)) + \mathbf{n}(t),
\end{equation}
where $\mathbf{n}(t)$ represents the stochastic noise. However, $\mathbf{n}(t)$ is not differentiable in the classical sense, requiring a rigorous mathematical treatment through stochastic calculus.

Given the standard Wiener process $\mathbf{w}(t)$ with independent Gaussian increments $\mathbf{w}(t + \Delta t) - \mathbf{w}(t) \sim \mathcal{N}(\mathbf{0}, \Delta t \mathbf{I})$, the noise term can be formally expressed as $\mathbf{n}(t) = \frac{\mathrm{d}\mathbf{w}(t)}{\mathrm{d}t}$. Although this derivative does not exist in the ordinary sense, the integral form is well-defined. Multiplying both sides by $\mathrm{d}t$ and recognizing that $\mathbf{n}(t)\mathrm{d}t = \mathrm{d}\mathbf{w}(t)$ in the It\^{o} calculus framework, we obtain:
\begin{equation}
	\mathrm{d}\mathbf{x}(t) = -\beta(t) \nabla f(\mathbf{x}(t)) \mathrm{d}t + \mathrm{d}\mathbf{w}(t).
\end{equation}

While this equation provides intuition from gradient descent, diffusion models require more general dynamics to effectively transform data distributions into tractable priors. The forward diffusion process $\{\mathbf{x}(t)\}_{t\in[0,T]}$ over the time interval $[0,T]$ is therefore modeled as the solution to the It\^{o} SDE:
\begin{equation}
	\mathrm{d}\mathbf{x} = \boldsymbol{f}(\mathbf{x}, t) \mathrm{d}t + g(t) \mathrm{d}\mathbf{w},  \label{eq:diffusion-forward-sde}
\end{equation}
where the vector-valued drift coefficient $\boldsymbol{f}(\cdot, t): \mathbb{R}^D \to \mathbb{R}^D$ captures the deterministic drift of particles under external forces, pulling them toward a target prior distribution. The scalar-valued diffusion coefficient $g(t): \mathbb{R}_{\geqslant 0} \to \mathbb{R}_{>0}$ quantifies the stochastic diffusion from molecular Brownian motion, which controls the intensity of noise injection at each time step, determining how strongly particles deviate from their deterministic drift trajectories\footnote{This drift-diffusion decomposition directly connects to the Fokker-Planck equation (\aka Kolmogorov's forward equation), which governs how the probability density $p_t(\mathbf{x})$ of the entire ensemble evolves over time.}. Here, each ``particle'' represents the state of a data sample at time $t$, with its position $\mathbf{x}(t)$ in the data space. This generalization allows flexible design of the forward process to ensure that $\mathbf{x}(T)$ converges to a desired prior distribution, typically an isotropic Gaussian.

\paragraph{Sampling Data with Reverse SDEs} Data sample generation proceeds by reversing the forward diffusion process, starting from samples $\mathbf{x}(T) \sim p_T$ and obtaining samples $\mathbf{x}(0) \sim p_0$. The reverse-time diffusion process (\ie sampling process) running backwards in time follows the reverse-time SDE~\cite{song2020score}:
\begin{equation}
	\mathrm{d}\mathbf{x} = [\boldsymbol{f}(\mathbf{x}, t) - g^2(t) \nabla_{\mathbf{x}} \log p_t(\mathbf{x})] \mathrm{d}t + g(t) \mathrm{d}\bar{\mathbf{w}},  \label{eq:diffusion-reverse-sde}
\end{equation}
where $\bar{\mathbf{w}}$ represents a standard Wiener process when time flows in reverse from $T$ to $0$, and $\mathrm{d}t$ is an infinitesimal negative time step. The score function $\nabla_{\mathbf{x}} \log p_t(\mathbf{x})$ of each marginal distribution plays a crucial role in the reverse drift, correcting the forward drift to enable backward evolution. Once the score is known for all $t$, the reverse process can be simulated based on this equation to generate samples from $p_0$.

\begin{figure*}[t]
	\centering
	\includegraphics[width=\textwidth]{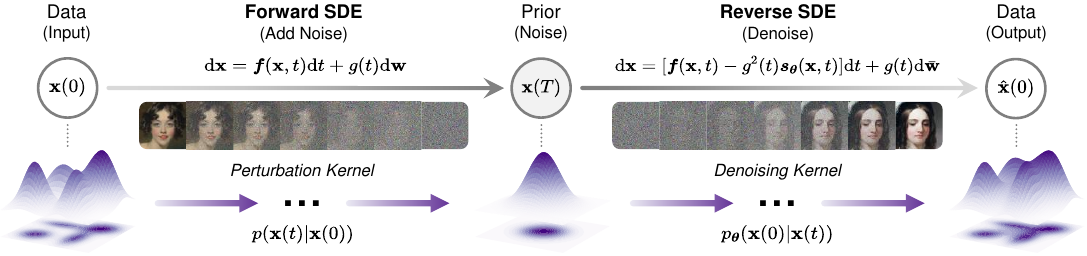}
	\caption{Forward-reverse SDE pipeline for score-based diffusion models. The forward SDE progressively corrupts the input $\mathbf{x}(0)$ into (approximately) Gaussian noise $\mathbf{x}(T)$ drawn from a chosen prior by composing perturbation kernels $p(\mathbf{x}(t) | \mathbf{x}(0))$. The reverse SDE performs iterative denoising guided by the learned score $\boldsymbol{s}_{\boldsymbol{\theta}}(\mathbf{x},t)$, composing denoising kernels $p_{\boldsymbol{\theta}}(\mathbf{x}(0) | \mathbf{x}(t))$ to yield the output $\hat{\mathbf{x}}(0)$. Distributions at representative time slices and their schematic projections highlight the high-probability regions targeted during sampling.}
	\label{fig:sde}
\end{figure*}

Diffusion models exhibit a notable advantage through their stochastic sampling process, achieving better robustness, semantic consistency, and perceptual quality when compared to the deterministic strategies inherent to VAEs and GANs~\cite{ohayon2023reasons}. The overall pipeline of score-based modeling with forward and reverse SDEs in diffusion models is illustrated in Fig.~\ref{fig:sde}.

\begin{table*}[t]
	\centering
	\newcolumntype{M}[1]{>{\centering\arraybackslash}m{#1}}
	\caption{Unified overview of diffusion models through VE and VP SDEs. Both continuous-time SDE formulations and discrete-time recursions are presented. \emph{Notations:} $\mathbf{w}$, $\bar{\mathbf{w}}$: forward and reverse Wiener processes; $\boldsymbol{\epsilon} \sim \mathcal{N}(\mathbf{0}, \mathbf{I})$: Gaussian noise; $\boldsymbol{s}(\mathbf{x}, t) = \nabla_{\mathbf{x}} \log p_t(\mathbf{x})$: score function; $\sigma(t)$, $\sigma_i$: noise schedule for VE SDE; $\beta(t)$, $\beta_i$: noise schedule for VP SDE.}
	\label{tab:sde}
	\renewcommand{\arraystretch}{1.5}
	
	\begin{tabular}{M{2cm} | M{1.6cm} | m{5.9cm} | m{6.9cm}}
		\toprule
		\textbf{Model} & \textbf{Process} & \textbf{Continuous-time SDE} & \textbf{Discrete-time Recursion} \\
		\midrule\midrule
		
		\multirow{2.5}{*}{\makecell[c]{VE SDE\\[0.2em]{(\eg SMLD)}}} 
		& Forward & 
		$\mathrm{d} \mathbf{x} = \sqrt{\frac{\mathrm{d}[\sigma^2(t)]}{\mathrm{d}t}} \mathrm{d} \mathbf{w}$ & 
		$\mathbf{x}_i = \mathbf{x}_{i-1} + \sqrt{\sigma_i^2 - \sigma_{i-1}^2} \boldsymbol{\epsilon}$ \\
		\cmidrule{2-4}
		& Reverse & 
		$\mathrm{d} \mathbf{x} = -\frac{\mathrm{d}[\sigma^2(t)]}{\mathrm{d}t} \boldsymbol{s}(\mathbf{x}, t) \mathrm{d}t + \sqrt{\frac{\mathrm{d}[\sigma^2(t)]}{\mathrm{d}t}} \mathrm{d} \bar{\mathbf{w}}$ & 
		$\mathbf{x}_{i-1} = \mathbf{x}_i + (\sigma_i^2 - \sigma_{i-1}^2) \boldsymbol{s}(\mathbf{x}_i) + \sqrt{\sigma_i^2 - \sigma_{i-1}^2} \boldsymbol{\epsilon}$ \\
		
		\midrule
		
		\multirow{2.5}{*}{\makecell[c]{VP SDE\\[0.2em]{(\eg DDPM)}}} 
		& Forward & 
		$\mathrm{d} \mathbf{x} = -\frac{1}{2} \beta(t) \mathbf{x} \mathrm{d}t + \sqrt{\beta(t)} \mathrm{d} \mathbf{w}$ & 
		$\mathbf{x}_i = \sqrt{1-\beta_i} \mathbf{x}_{i-1} + \sqrt{\beta_i} \boldsymbol{\epsilon}$ \\
		\cmidrule{2-4}
		& Reverse & 
		$\mathrm{d} \mathbf{x} = -\beta(t) \left[\frac{1}{2} \mathbf{x} + \boldsymbol{s}(\mathbf{x}, t) \right] \mathrm{d}t + \sqrt{\beta(t)} \mathrm{d} \bar{\mathbf{w}}$ & 
		$\mathbf{x}_{i-1} = \frac{1}{\sqrt{1-\beta_i}} \left[\mathbf{x}_i + \frac{\beta_i}{2} \boldsymbol{s}(\mathbf{x}_i) \right] + \sqrt{\beta_i} \boldsymbol{\epsilon}$ \\
		
		\bottomrule
	\end{tabular}
\end{table*}

\paragraph{Unifying Diffusion Models with VE and VP SDEs} Further, by extending the denoising score matching objective in Eq.~\eqref{eq:dsm} to the continuous-time setting, we can unify diffusion models through two principal SDE formulations: Variance Exploding (VE) and Variance Preserving (VP) SDEs~\cite{song2020score}.

In particular, the VE formulation corresponds to Denoising Score Matching with Langevin Dynamics (SMLD)~\cite{song2019generative}, while the VP formulation underlies Denoising Diffusion Probabilistic Models (DDPMs)~\cite{ho2020denoising}. Both approaches can be understood as discretizations of their respective continuous-time SDEs. Table~\ref{tab:sde} presents a unified overview of diffusion models through VE and VP SDEs. The key distinction between VE and VP SDEs lies in their treatment of variance evolution. While VE SDEs allow unbounded variance growth through pure diffusion, VP SDEs maintain unit variance asymptotically through the balance between contraction and injection.

To help readers bridge the discrete recursions summarized in Table~\ref{tab:sde} with their continuous-time counterparts, we provide derivations for both VE and VP cases in the following remarks. Both derivations proceed in four steps: \emph{(i)}~re-parameterize the discrete noise schedule as continuous functions of time; \emph{(ii)}~Taylor-expand the discrete forward recursion and take the step size $\Delta t \to 0$ to identify the drift and diffusion coefficients of the forward SDE; \emph{(iii)}~substitute these coefficients into reverse-time SDE to obtain the corresponding reverse dynamics; and (iv)~re-discretize the resulting reverse SDE to verify that the original discrete sampling recursion is recovered.

\begin{remark}[Deriving Continuous-time VE SDE from SMLD]
	\hspace{0.5em} SMLD~\cite{song2019generative} estimates scores at each noise scale and samples via Langevin dynamics with progressively decreasing noise. Consider $N$ noise levels $\{\sigma_i\}_{i=1}^N$; the discrete forward recursion follows a Markov chain, for $i = 1, 2, \ldots, N$:
	\begin{equation}
		\mathbf{x}_i = \mathbf{x}_{i-1} + \sqrt{\sigma_i^2 - \sigma_{i-1}^2}\, \boldsymbol{\epsilon}, \quad \boldsymbol{\epsilon} \sim \mathcal{N}(\mathbf{0}, \mathbf{I}).
	\end{equation}
	
	\hspace{0.5em} In the continuous-time limit, let $\{\sigma_i\}_{i=1}^N$ become $\sigma(t)$ for $t \in [0,1]$, and $\{\mathbf{x}_i\}_{i=1}^N$ become $\mathbf{x}(t)$ with $\mathbf{x}_i = \mathbf{x}(i/N)$. The increment then reads $\mathbf{x}(t+\Delta t) - \mathbf{x}(t) = \sqrt{\sigma^2(t+\Delta t) - \sigma^2(t)}\,\boldsymbol{\epsilon} \approx \sqrt{\frac{\mathrm{d}[\sigma^2(t)]}{\mathrm{d}t}\Delta t}\,\boldsymbol{\epsilon}$. As $\Delta t \to 0$, this yields the VE forward SDE:
	\begin{equation}
		\mathrm{d}\mathbf{x} = \underbrace{\sqrt{\frac{\mathrm{d}[\sigma^2(t)]}{\mathrm{d}t}}}_{g(t)}\,\mathrm{d}\mathbf{w},
	\end{equation}
	with zero drift $\boldsymbol{f}(\mathbf{x},t) = \mathbf{0}$. Mapping into the reverse-time SDE in Eq.~\eqref{eq:diffusion-reverse-sde}:
	\begin{equation}
		\mathrm{d}\mathbf{x} = -\frac{\mathrm{d}[\sigma^2(t)]}{\mathrm{d}t}\,\boldsymbol{s}(\mathbf{x},t)\,\mathrm{d}t + \sqrt{\frac{\mathrm{d}[\sigma^2(t)]}{\mathrm{d}t}}\,\mathrm{d}\bar{\mathbf{w}}.
	\end{equation}
	
	\hspace{0.5em} To verify consistency, define $\alpha(t) = \frac{\mathrm{d}[\sigma^2(t)]}{\mathrm{d}t}$ and discretize this reverse SDE with $\Delta t = 1/N$:
	\begin{align}
		\mathbf{x}(t+\Delta t) - \mathbf{x}(t) &= -\alpha(t)\,\boldsymbol{s}(\mathbf{x},t)\,\Delta t - \sqrt{\alpha(t)\Delta t}\,\boldsymbol{\epsilon} \\
		\implies\quad \mathbf{x}_{i-1} &= \mathbf{x}_i + (\sigma_i^2 - \sigma_{i-1}^2)\,\boldsymbol{s}(\mathbf{x}_i) + \sqrt{\sigma_i^2 - \sigma_{i-1}^2}\,\boldsymbol{\epsilon},\nonumber
	\end{align}
	which is identical to the SMLD reverse recursion (ancestral sampling) listed in Table~\ref{tab:sde}. Since $\sigma(t) \to \infty$ as $t \to \infty$, the variance of the diffusion process grows without bound, hence the name VE SDE.
\end{remark}

\begin{remark}[Deriving Continuous-time VP SDE from DDPM]
	\hspace{0.5em} DDPM~\cite{ho2020denoising} learns to reverse each noise perturbation step using the functional form of the reverse distributions. The discrete forward Markov chain with perturbation kernels $\{p(\mathbf{x}_i|\mathbf{x}_0)\}_{i=1}^N$ reads, for $i = 1, 2, \ldots, N$:
	\begin{equation}
		\mathbf{x}_i = \sqrt{1-\beta_i}\,\mathbf{x}_{i-1} + \sqrt{\beta_i}\,\boldsymbol{\epsilon}, \quad \boldsymbol{\epsilon} \sim \mathcal{N}(\mathbf{0}, \mathbf{I}).
	\end{equation}
	
	\hspace{0.5em} Define $\Delta t = 1/N$ and an auxiliary noise schedule $\{\bar{\beta}_i\}_{i=1}^N$ such that $\beta_i = \bar{\beta}_i \Delta t = \beta(t+\Delta t)\Delta t$. As $N \to \infty$, $\bar{\beta}_i \to \beta(t)$ becomes a continuous function for $t \in [0,1]$. Setting $\mathbf{x}_i = \mathbf{x}(t+\Delta t)$ and applying the first-order Taylor expansion $\sqrt{1-\beta(t)\Delta t} \approx 1 - \frac{1}{2}\beta(t)\Delta t$, we obtain:
	\begin{equation}
		\mathbf{x}(t+\Delta t) - \mathbf{x}(t) = -\frac{1}{2}\beta(t)\,\mathbf{x}(t)\,\Delta t + \sqrt{\beta(t)\Delta t}\,\boldsymbol{\epsilon}.
	\end{equation}
	
	\hspace{0.5em} Taking $\Delta t \to 0$ yields the VP forward SDE:
	\begin{equation}
		\mathrm{d}\mathbf{x} = \underbrace{-\frac{1}{2}\beta(t)\,\mathbf{x}}_{\boldsymbol{f}(\mathbf{x},t)}\,\mathrm{d}t + \underbrace{\sqrt{\beta(t)}}_{g(t)}\,\mathrm{d}\mathbf{w}.
	\end{equation}
	
	\hspace{0.5em} Mapping into the reverse-time SDE in Eq.~\eqref{eq:diffusion-reverse-sde}:
	\begin{equation}
		\mathrm{d}\mathbf{x} = -\beta(t)\left[\frac{1}{2}\mathbf{x} + \boldsymbol{s}(\mathbf{x},t)\right]\mathrm{d}t + \sqrt{\beta(t)}\,\mathrm{d}\bar{\mathbf{w}}.
	\end{equation}
	
	\hspace{0.5em} To verify consistency, discretize with $\mathbf{x}(t-\Delta t) = \mathbf{x}_{i-1}$, $\mathbf{x}(t) = \mathbf{x}_i$, and $\beta(t)\Delta t = \beta_i \ll 1$:
	\begin{align}
		\mathbf{x}_{i-1} &= \left(1 + \frac{\beta_i}{2}\right)\left[\mathbf{x}_i + \frac{\beta_i}{2}\,\boldsymbol{s}(\mathbf{x}_i)\right] + \sqrt{\beta_i}\,\boldsymbol{\epsilon}\nonumber \\
		&\approx \frac{1}{\sqrt{1-\beta_i}}\left[\mathbf{x}_i + \frac{\beta_i}{2}\,\boldsymbol{s}(\mathbf{x}_i)\right] + \sqrt{\beta_i}\,\boldsymbol{\epsilon},
	\end{align}
	which is identical to the DDPM reverse recursion (ancestral sampling) listed in Table~\ref{tab:sde}. Unlike VE SDEs, the linear contraction term $-\frac{1}{2}\beta(t)\mathbf{x}$ counteracts noise injection, so that when the initial distribution has unit variance and $t \to \infty$, the marginal variance remains bounded at one, hence the name VP.
\end{remark}

Both VE and VP SDEs have affine drift coefficients, so their perturbation kernels $p(\mathbf{x}(t)|\mathbf{x}(0))$ are Gaussian and admit closed-form expressions, making training particularly efficient.

\subsubsection{Solving Reverse Process with Probability Flow ODEs} \label{sec:ode-solver}
The reverse-time SDE in Eq.~\eqref{eq:diffusion-reverse-sde} generates samples through iterative stochastic denoising. While stochasticity ensures sample diversity, it introduces computational overhead and limits controllability for tasks requiring determinism (\eg data compression, semantic manipulation). This motivates a deterministic alternative.

For every score-based reverse SDE, there exists a corresponding probability flow ordinary differential equation (PF ODE) that shares the same marginal densities $\{p_t(\mathbf{x})\}_{t\in[0,T]}$:
\begin{equation}
	\mathrm{d} \mathbf{x} = \left[ \boldsymbol{f}(\mathbf{x}, t) - \frac{1}{2} g^2(t) \boldsymbol{s}(\mathbf{x}, t) \right] \mathrm{d}t.  \label{eq:pf-ode}
\end{equation}

With a learned score $\boldsymbol{s}_{\boldsymbol{\theta}}(\mathbf{x}, t)$, Eq.~\eqref{eq:pf-ode} becomes a neural ODE~\cite{chen2018neural} solvable via numerical methods. Three families of solvers are commonly used, illustrated on the scalar ODE:
\begin{equation}
	\frac{\mathrm{d}x(t)}{\mathrm{d}t} = f(x(t), t), \quad x(t_0) = x_0. \label{eq:ref-scalar}
\end{equation}

\paragraph{Euler-Maruyama Method} The simplest first-order approach iterates as:
\begin{equation}
	x_{i+1} = x_i + \eta f(x_i, t_i), \quad i = 0, 1, \ldots, N-1,
\end{equation}
where $\eta = t_{i+1} - t_i$ is the step size, yielding local error $\mathcal{O}(\eta^2)$. For diffusion models, this corresponds to ancestral sampling in SMLD and DDPM, offering computational efficiency.

\paragraph{Runge-Kutta Method} Higher-order RK methods (\eg second-order RK, \aka Heun's method) improve accuracy via multiple intermediate evaluations. The classical RK-4 iterates as:
\begin{equation}
	x_{i+1} = x_i + \frac{\eta}{6} \left( k_1 + 2 k_2 + 2 k_3 + k_4 \right),
\end{equation}
where $k_1, k_2, k_3, k_4$ are intermediate slopes with respect to $\eta$. RK-4 achieves local error $\mathcal{O}(\eta^5)$, providing higher sample quality at the cost of additional score evaluations.

\begin{figure}[t]
	\centering
	\includegraphics[width=\columnwidth]{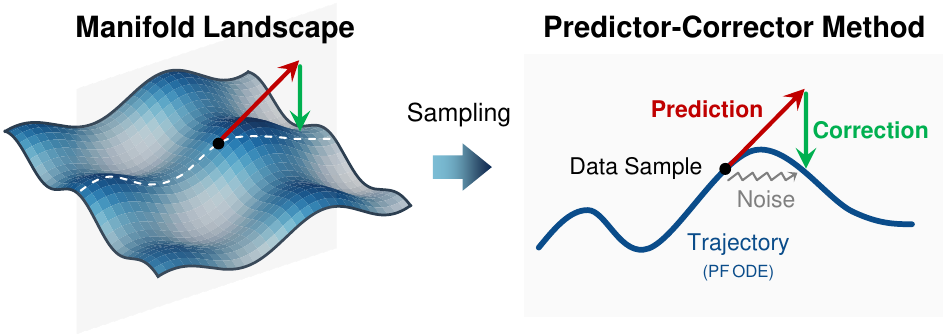}
	\caption{Solving probability flow ODEs with the Predictor-Corrector method. The method follows a two-phase philosophy: prediction provides a coarse estimate as prior knowledge, while correction performs score-based refinement for the estimates.}
	\label{fig:pc}
\end{figure}

\paragraph{Predictor-Corrector Method} This method alternates between a prediction step (any numerical solver for the reverse-time SDE from Table~\ref{tab:sde}) and a correction step (any score-based MCMC approach), balancing efficiency and quality as illustrated in Fig.~\ref{fig:pc}.

\begin{remark}[Sampling as Numerical Solving]
	\hspace{0.5em} The discussion above reveals a unifying perspective: the sampling process of diffusion models is fundamentally the numerical solution of a differential equation, whether the stochastic reverse SDE in Eq.~\eqref{eq:diffusion-reverse-sde} or the deterministic PF ODE in Eq.~\eqref{eq:pf-ode}. Each sampler is essentially a numerical solver, and different samplers correspond to different discretization schemes with distinct accuracy and efficiency trade-offs.
	
	\hspace{0.5em} This viewpoint clarifies why early diffusion models such as DDPM require a thousand steps: their ancestral sampling implements a first-order Euler discretization, which demands small step sizes to keep truncation errors manageable. Subsequent advances in fast sampling, including DDIM~\cite{song2020denoising} and DPM-Solver~\cite{lu2022dpm}, can be understood as importing mature techniques from numerical analysis, such as higher-order solvers and adaptive step sizes, into the diffusion ODE/SDE framework. By improving the accuracy per step, these methods achieve comparable generation quality with far fewer function evaluations, striking a more favorable balance between step count and approximation error.
	
	\hspace{0.5em} From this ``sampler as solver'' perspective, reducing the number of sampling steps introduces three categories of error that jointly govern output quality: \emph{(i)}~\emph{discretization error} from finite step sizes, which grows with step length as the solver can only approximate the continuous trajectory; \emph{(ii)}~\emph{score estimation error} from the neural network's imperfect approximation of the true score, which amplifies when fewer steps leave less opportunity for self-correction; and \emph{(iii)}~\emph{stochastic error} arising from the random noise injected at each SDE step, whose variance scales with step size as $g(t)\sqrt{\Delta t}\,\boldsymbol{\epsilon}$, making large-step SDE sampling inherently more volatile. The ODE formulation eliminates the third source entirely by removing stochasticity, which is precisely why ODE-based samplers tolerate larger step sizes and achieve effective acceleration. The goal of all acceleration techniques surveyed in Section~\ref{sec:efficient} can therefore be understood as controlling these three error sources under a reduced computational budget.
	
	\hspace{0.5em} A related question concerns \emph{the role of stochasticity in SDE sampling}. A common misconception is that the noise injected at each reverse SDE step is the primary source of sample diversity. In fact, the fundamental source of diversity is the random initial sample $\mathbf{x}(T) \sim p_T$: different starting points traverse different trajectories regardless of whether the solver is stochastic or deterministic. The per-step noise in SDE sampling instead serves a corrective function, helping the trajectory explore nearby probability mass and compensate for accumulated score estimation errors.
\end{remark}

\subsection{Conditional Diffusion Models} \label{sec:conditional}

Having established the mathematical foundations of score-based diffusion models, we now address the central question of \emph{controllability}: how to steer the reverse process of diffusion models so that generated samples conform to the task intent. Unconditional diffusion models excel at perceptual quality and diversity, yet they do not guarantee semantic or structural alignment with a target. Ironically, what these models lack (\ie controllability) is precisely the requirement of semantic communications. In semantic communications, reconstructions at the receiver must be semantically equivalent to the transmitter’s source, conditioned on side information that governs what is essential. This supply-demand mismatch underscores the necessity of introducing conditional diffusion models.

We classify conditioning mechanisms in diffusion models by injection time, namely, by when the conditioning signal $\mathbf{y}$ enters the generative pipeline. Table~\ref{tab:conditional-diffusion-models} presents a comprehensive taxonomy with representative methods from both categories. \emph{Inference-time conditioning} injects $\mathbf{y}$ only during sampling through external guidance fields, preserving the pre-trained unconditional model while enabling plug-and-play adaptation to diverse downstream tasks. \emph{Training-time conditioning} incorporates $\mathbf{y}$ during learning by jointly optimizing conditional and unconditional scores, yielding tighter control at the cost of computational investment. We examine each category in detail.

\begin{table*}[t]
	\caption{Taxonomy of conditional diffusion models categorized by the timing of condition injection. Inference-time methods introduce external guidance only during the sampling process, avoiding model retraining, while training-time methods incorporate conditioning directly into model learning, thereby reshaping the inductive bias of diffusion models toward the target distribution.}
	\label{tab:conditional-diffusion-models}
	\centering
	\renewcommand{\arraystretch}{1.5}
	\begin{tabularx}{\textwidth}{C{1.5cm}|C{0.8cm}|r|R|C{7.2cm}|C{1.5cm}}
		\toprule
		\textbf{Category} & \textbf{\#} & \textbf{Related Work} & \textbf{Venue} & \textbf{Illustration \& Canonical Formula} & \textbf{Links} \\
		\midrule\midrule
		
		\multirow{15}{*}{\rotatebox{90}{\makecell{\textsc{Inference-time}\\\textsc{Conditional Diffusion Models}}}}
		& \cellcolor{black!3!white} 1  & \cellcolor{black!3!white} CG \cite{dhariwal2021diffusion} & \cellcolor{black!3!white} NeurIPS'21 &
		\multirow{15}{*}{\begin{minipage}[c][6.5cm][c]{\linewidth}\centering
				\includegraphics[height=6cm]{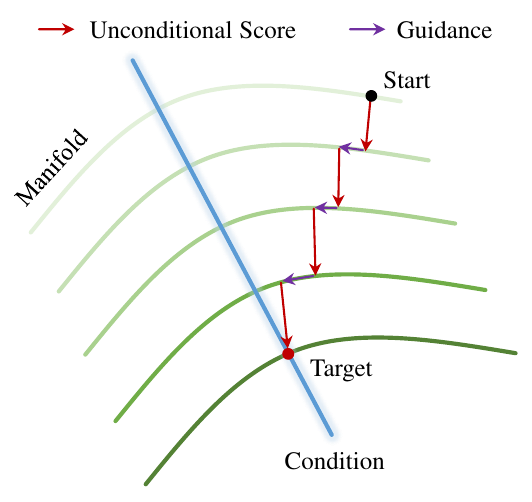}\\[0.2em]
				$\displaystyle
				\boldsymbol{s}(\mathbf{x}|\mathbf{y},t)\approx
				\boldsymbol{s}_{\boldsymbol{\theta}}(\mathbf{x},t)
				+\gamma\boldsymbol{g}(\mathbf{y}|\mathbf{x},t)$
		\end{minipage}}
		& \cellcolor{black!3!white} \href{https://github.com/openai/guided-diffusion}{\github} \\
		& 2  & ILVR \cite{choi2021ilvr} & ICCV'21 & & \href{https://github.com/jychoi118/ilvr_adm}{\github} \\
		& \cellcolor{black!3!white} 3  & \cellcolor{black!3!white} SDEdit \cite{meng2022sdedit} & \cellcolor{black!3!white} ICLR'22 & & \cellcolor{black!3!white} \href{https://github.com/ermongroup/SDEdit}{\github}\ \href{https://sde-image-editing.github.io}{\link} \\
		& 4  & RePaint \cite{lugmayr2022repaint} & CVPR'22 & & \href{https://github.com/andreas128/RePaint}{\github} \\
		& \cellcolor{black!3!white} 5  & \cellcolor{black!3!white} DDRM \cite{kawar2022denoising} & \cellcolor{black!3!white} NeurIPS'22 & & \cellcolor{black!3!white} \href{https://github.com/bahjat-kawar/ddrm}{\github}\ \href{https://ddrm-ml.github.io/}{\link} \\
		& 6  & MCG \cite{chung2022improving} & NeurIPS'22 & & \href{https://github.com/hyungjin-chung/MCG_diffusion}{\github} \\
		& \cellcolor{black!3!white} 7  & \cellcolor{black!3!white} FreeDoM \cite{yu2023freedom} & \cellcolor{black!3!white} ICCV'23 & & \cellcolor{black!3!white} \href{https://github.com/vvictoryuki/FreeDoM}{\github} \\
		& 8  & DG \cite{kim2023dg} & ICML'23 & & \href{https://github.com/alsdudrla10/DG}{\github} \\
		& \cellcolor{black!3!white} 9  & \cellcolor{black!3!white} DPS \cite{chung2022diffusion} & \cellcolor{black!3!white} ICLR'23 & & \cellcolor{black!3!white} \href{https://github.com/DPS2022/diffusion-posterior-sampling}{\github}\ \href{https://dps2022.github.io/diffusion-posterior-sampling-page/}{\link} \\
		& 10 & $\Pi$GDM \cite{song2023pseudoinverseguided} & ICLR'23 & & \href{https://github.com/HatimRabet/PiGDM}{\github} \\
		& \cellcolor{black!3!white} 11 & \cellcolor{black!3!white} PSLD \cite{rout2024solving} & \cellcolor{black!3!white} NeurIPS'23 & & \cellcolor{black!3!white} \href{https://github.com/LituRout/PSLD}{\github} \\
		& 12 & RED-diff \cite{mardani2023variational} & ICLR'24 & & \href{https://github.com/NVlabs/RED-diff}{\github} \\
		& \cellcolor{black!3!white} 13 & \cellcolor{black!3!white} DeqIR \cite{cao2024deep} & \cellcolor{black!3!white} CVPR'24 & & \cellcolor{black!3!white} \href{https://github.com/caojiezhang/DeqIR}{\github} \\
		& 14 & DAPS \cite{zhang2025improving} & CVPR'25 & & \href{https://github.com/zhangbingliang2019/DAPS}{\github}\ \href{https://daps-inverse-problem.github.io/}{\link} \\
		& \cellcolor{black!3!white} 15 & \cellcolor{black!3!white} SITCOM \cite{alkhouri2024sitcom} & \cellcolor{black!3!white} ICML'25 & & \cellcolor{black!3!white} \href{https://github.com/sjames40/SITCOM}{\github} \\
		\midrule
		
		\multirow{15}{*}{\rotatebox{90}{\makecell{\textsc{Training-time}\\\textsc{Conditional Diffusion Models}}}}
		& \cellcolor{black!3!white} 1  & \cellcolor{black!3!white} CFG \cite{ho2022classifier} & \cellcolor{black!3!white} NeurIPS'21 &
		\multirow{15}{*}{\begin{minipage}[c][6.5cm][c]{\linewidth}\centering
				\includegraphics[height=6cm]{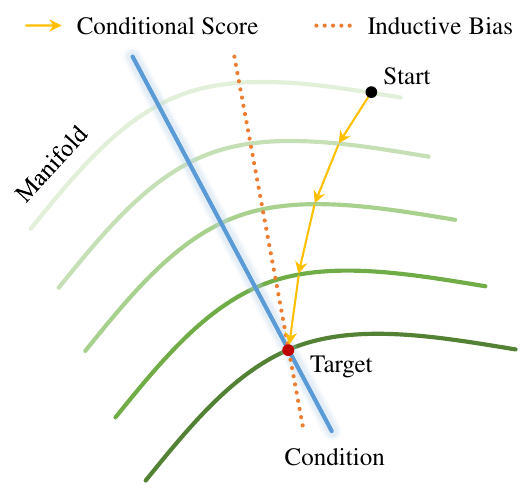}\\[0.2em]
				$\displaystyle
				\boldsymbol{s}(\mathbf{x}|\mathbf{y},t)\approx
				(1-\gamma)\boldsymbol{s}_{\boldsymbol{\theta}}(\mathbf{x},t)
				+\gamma\boldsymbol{s}_{\boldsymbol{\theta}}(\mathbf{x}|\mathbf{y},t)$
		\end{minipage}}
		& \cellcolor{black!3!white} \href{https://github.com/lucidrains/classifier-free-guidance-pytorch}{\github} \\
		& 2  & LDM \cite{rombach2022high} & CVPR'22 & & \href{https://github.com/CompVis/latent-diffusion}{\github}\ \href{https://huggingface.co/CompVis}{\huggingface} \\
		& \cellcolor{black!3!white} 3  & \cellcolor{black!3!white} GLIGEN \cite{li2023gligen} & \cellcolor{black!3!white} CVPR'23 & & \cellcolor{black!3!white} \href{https://github.com/gligen/GLIGEN}{\github}\ \href{https://gligen.github.io/}{\link} \\
		& 4  & InstructPix2Pix \cite{brooks2023instruct} & CVPR'23 & & \href{https://github.com/timothybrooks/instruct-pix2pix}{\github}\ \href{https://www.timothybrooks.com/instruct-pix2pix}{\link} \\
		& \cellcolor{black!3!white} 5  & \cellcolor{black!3!white} Shap-E \cite{jun2023shape} & \cellcolor{black!3!white} arXiv'23 & & \cellcolor{black!3!white} \href{https://github.com/openai/shap-e}{\github}\ \href{https://huggingface.co/openai/shap-e}{\huggingface} \\
		& 6  & DiT \cite{peebles2023dit} & ICCV'23 & & \href{https://github.com/facebookresearch/DiT}{\github}\ \href{https://www.wpeebles.com/DiT}{\link} \\
		& \cellcolor{black!3!white} 7  & \cellcolor{black!3!white} MDT \cite{gao2023mdtv2} & \cellcolor{black!3!white} ICCV'23 & & \cellcolor{black!3!white} \href{https://github.com/sail-sg/MDT}{\github} \\
		& 8  & ControlNet \cite{zhang2023adding} & ICCV'23 & & \href{https://github.com/lllyasviel/ControlNet}{\github}\ \href{https://huggingface.co/lllyasviel/ControlNet}{\huggingface} \\
		& \cellcolor{black!3!white} 9  & \cellcolor{black!3!white} T2I-Adapter \cite{mou2024t2i} & \cellcolor{black!3!white} AAAI'24 & & \cellcolor{black!3!white} \href{https://github.com/TencentARC/T2I-Adapter}{\github}\ \href{https://huggingface.co/TencentARC}{\huggingface} \\
		& 10 & AnimateDiff \cite{guo2023animatediff} & ICLR'24 & & \href{https://github.com/guoyww/AnimateDiff}{\github}\ \href{https://animatediff.github.io/}{\link} \\
		& \cellcolor{black!3!white} 11 & \cellcolor{black!3!white} LVD \cite{lian2023llmgroundedvideo} & \cellcolor{black!3!white} ICLR'24 & & \cellcolor{black!3!white} \href{https://github.com/TonyLianLong/LLM-groundedVideoDiffusion}{\github}\ \href{https://llm-grounded-video-diffusion.github.io/}{\link} \\
		& 12 & SEINE \cite{chen2023seine} & ICLR'24 & & \href{https://github.com/Vchitect/SEINE}{\github}\ \href{https://huggingface.co/Vchitect/SEINE/tree/main}{\huggingface}\
		\href{https://vchitect.github.io/SEINE-project}{\link} \\
		& \cellcolor{black!3!white} 13 & \cellcolor{black!3!white} PixArt-$\Sigma$ \cite{chen2024pixart} & \cellcolor{black!3!white} ECCV'24 & & \cellcolor{black!3!white} \href{https://github.com/PixArt-alpha/PixArt-sigma}{\github}\ \href{https://huggingface.co/PixArt-alpha/pixart_sigma_sdxlvae_T5_diffusers}{\huggingface}\
		\href{https://pixart-alpha.github.io/PixArt-sigma-project}{\link} \\
		& 14 & DDO \cite{zheng2025direct} & ICML'25 & & \href{https://github.com/NVlabs/DDO}{\github}\ \href{https://huggingface.co/nvidia/DirectDiscriminativeOptimization}{\huggingface}\ \href{https://research.nvidia.com/labs/dir/ddo}{\link} \\
		& \cellcolor{black!3!white} 15 & \cellcolor{black!3!white} T2V-Turbo-v2 \cite{li2024t2v} & \cellcolor{black!3!white} ICLR'25 & & \cellcolor{black!3!white} \href{https://github.com/Ji4chenLi/t2v-turbo}{\github}\ \href{https://t2v-turbo-v2.github.io/}{\link} \\
		\bottomrule
		\end{tabularx}
\end{table*}

\subsubsection{Inference-time Conditioning}
Conditional diffusion models extend the unconditional baseline by learning the conditional score $\boldsymbol{s}(\mathbf{x}|\mathbf{y},t)$ that characterizes the conditional distribution $p_t(\mathbf{x}|\mathbf{y})$. A natural strategy emerges: rather than retraining the entire model, can we transform a pre-trained unconditional diffusion model into a conditional one during sampling? This insight motivates inference-time conditioning, which steers unconditional generation toward condition-specified targets without modifying model parameters. Bayes' theorem\footnote{For scalar variables, Bayes' theorem states $p(x|y) = \frac{p(y|x)p(x)}{p(y)}$, where $p(x)$ is the prior, $p(y|x)$ is the likelihood, and $p(x|y)$ is the posterior.} reveals how conditional and unconditional scores relate. Taking the score with respect to $\mathbf{x}$, we obtain
\begin{equation}
	\underbrace{\nabla_{\mathbf{x}} \log p_t(\mathbf{x}|\mathbf{y})}_{\text{cond. score}} = \underbrace{\nabla_{\mathbf{x}} \log p_t(\mathbf{x})}_{\text{uncond. score}} + \underbrace{\nabla_{\mathbf{x}} \log p_t(\mathbf{y}|\mathbf{x})}_{\text{guidance}}.
\end{equation}

The first term on the right-hand side is the unconditional score, readily available from the pre-trained model $\boldsymbol{s}_{\boldsymbol{\theta}}(\mathbf{x},t)$. The second term, the log-likelihood gradient, presents the core of conditioning: it acts as an ``external force'' that regularizes the unconditional score toward conditional generation. Denoting this guidance field as $\boldsymbol{g}(\mathbf{y}|\mathbf{x},t) = \nabla_{\mathbf{x}} \log p_t(\mathbf{y}|\mathbf{x})$, we obtain the canonical equation for inference-time conditioning:
\begin{equation}
	\boldsymbol{s}(\mathbf{x}|\mathbf{y},t) \approx \boldsymbol{s}_{\boldsymbol{\theta}}(\mathbf{x},t) + \gamma\boldsymbol{g}(\mathbf{y}|\mathbf{x},t),
	\label{eq:inference-canonical}
\end{equation}
where $\gamma \geqslant 0$ modulates the guidance strength.

The technical realization of $\boldsymbol{g}(\mathbf{y}|\mathbf{x},t)$ admits two principal approaches, distinguished by whether the ground-truth target is directly accessible.

\paragraph{Classifier Guidance} When ground-truth labels or targets are available, a natural insight is to intervene in the generated results of unconditional diffusion models with ground truths (conditional signals) at each time step. Dhariwal and Nichol propose the Classifier Guidance (CG, \aka Guided Diffusion or GD)~\cite{dhariwal2021diffusion}, training a time-dependent classifier $p_{\boldsymbol{\phi}}(\mathbf{y}|\mathbf{x},t)$ parameterized by $\boldsymbol{\phi}$ on noisy data across the diffusion trajectory. Notably, this classifier is specifically designed to operate on noise-corrupted data at arbitrary noise levels, distinguishing it from standard classifiers trained on clean data. The guidance field thus becomes $\boldsymbol{g}(\mathbf{y}|\mathbf{x},t) = \nabla_{\mathbf{x}} \log p_{\boldsymbol{\phi}}(\mathbf{y}|\mathbf{x},t)$ yielding the conditional score
\begin{equation}
	\boldsymbol{s}(\mathbf{x}|\mathbf{y},t) = \boldsymbol{s}_{\boldsymbol{\theta}}(\mathbf{x},t) + \gamma\nabla_{\mathbf{x}} \log p_{\boldsymbol{\phi}}(\mathbf{y}|\mathbf{x},t).
\end{equation}

In the $\boldsymbol{\epsilon}$-parameterization\footnote{Diffusion models such as DDPM \cite{ho2020denoising} and DDIM \cite{song2020denoising} employ the $\boldsymbol{\epsilon}$-parameterization, where the neural network directly predicts the noise $\boldsymbol{\epsilon}$ to be removed rather than the score function. The equivalence follows from  $\boldsymbol{\epsilon}_{\boldsymbol{\theta}}(\mathbf{x},t) = -\sqrt{1-\bar{\alpha}_t}\boldsymbol{s}_{\boldsymbol{\theta}}(\mathbf{x},t)$, allowing the denoising network to implicitly estimate the score. This parameterization proves more stable during training as the noise prediction target has unit variance across all time steps.}, this equivalently becomes
\begin{equation}
	\boldsymbol{\epsilon}_{\boldsymbol{\theta}}(\mathbf{x}|\mathbf{y},t) = \boldsymbol{\epsilon}_{\boldsymbol{\theta}}(\mathbf{x},t) - \sqrt{1 - \bar{\alpha}_t}\gamma\nabla_{\mathbf{x}} \log p_{\boldsymbol{\phi}}(\mathbf{y}|\mathbf{x},t),
\end{equation}
where $\bar{\alpha}_t = \prod_{s=1}^{t}(1-\beta_s)$ with $\beta_s$ denoting the variance schedule that controls the amount of noise added at each diffusion step $s$, and $\boldsymbol{\epsilon}_{\boldsymbol{\theta}}(\mathbf{x},t)$ represents the denoising network.

Generally speaking, CG reshapes the probability distribution by amplifying the influence of conditional signals, making the generation process more focused on target modes. This means that in the face of different downstream tasks, we can pre-train an unconditional diffusion model and then use specific classifiers during inference, fine-tuning classifiers according to task requirements to achieve specific generation goals.

However, CG encounters two inherent problems: noise adversity and optimization failure. The former means any classifiers applied for unconditional diffusion models must adapt to multi-level noise, ensuring the effects of conditional guidance during the sampling process. The latter means the given condition $\mathbf{y}$ may correlate weakly with the input $\mathbf{x}$, leading to classifier gradients with respect to $\mathbf{x}$ being along any or even adversarial optimization directions.

\paragraph{Estimator Guidance} For various scientific and engineering problems, we have partial measurements derived from the source rather than direct access to ground truth. For example in medical imaging, only partial $k$-space measurements are available in MRI reconstruction~\cite{lustig2007sparse}. These applications naturally formulate as inverse problems\footnote{An inverse problem seeks to recover an unknown source $\mathbf{x}$ from its measurement $\mathbf{y}$ through a forward model $\mathbf{y} = \mathcal{A}(\mathbf{x}) + \mathbf{n}$. The problem is typically ill-posed: multiple sources may explain the same measurement, making unique recovery impossible without prior knowledge.}, where measurements serve as conditions for guiding generation.

When the score-based diffusion model serves as the prior, we need to specify how the measurement $\mathbf{y} \in \mathbb{R}^m$ relates to the ``clean'' source $\mathbf{x}_0$. The forward model takes the form
\begin{equation}
	\mathbf{y} = \mathcal{A}(\mathbf{x}_0) + \mathbf{n},
	\label{eq:forward-model}
\end{equation}
where $\mathbf{n} \sim \mathcal{N}(\mathbf{0}, \sigma_{\mathbf{n}}^2\mathbf{I})$ is additive Gaussian noise with variance $\sigma_{\mathbf{n}}^2$ for simplicity. The forward operator $\mathcal{A}(\cdot): \mathbb{R}^D \to \mathbb{R}^m$ encodes the degradation process.

Based on Eq.~\eqref{eq:inference-canonical}, when a diffusion model is employed as the prior, the unconditional score $\boldsymbol{s}_{\boldsymbol{\theta}}(\mathbf{x},t)$ is already known, so estimating the conditional score $\boldsymbol{s}(\mathbf{x}|\mathbf{y},t)$ essentially amounts to recovering the log-likelihood gradient $\nabla_{\mathbf{x}} \log p_t(\mathbf{y}|\mathbf{x})$. However, this gradient is not directly accessible, making the task inherently difficult. The measurement $\mathbf{y}$ depends explicitly on the clean data through Eq.~\eqref{eq:forward-model}. Yet, during the reverse sampling procedure at time step $t$, only the intermediate noisy state $\mathbf{x}_t$ is available, while the original signal $\mathbf{x}_0$ is not.

The remedy is to follow the indirect path $\mathbf{x}_t \to \mathbf{x}_0 \to \mathbf{y}$: first estimate $\mathbf{x}_0$ from $\mathbf{x}_t$ using the score model $\boldsymbol{s}_{\boldsymbol{\theta}}$, and then exploit the explicit dependency between $\mathbf{x}_0$ and $\mathbf{y}$ to approximate the otherwise intractable likelihood.

As such, Chung \etal~propose the Diffusion Posterior Sampling (DPS)~\cite{chung2022diffusion}, inferring $\mathbf{x}_0$ from $\mathbf{x}_t$ with $\boldsymbol{s}_{\boldsymbol{\theta}}$, yielding the \emph{posterior mean} (\aka denoised estimate) given by Tweedie’s formula~\cite{efron2011tweedie}:
\begin{equation}
	\hat{\mathbf{x}}_{0|t} = \mathbb{E}\left[\mathbf{x}_0|\mathbf{x}_t\right] = \frac{1}{\alpha_t}\left(\mathbf{x}_t + \sigma_t^2\boldsymbol{s}_{\boldsymbol{\theta}}(\mathbf{x},t)\right),
	\label{eq:posterior-mean}
\end{equation}
where $\alpha_t = \sqrt{\bar{\alpha}_t}$ and $\sigma_t = \sqrt{1-\bar{\alpha}_t}$. This estimator, rooted in empirical Bayes\footnote{In empirical Bayes, a common prior is estimated from the entire dataset to inform and regularize each individual estimate.}, provides a principled means of reconstructing $\mathbf{x}_0$ from its noisy counterpart $\mathbf{x}_t$.

\begin{remark}[Validity Conditions of Tweedie's Formula]
	\hspace{0.5em} The posterior mean estimator $\hat{\mathbf{x}}_{0|t} = \mathbb{E}[\mathbf{x}_0|\mathbf{x}_t]$ is exact under two conditions: \emph{(i)}~the score model perfectly approximates the true score function, \ie $\boldsymbol{s}_{\boldsymbol{\theta}}(\mathbf{x},t) = \nabla_{\mathbf{x}}\log p_t(\mathbf{x})$ with zero score estimation error $\varepsilon_{\mathrm{score}}=0$, and \emph{(ii)}~the forward process noise is Gaussian so that the perturbation kernel $p(\mathbf{x}_t|\mathbf{x}_0)$ retains a known Gaussian form. Formally, the score estimation error $\varepsilon_{\mathrm{score}}$ (\aka score matching loss) quantifies the expected squared distance between the parameterized score model and the true score of the marginal distribution, typically defined as $\varepsilon_{\mathrm{score}} :=  \mathbb{E}_{t \sim \mathcal{U}[0,1], \mathbf{x} \sim p_t(\mathbf{x})} \left[ \| \nabla_{\mathbf{x}} \log p_t(\mathbf{x}) - \boldsymbol{s}_{\boldsymbol{\theta}}(\mathbf{x}, t) \|_2^2 \right]$. In realistic communication scenarios, both conditions are violated to varying degrees. Score network mismatch (\ie $\varepsilon_{\mathrm{score}}>0$) induces a persistent bias in $\hat{\mathbf{x}}_{0|t}$, as the learned score systematically deviates from the true data score in regions of low training density. Non-Gaussian channel residuals, such as those arising from impulsive interference or residual equalization errors, further distort the posterior mean by invalidating the Gaussian perturbation kernel assumption. To mitigate error accumulation across the reverse sampling trajectory, a practical strategy is to use $\hat{\mathbf{x}}_{0|t}$ as a \emph{soft} guiding direction scaled by the guidance weight $\gamma$ (as in Eq.~\eqref{eq:dps}) rather than enforcing a hard projection at each step. This soft-guidance approach prevents small per-step biases from compounding into large reconstruction errors over hundreds of denoising iterations.
\end{remark}

Building upon this posterior mean, DPS proposes to approximate the time-dependent log-likelihood gradient by replacing the intractable conditional term with its denoised surrogate: $\nabla_{\mathbf{x}} \log p_t(\mathbf{y} | \mathbf{x}) 
\approx \nabla_{\mathbf{x}} \log p_t(\mathbf{y} | \hat{\mathbf{x}}_{0|t})$. This substitution bridges the gap between the noisy state $\mathbf{x}_t$ and the measurement $\mathbf{y}$, enabling a tractable yet principled approximation that extends naturally to general noisy inverse problems.

The general form of estimator guidance-based inference-time conditioning can be concluded that
\begin{equation}
	\boldsymbol{s}(\mathbf{x}|\mathbf{y},t) \approx \boldsymbol{s}_{\boldsymbol{\theta}}(\mathbf{x},t) + \gamma \nabla_{\mathbf{x}} \mathcal{R}\left(\mathbf{y}, \mathcal{A}(\hat{\mathbf{x}}_{0|t})\right),  \label{eq:dps}
\end{equation}
where the guidance field takes $\boldsymbol{g}(\mathbf{y}|\mathbf{x},t) = \nabla_{\mathbf{x}} \mathcal{R}\left(\mathbf{y}, \mathcal{A}(\hat{\mathbf{x}}_{0|t})\right)$, serving as the regularization with posterior means for measurement consistency, and is typically set as $\nabla_{\mathbf{x}} \| \mathbf{y} - \mathcal{A}(\hat{\mathbf{x}}_{0|t}) \|_2^2$ when the measurement noise $\mathbf{n}$ is Gaussian.

The discussion so far has assumed that the forward operator $\mathcal{A}$ in Eq.~\eqref{eq:forward-model} is known, which corresponds to the classical, non-blind setting of inverse problems. Yet, in many real-world scenarios the definite degradation is unavailable or only partially characterized. This gives rise to \emph{blind inverse problems}, where both the underlying data $\mathbf{x}$ and the operator parameters $\boldsymbol{\vartheta}$ must be inferred from the measurement $\mathbf{y}$.

Assuming the functional form of $\mathcal{A}_{\boldsymbol{\vartheta}}$ is known, while its parameters $\boldsymbol{\vartheta}$ are unknown. Formally, the blind setting can be described as
\begin{equation}
	\mathbf{y} = \mathcal{A}_{\boldsymbol{\vartheta}}(\mathbf{x}_0) + \mathbf{n},
\end{equation}
where $\boldsymbol{\vartheta}$ parameterizes the forward operator and is itself unknown. A natural extension of estimator guidance is therefore to construct estimators not only for $\mathbf{x}$ but also for $\boldsymbol{\vartheta}$, enabling simultaneous recovery of source and operator.

BlindDPS~\cite{chung2023parallel} realizes this idea with parallel priors and joint estimator guidance: one diffusion prior $\boldsymbol{s}_{\boldsymbol{\theta}}(\mathbf{x},t)$ for data and another $\boldsymbol{s}_{\boldsymbol{\theta}}(\boldsymbol{\rho},t)$ for the operator parameters, updated in parallel along the reverse process. The dimensionality of the operator parameter vector $\boldsymbol{\rho}$ is typically far smaller than that of $\mathbf{x}$, making the second prior lightweight to train by score matching.

The joint conditional score can then be decomposed as
\begin{equation}
	\nabla_{\mathbf{x}} \log p_t(\mathbf{x},\boldsymbol{\rho}|\mathbf{y}) 
	= \nabla_{\mathbf{x}} \log p_t(\mathbf{x}) + \nabla_{\mathbf{x}} \log p_t(\mathbf{y}|\mathbf{x},\boldsymbol{\rho}).
\end{equation}

Since the second term on the right-hand side is again intractable, BlindDPS approximates it through denoised surrogates $(\hat{\mathbf{x}}_{0|t},\hat{\boldsymbol{\rho}}_{0|t})$. For general noisy measurements with Gaussian noise the conditional score reduces to
\begin{equation}
	\boldsymbol{s}(\mathbf{x}, \boldsymbol{\rho}|\mathbf{y},t) \approx 
	\boldsymbol{s}_{\boldsymbol{\theta}}(\mathbf{x},t)
	+ \gamma \nabla_{\mathbf{x}} \mathcal{R}\left(\mathbf{y},\mathcal{A}_{\hat{\boldsymbol{\vartheta}}}(\hat{\mathbf{x}}_{0|t})\right),
	\label{eq:blind-dps}
\end{equation}
where $\hat{\boldsymbol{\vartheta}}$ is parameterized by $\hat{\boldsymbol{\rho}}_{0|t}$, and $\gamma$ controls the guidance strength. Through parallel refinement of both $\mathbf{x}$ and $\boldsymbol{\vartheta}$ along the reverse diffusion trajectory, BlindDPS effectively converts an unknown operator into an approximate known operator, after which standard DPS principles apply.

In essence, BlindDPS extends DPS by coupling two diffusion priors through measurement consistency, enabling data to be denoised while the unknown operator is estimated in parallel within the same reverse process.

Compared with classifier guidance, estimator guidance avoids the noise adversity and optimization failure problems by not relying on a separately trained classifier. Instead, it derives the guidance signal directly from the forward model, which ties the sampling trajectory to the measurement.

However, estimator guidance faces two main limitations. First, it is computationally expensive because it uses a pre-trained diffusion model and needs to inject measurement consistency during gradient-based sampling. This makes it slow and sometimes unstable~\cite{du2023reduce}. Second, when measurements are heavily degraded, the guidance becomes less useful. In extreme cases, it can actually push the reverse trajectory away from the true data manifold, causing out-of-distribution artifacts and poor perceptual quality~\cite{daras2024survey}.

A natural concern arises from the second limitation: if the diffusion prior is trained on data that do not exactly match the target signal domain, does the estimator guidance remain viable? This question is especially pertinent for generative semantic communications, where the receiver's diffusion prior may have been trained on a generic data corpus rather than on the specific content being transmitted. Surprisingly, recent work by Jia~\etal~\cite{jia2026weak} provides both empirical and theoretical evidence that such ``weak'' diffusion priors, whether arising from domain mismatch or inference-time truncation, can still achieve strong reconstruction performance when the measurements are sufficiently informative. Their analysis, grounded in Bayesian posterior consistency, shows that high-dimensional measurements can effectively dominate the prior, causing the posterior to concentrate near the true signal regardless of prior fidelity. We defer a detailed treatment of the conditions under which this robustness holds, and its implications for semantic communication system design, to Section~\ref{subsec:potential_solutions}.

\subsubsection{Training-time Conditioning}
While inference-time conditioning offers flexibility by adapting pre-trained models, it fundamentally relies on external guidance signals that may not align well with the diffusion process. An alternative paradigm emerges: can we directly train diffusion models to internalize conditional information? This insight motivates training-time conditioning, where the model learns both conditional and unconditional distributions within a unified framework.

The core idea builds upon a simple observation. Rather than decomposing the conditional score through Bayes' theorem as in Eq.~\eqref{eq:inference-canonical}, we can directly interpolate between unconditional and conditional scores:
\begin{equation}
	\boldsymbol{s}(\mathbf{x}|\mathbf{y},t) \approx (1-\gamma)\boldsymbol{s}_{\boldsymbol{\theta}}(\mathbf{x},t) + \gamma\boldsymbol{s}_{\boldsymbol{\theta}}(\mathbf{x}|\mathbf{y},t),
	\label{eq:training-canonical}
\end{equation}
where both $\boldsymbol{s}_{\boldsymbol{\theta}}(\mathbf{x},t)$ and $\boldsymbol{s}_{\boldsymbol{\theta}}(\mathbf{x}|\mathbf{y},t)$ are learned by the same neural network, and $\gamma \geqslant 0$ controls the conditioning strength. This formulation shifts the burden from inference-time guidance to training-time learning.

\paragraph{Classifier-Free Guidance} Ho and Salimans~\cite{ho2022classifier} realize this idea in Eq.~\eqref{eq:training-canonical} through Classifier-Free Guidance (CFG), which trains a single diffusion model to handle both conditional and unconditional generation. During training, the model learns to denoise with probability $p$ using the condition $\mathbf{y}$, and with probability $1-p$ using a null token $\varnothing$ (effectively learning the unconditional distribution).

In the $\boldsymbol{\epsilon}$-parameterization, the interpolation becomes:
\begin{equation}
	\boldsymbol{\epsilon}_{\boldsymbol{\theta}}(\mathbf{x}|\mathbf{y},t) = (1-\gamma)\boldsymbol{\epsilon}_{\boldsymbol{\theta}}(\mathbf{x}|\varnothing,t) + \gamma\boldsymbol{\epsilon}_{\boldsymbol{\theta}}(\mathbf{x}|\mathbf{y},t),
\end{equation}
where $\boldsymbol{\epsilon}_{\boldsymbol{\theta}}(\mathbf{x}|\varnothing,t)$ denotes the unconditional denoising network learned through null conditioning. When $\gamma=0$, the diffusion model generates unconditionally; when $\gamma=1$, it follows standard conditional generation; and when $\gamma>1$, it amplifies the conditional signal, often improving sample quality at the cost of diversity.

The elegance of CFG lies in its unified training scheme. The model naturally adapts to different noise levels for both conditional and unconditional generation, avoiding the noise adversity problem of classifier guidance. Since the conditional information directly influences the denoising predictions during training, the gradient directions inherently align with generation objectives, circumventing optimization failure. Besides, since CFG does not depend on external measurements, it avoids the computational burden and instability issues that arise from degraded measurements in estimator guidance, enabling efficient and direct sample generation.

However, CFG introduces its own challenges. The joint optimization of conditional and unconditional distributions creates a complex learning landscape. As $\gamma$ increases, sample diversity typically decreases, potentially causing mode collapse where the model converges to a limited set of outputs. The need to run the model twice during inference (once with and once without conditioning) doubles the computational cost compared to standard conditional models. Moreover, the optimal choice of $\gamma$ varies across different conditions and datasets, requiring careful tuning.

\subsection{Efficient Diffusion Models} \label{sec:efficient}
While diffusion models generate high-quality content, their iterative sampling process, which typically requires hundreds to thousands of neural network evaluations, presents significant computational challenges. Each step involves solving reverse SDEs or PF ODEs, leading to substantial inference time and memory consumption that hinder real-time edge deployment.

Five primary acceleration strategies have been developed to address these bottlenecks:
\begin{itemize}
	\item \textbf{Dimensionality Reduction}: Diffusion in compressed latent spaces rather than high-dimensional data space.
	\item \textbf{Knowledge Distillation}: Training efficient students to replicate teacher behavior with fewer steps or reduced complexity.
	\item \textbf{Structure Pruning}: Removing redundant components while preserving generative capabilities.
	\item \textbf{Cache Reuse}: Reusing intermediate features across sampling steps to reduce redundancy.
	\item \textbf{Flow Matching}: Learning optimal transport paths for deterministic, efficient generation.
\end{itemize}

Table~\ref{tab:efficient-diffusion-models} taxonomizes these acceleration methods for diffusion sampling with representative works.

\begin{table*}[t]
	\caption{Taxonomy of efficient diffusion models categorized by the ways of sampling acceleration. All the five categories aim to eliminate computational redundancy or reduce the inherent stochasticity of the generation process.}
	\label{tab:efficient-diffusion-models}
	\centering
	\renewcommand{\arraystretch}{1.5}
	\begin{tabularx}{\textwidth}{C{1.5cm}|C{0.8cm}|r|R|C{7.8cm}|C{1.5cm}}
		\toprule
		\textbf{Category} & \textbf{\#} & \textbf{Related Work} & \textbf{Venue} & \textbf{Illustration} & \textbf{Links} \\
		\midrule\midrule
		
		\multirow{5}{*}{\rotatebox{90}{\makecell{\textsc{Dimensionality}\\\textsc{Reduction}}}}
		& \cellcolor{black!3!white} 1  & \cellcolor{black!3!white} LDM \cite{rombach2022high} & \cellcolor{black!3!white} CVPR'22 & \multirow{5}{*}{\begin{minipage}[c][2.3cm][c]{\linewidth}\centering
				\includegraphics[height=2.5cm]{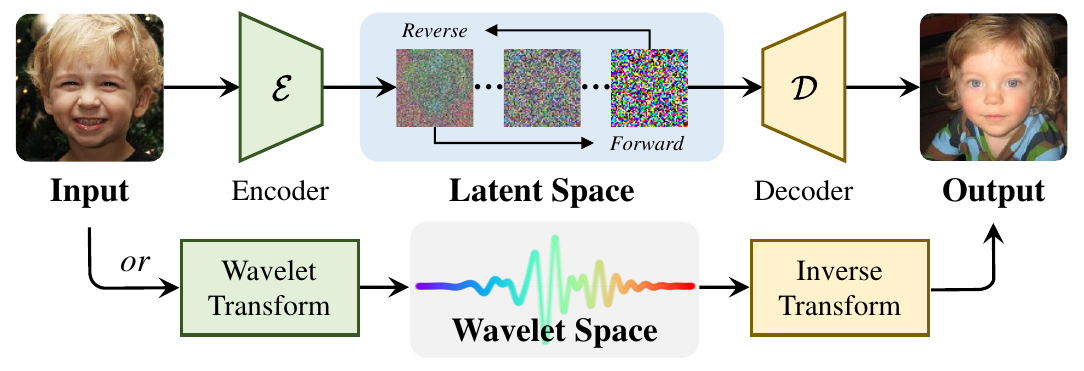}
		\end{minipage}}
		& \cellcolor{black!3!white} \href{https://github.com/CompVis/latent-diffusion}{\github}\ \href{https://huggingface.co/CompVis}{\huggingface} \\
		& 2  & WSGM \cite{guth2022wavelet} & NeurIPS'22 & & \href{https://openreview.net/forum?id=xZmjH3Pm2BK}{\link} \\
		& \cellcolor{black!3!white} 3  & \cellcolor{black!3!white} DiT \cite{peebles2023dit} & \cellcolor{black!3!white} ICCV'23 & & \cellcolor{black!3!white} \href{https://github.com/facebookresearch/DiT}{\github}\
		\href{https://www.wpeebles.com/DiT}{\link} \\
		& 4  & WaveDiff \cite{phung2023wavelet} &  CVPR'23 & & \href{https://github.com/VinAIResearch/WaveDiff}{\github} \\
		& \cellcolor{black!3!white} 5  & \cellcolor{black!3!white} LMD \cite{ma2024lmd} & \cellcolor{black!3!white} AAAI'24 & & \cellcolor{black!3!white} \href{https://github.com/AnonymousPony/lmd}{\github} \\
		\midrule
		
		\multirow{5}{*}{\rotatebox{90}{\makecell{\textsc{Knowledge}\\\textsc{Distillation}}}}
		& \cellcolor{black!3!white} 1  & \cellcolor{black!3!white} PD \cite{salimans2022progressive} & \cellcolor{black!3!white} ICLR'22 & \multirow{5}{*}{\begin{minipage}[c][2.3cm][c]{\linewidth}\centering
				\includegraphics[height=2.05cm]{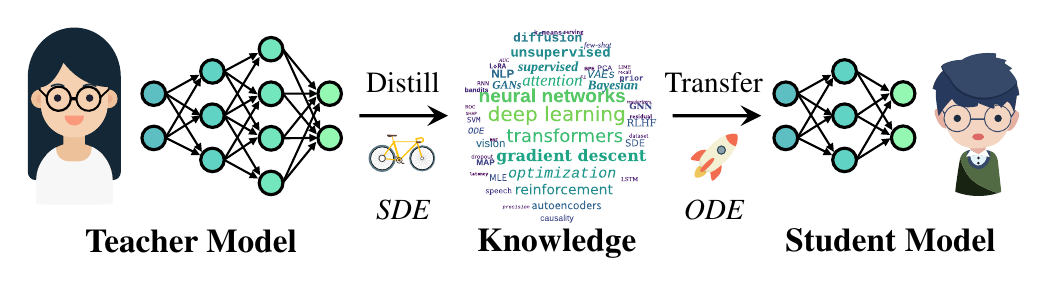}
		\end{minipage}}
		& \cellcolor{black!3!white} \href{https://github.com/google-research/google-research/tree/master/diffusion_distillation}{\github} \\
		& 2  & CM \cite{song2023consistency} & ICML'23 & & \href{https://github.com/openai/consistency_models}{\github} \\
		& \cellcolor{black!3!white} 3  & \cellcolor{black!3!white} LCM \cite{luo2023latent} & \cellcolor{black!3!white} arXiv'23 & & \cellcolor{black!3!white} \href{https://github.com/luosiallen/latent-consistency-model}{\github}\ \href{https://huggingface.co/SimianLuo/LCM_Dreamshaper_v7}{\huggingface}\
		\href{https://latent-consistency-models.github.io}{\link} \\
		& 4  & DMD2 \cite{yin2024improved} & NeurIPS'24 & & \href{https://github.com/tianweiy/DMD2}{\github}\
		\href{https://huggingface.co/tianweiy/DMD2}{\huggingface}\
		\href{https://tianweiy.github.io/dmd2/}{\link} \\
		& \cellcolor{black!3!white} 5  & \cellcolor{black!3!white} CTM \cite{kim2023consistency} & \cellcolor{black!3!white} ICLR'24 & & \cellcolor{black!3!white} \href{https://github.com/sony/ctm}{\github}\
		\href{https://consistencytrajectorymodel.github.io/CTM}{\link} \\
		\midrule

		\multirow{5}{*}{\rotatebox{90}{\makecell{\textsc{Structure}\\\textsc{Pruning}}}}
		& \cellcolor{black!3!white} 1  & \cellcolor{black!3!white} Diff-Pruning \cite{fang2023structural} & \cellcolor{black!3!white} NeurIPS'23 & \multirow{5}{*}{\begin{minipage}[c][2.3cm][c]{\linewidth}\centering
				\includegraphics[height=2.5cm]{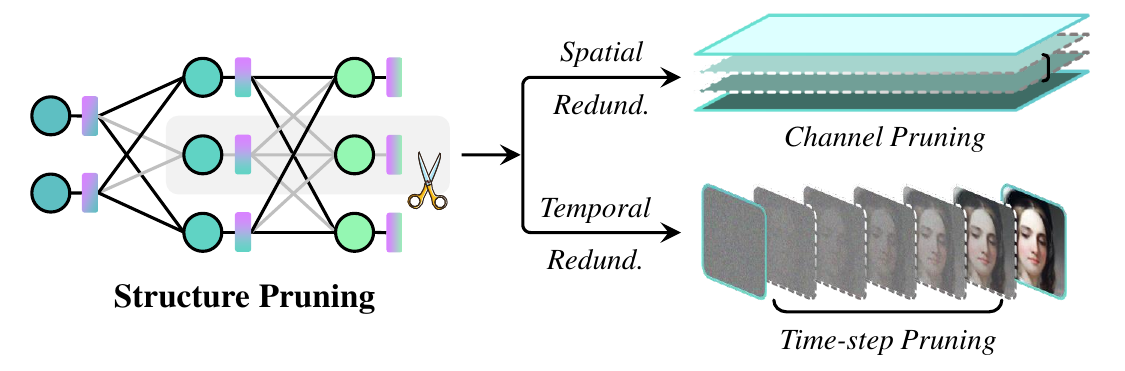}
		\end{minipage}} & \cellcolor{black!3!white} \href{https://github.com/VainF/Diff-Pruning}{\github} \\
		& 2  & TDPM \cite{zheng2022truncated} & ICLR'23 & & \href{https://github.com/JegZheng/truncated-diffusion-probabilistic-models}{\github} \\
		& \cellcolor{black!3!white} 3  & \cellcolor{black!3!white} LD-Pruner \cite{castells2024ld} & \cellcolor{black!3!white} CVPR'24 & 
		& \cellcolor{black!3!white} \href{https://openaccess.thecvf.com/content/CVPR2024W/EDGE/html/Castells_LD-Pruner_Efficient_Pruning_of_Latent_Diffusion_Models_using_Task-Agnostic_Insights_CVPRW_2024_paper.html}{\link} \\
		& 4  & DiP-GO \cite{zhu2024dip} & NeurIPS'24 & & \href{https://github.com/haoweiz23/dip-go}{\github} \\
		& \cellcolor{black!3!white} 5  & \cellcolor{black!3!white} AdaDiff \cite{zhang2024adadiff} & \cellcolor{black!3!white} ECCV'24 & & \cellcolor{black!3!white} \href{https://github.com/Tangshengku/AdaDiff}{\github} \\
		\midrule
		
		\multirow{5}{*}{\rotatebox{90}{\makecell{\textsc{Cache}\\\textsc{Reuse}}}}
		& \cellcolor{black!3!white} 1  & \cellcolor{black!3!white} DeepCache \cite{ma2024deepcache} & \cellcolor{black!3!white} CVPR'24 & \multirow{5}{*}{\begin{minipage}[c][2.3cm][c]{\linewidth}\centering
				\includegraphics[height=2.5cm]{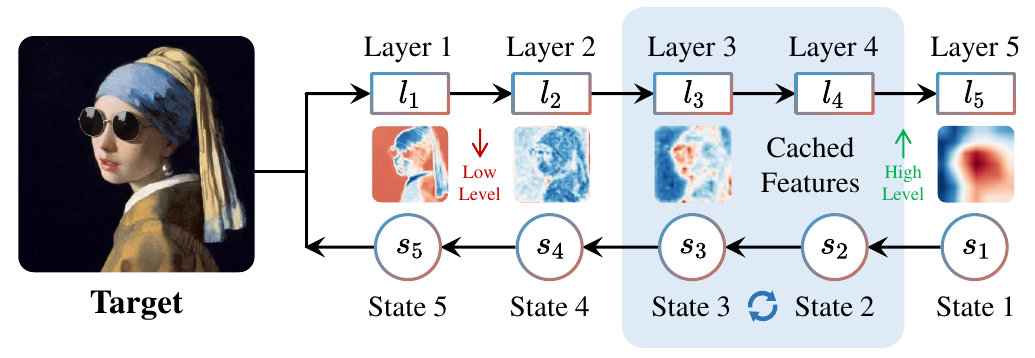}
		\end{minipage}}
		& \cellcolor{black!3!white} \href{https://github.com/horseee/DeepCache}{\github}\ \href{https://horseee.github.io/Diffusion_DeepCache/}{\link} \\
		& 2  & BlockCaching \cite{wimbauer2024cache} & CVPR'24 & & \href{https://fwmb.github.io/blockcaching}{\link} \\
		& \cellcolor{black!3!white} 3  & \cellcolor{black!3!white} L2C \cite{ma2024learning} & \cellcolor{black!3!white} NeurIPS'24 & & \cellcolor{black!3!white} \href{https://github.com/horseee/learning-to-cache}{\github} \\
		& 4  & ToCa \cite{zou2025accelerating} & ICLR'25 & & \href{https://github.com/Shenyi-Z/ToCa}{\github}\
		\href{https://toca2024.github.io/ToCa/}{\link} \\
		& \cellcolor{black!3!white} 5  & \cellcolor{black!3!white} ClusCa \cite{zheng2025compute} & \cellcolor{black!3!white} MM'25 & & \cellcolor{black!3!white} \href{https://github.com/Shenyi-Z/Cache4Diffusion}{\github} \\
		\midrule
		
		\multirow{5}{*}{\rotatebox{90}{\makecell{\textsc{Flow}\\\textsc{Matching}}}}
		& \cellcolor{black!3!white} 1  & \cellcolor{black!3!white}Flow Matching \cite{lipman2023flow} & \cellcolor{black!3!white}ICLR'23 & \multirow{5}{*}{\begin{minipage}[c][2.3cm][c]{\linewidth}\centering
				\includegraphics[height=2.4cm]{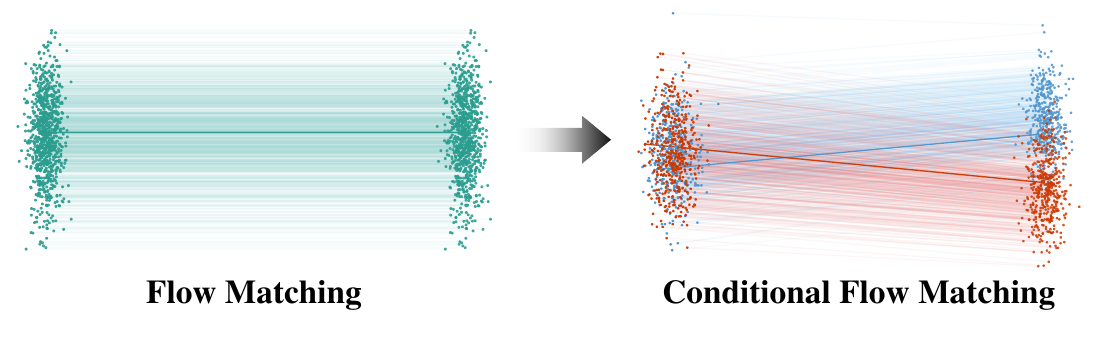}
		\end{minipage}} & \cellcolor{black!3!white}\href{https://github.com/facebookresearch/flow_matching}{\github}\
	    \href{https://facebookresearch.github.io/flow_matching}{\link} \\
		& 2  & Rectified Flow \cite{liu2023flow} & ICLR'23 & &  \href{https://github.com/gnobitab/RectifiedFlow}{\github} \\
		& \cellcolor{black!3!white} 3  & \cellcolor{black!3!white}PeRFlow \cite{yan2024perflow} & \cellcolor{black!3!white}NeurIPS'24 & & \cellcolor{black!3!white}\href{https://github.com/magic-research/piecewise-rectified-flow}{\github}\
		\href{https://huggingface.co/hansyan}{\huggingface}\
		\href{https://piecewise-rectified-flow.github.io}{\link} \\
		& 4  & InstaFlow \cite{liu2023instaflow} & ICLR'24 &
		& \href{https://github.com/gnobitab/InstaFlow}{\github} \\
		& \cellcolor{black!3!white} 5  & \cellcolor{black!3!white} MeanFlow \cite{geng2025mean} & \cellcolor{black!3!white} NeurIPS'25 & & \cellcolor{black!3!white} \href{https://github.com/Gsunshine/meanflow}{\github} \\
		
		\bottomrule
	\end{tabularx}
\end{table*}

\subsubsection{Dimensionality Reduction}
Dimensionality reduction accelerates sampling by performing diffusion in lower-dimensional spaces, reducing computational complexity.

\paragraph{Latent-based Sampling}
Latent-based methods map high-dimensional data to compact latent spaces. Latent Diffusion Models (LDM, \aka Stable Diffusion)~\cite{rombach2022high} employ a VAE~\cite{van2017neural} to compress images before applying diffusion.

Early models like DDPM~\cite{ho2020denoising} operate in pixel space, introducing significant overhead as resolution increases. LDM addresses this by encoding images into a compact latent space with spatial downsampling factor $k$. For instance, Stable Diffusion uses $k=8$, transforming $(512, 512, 3)$ images to $(64, 64, 4)$ latents, which is a $64\times$ spatial compression. This decouples computational cost from output resolution, enabling efficient high-resolution generation.

\paragraph{Wavelet-based Sampling}
Wavelet-based methods exploit frequency-domain sparsity~\cite{phung2023wavelet, mallat1999wavelet}. Wavelet transforms decompose signals into frequency components, with most energy concentrated in few coefficients. During sampling, the model selectively updates important wavelet coefficients while neglecting lower-priority components. This allows focusing computational resources on perceptually significant frequencies, achieving substantial speedup while maintaining visual quality. Beyond sampling acceleration, wavelet analysis has also been applied to accelerate auxiliary diffusion pipelines; for instance, Koo~\etal~\cite{koo2024wavelet} leverage wavelet-based frequency characterization to adaptively determine the optimization stopping point in text inversion for diffusion-based image editing.

\subsubsection{Knowledge Distillation}
Knowledge distillation transfers knowledge from complex teacher models to simpler students, enabling efficient models that replicate teacher performance with fewer resources.

\paragraph{Model Compression Distillation}
Model compression trains compact students to emulate larger teachers through architectural modifications (\eg reduced channels, fewer blocks, lightweight attention). Distillation employs multi-level transfer: output-level matching replicates teacher predictions across noise levels, while feature-level alignment matches intermediate representations via additional losses, preserving generative capabilities despite parameter reduction.

Recent work has demonstrated effective distillation strategies for reducing inference cost. For example, Plug-and-Play Diffusion Distillation~\cite{hsiao2024plug} trains a lightweight external guide model to replace the dual forward pass required by CFG, reducing per-step inference computation by nearly half while keeping the base diffusion model frozen and requiring only $1\%$ additional trainable parameters. This plug-and-play design allows the distilled guide to be directly applied to various fine-tuned variants of the base model without retraining, facilitating deployment in resource-constrained environments.

\paragraph{Time-step Reduction Distillation}
Time-step reduction trains students to replicate multiple teacher steps in fewer evaluations. Progressive Distillation (PD)~\cite{salimans2022progressive} enables students to produce outputs equivalent to two consecutive teacher steps in one evaluation, iteratively reducing step counts to $4$--$8$.

Consistency Models (CMs)~\cite{song2023consistency} enforce \emph{self-consistency} across diffusion sampling trajectories. For Eq.~\eqref{eq:pf-ode}, using the parameterization from~\cite{karras2022elucidating} with $\boldsymbol{f}(\mathbf{x}, t) = \mathbf{0}$ and $g(t) = \sqrt{2 t}$ yields:
\begin{equation}
	\frac{\mathrm{d} \mathbf{x}_t}{\mathrm{d} t} = - t \boldsymbol{s}_{\boldsymbol{\theta}}(\mathbf{x}, t),
\end{equation}
initialized from ${\mathbf{x}}_T \sim \mathcal{N}(\mathbf{0}, T^2\mathbf{I})$ and solved backwards to approximate ${\mathbf{x}}_0$.
The consistency function (\aka endpoint mapping) $\boldsymbol{c}:(\mathbf{x}, t) \to \mathbf{x}_\xi$ with the endpoint $\mathbf{x}_\xi$ ensures self-consistency $\boldsymbol{c}(\mathbf{x}, s) = \boldsymbol{c}(\mathbf{x}, t)$ for all points on the same ODE trajectory. The consistency model is parameterized as:
\begin{equation}
	\boldsymbol{c}_{\boldsymbol{\theta}}(\mathbf{x},t) = \boldsymbol{c}_{\mathrm{skip}}(t) \mathbf{x} + \boldsymbol{c}_{\mathrm{out}}(t) F_{\boldsymbol{\theta}}(\mathbf{x}, t),
\end{equation}
where $\boldsymbol{c}_{\mathrm{skip}}(\xi) = 1$ and $\boldsymbol{c}_{\mathrm{out}}(\xi) = 0$ ensure boundary conditions with $\xi$ being a small positive constant. The parameterized function $F_{\boldsymbol{\theta}}(\mathbf{x}, t)$ is a supposed free-form DNN, once which is differentiable, continuous-time consistency models can be trained by enforcing the self-consistency constraint along the ODE trajectory.

To train the consistency model, one exploits a pre-trained diffusion model to first trace an actual ODE trajectory from a data sample, obtaining noisy samples at two distinct time steps $s$ and $t$ along the same trajectory. Since both originate from the same clean data point, their consistency function outputs must agree, which yields the training objective:
\begin{equation}
	\mathcal{L}(\boldsymbol{\theta})\! =\! \mathbb{E}_{t \sim \mathcal{U}[0,1], s \sim \mathcal{U}[0,t), \mathbf{x} \sim p_{t}(\mathbf{x})}\!\left\| \boldsymbol{c}_{\boldsymbol{\theta}}(\mathbf{x}, s) \!-\! \boldsymbol{c}_{\boldsymbol{\theta}}(\mathbf{x}, t) \right\|_2^2\!.
\end{equation}

However, minimizing this loss alone admits a trivial solution where the network maps all inputs to the same constant, collapsing the model to a degenerate state (\ie mode collapse). The boundary condition $\boldsymbol{c}_{\boldsymbol{\theta}}(\mathbf{x}, \xi) = \mathbf{x}_\xi \approx \mathbf{x}_0$ prevents this collapse by anchoring the output: at time step $\xi$ (close to the data endpoint), the model must faithfully reproduce its input rather than projecting onto an arbitrary constant. Combined with techniques such as teacher guidance~\cite{karras2022elucidating} or stop-gradient~\cite{van2017neural} on one branch of the loss, this boundary anchor forces the consistency model to genuinely learn the trajectory-to-endpoint mapping dictated by the underlying probability flow. As training progresses, the network internalizes a time-invariant mapping: given any noisy sample from any point along the ODE trajectory, it directly recovers the corresponding clean data, enabling single-step generation without iterative denoising.

Viewed through the lens of numerical analysis, consistency models learn a direct mapping from any intermediate state to the trajectory endpoint (analogous to learning the analytical solution). This bypasses iterative numerical integration entirely, enabling single-step generation at the cost of a more demanding training objective~\cite{song2023improved}. This method extends to latent space through Latent Consistency Models (LCMs)~\cite{luo2023latent}, which apply consistency training within VAE-encoded representations.

More recent advances on distillation-based diffusion acceleration include Distribution Matching Distillation (DMD)~\cite{yin2024one}, which focuses on matching output distributions rather than individual samples, enabling robust one-step generation with competitive quality.

\subsubsection{Structure Pruning}
Structure pruning systematically removes redundant computational components while preserving essential generative functionality.

\paragraph{Channel Pruning}
Channel pruning reduces computational load by removing entire channels from convolutional layers based on their contribution to generation quality~\cite{he2017channel}. This approach requires careful analysis of channel importance across different diffusion time steps, as channels may contribute differently during various sampling phases.

SnapFusion~\cite{li2023snapfusion} demonstrates effective channel pruning for accelerating text-to-image diffusion models on mobile devices, targeting both the VAE decoder and the denoising U-Net. For the decoder, SnapFusion applies $50\%$ uniform channel pruning, reducing its size and multiply-accumulate operations (MACs) to approximately $1/4$ of the original, and recovers generation quality through a data distillation pipeline that trains the pruned decoder on synthetic latent-image pairs produced by the full-precision model. For the denoising U-Net~\cite{ronneberger2015u}, SnapFusion employs architecture evolution that evaluates each cross-attention and ResNet block by its joint impact on CLIP score degradation and latency improvement, systematically removing blocks with minimal quality loss and maximal speedup. This combined strategy enables sub-two-second text-to-image generation on mobile devices.

\paragraph{Time-step Pruning}
Time-step pruning reduces computation by selectively skipping or approximating diffusion time steps that contribute minimally to final generation quality. The method analyzes each time step's contribution to identify steps that can be safely skipped without significant quality loss~\cite{fang2023structural, zheng2022truncated}. Recent developments include adaptive computation methods~\cite{zhang2024adadiff} that dynamically allocate per-step computational resources through early-exit mechanisms, using time-step-aware uncertainty estimation to determine whether full network evaluation is necessary at each denoising step. Complementary approaches such as skip-step training~\cite{wang2024s2dms} introduce auxiliary loss terms that account for information lost during accelerated sampling, improving generation quality when fewer steps are used. Advanced scheduling optimization techniques~\cite{sabour2024align, xue2024accelerating} further enable flexible trade-offs between generation speed and quality by optimizing time step discretization schedules specific to different solvers and datasets.

\subsubsection{Cache Reuse}
Cache reuse techniques exploit temporal redundancy in diffusion processes to eliminate redundant computations. The key insight is that high-level features often remain stable across consecutive time steps, enabling reuse of previously computed results.

\paragraph{Feature Cache Reuse}
Feature cache reuse accelerates sampling by caching and reusing slowly-changing high-level features across multiple time steps. DeepCache~\cite{ma2024deepcache} analyzes U-Net architectures to identify layers producing temporally stable features that can be cached and reused.

The approach caches high-level features that exhibit minimal changes between adjacent diffusion steps while recomputing rapidly-changing low-level features. This selective computation maintains generation quality while substantially reducing computational overhead, particularly for high-resolution image generation.

\paragraph{Interval Cache Reuse}
While feature cache reuse with fixed intervals provides a straightforward acceleration strategy, the optimal caching behavior varies across the sampling trajectory and across different network components. Learning-to-Cache (L2C)~\cite{ma2024learning} addresses this by training a timestep-variant router that dynamically determines which transformer layers can reuse cached features and which require fresh computation at each denoising step. By formulating the layer selection as a differentiable optimization problem, L2C discovers that a substantial proportion of transformer layers exhibit sufficient temporal redundancy to be safely cached, with the caching ratio adapted according to the generation phase. In early diffusion stages where coarse structure is being established, a larger fraction of layers can rely on cached features across longer intervals, whereas later refinement stages allocate more active computation to preserve fine-grained details. This adaptive interval scheme produces a static computation graph that can be directly deployed without runtime overhead, achieving significant acceleration while largely maintaining generation quality.

\subsubsection{Flow Matching}
While score-based diffusion models have achieved remarkable success through their SDE and ODE formulations, they inherently require learning the score function at all time steps, which can be computationally intensive and sometimes unstable during training. Flow matching~\cite{lipman2023flow} emerges as an alternative paradigm that, while distinct from score-based modeling, is deeply influenced by its principles and serves as another effective framework for constructing and training efficient diffusion models. 

\begin{figure}[t]
	\centering
	\includegraphics[width=\columnwidth]{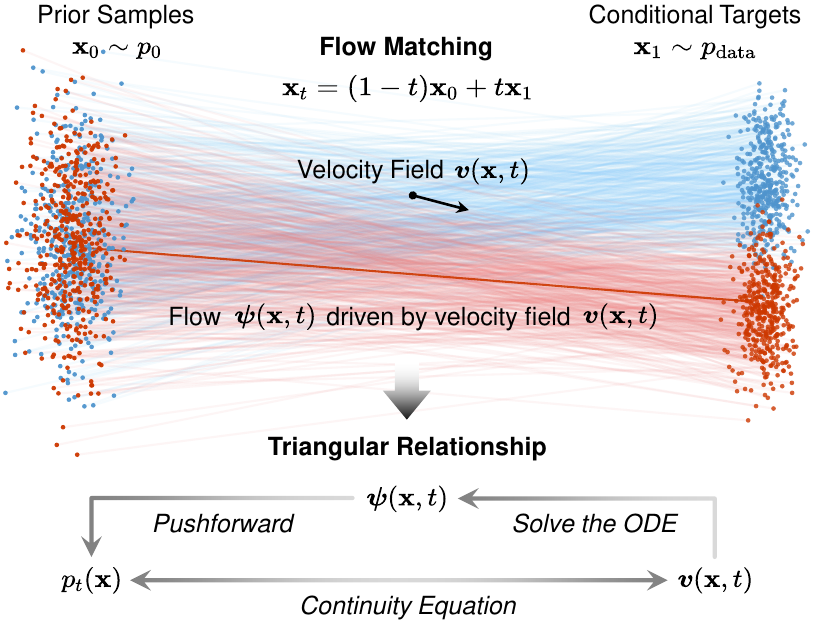}
	\caption{Underlying mechanism of flow-matching process. \emph{Top:} Samples from the prior $p_0$ are transported along straight-line conditional paths toward target points $\mathbf{x}_1 \sim p_{\mathrm{data}}$, with trajectories given by the flow $\boldsymbol{\psi}(\mathbf{x},t)$ under the velocity field $\boldsymbol{v}(\mathbf{x},t)$. The samples are colored in \textcolor[HTML]{CD3800}{red} and \textcolor[HTML]{4F94CD}{blue} for visualization, indicating two distinct data modes. \emph{Bottom:} Schematic summary of the triangular relationship among the probability path $p_t(\mathbf{x})$, the velocity field $\boldsymbol{v}(\mathbf{x},t)$, and the flow $\boldsymbol{\psi}(\mathbf{x},t)$. The probability path $p_t(\mathbf{x})$ reflects the macroscopic evolution of the distribution, while the velocity field $\boldsymbol{v}(\mathbf{x},t)$ characterizes the microscopic motions of data samples that drive this evolution. The flow $\boldsymbol{\psi}(\mathbf{x},t)$ links the two levels by tracing trajectories that reconcile local dynamics with global density change, establishing the generative mapping from prior to data.}
	\label{fig:flow}
\end{figure}

Specifically, following the trajectory of individual data points in the probability flow depicted in Continuous Normalizing Flows (CNFs)~\cite{chen2018neural}, each point $\mathbf{x}(t)$ evolves according to the ODE as follows:
\begin{equation}
	\frac{\mathrm{d}\mathbf{x}}{\mathrm{d}t} = \boldsymbol{v}(\mathbf{x}, t), \label{eq:cnf-ode}
\end{equation}
with initial condition $\mathbf{x}_0 \sim p_0$. Clearly, Eq.~\eqref{eq:cnf-ode} can be regarded as the ODE counterpart of Eq.~\eqref{eq:pf-ode}, by replacing the combined term $\boldsymbol{f}(\mathbf{x}, t) - \frac{1}{2} g^2(t)\boldsymbol{s}(\mathbf{x}, t)$ with the velocity field $\boldsymbol{v}(\mathbf{x}, t)$. This ODE defines a time-dependent flow map $\boldsymbol{\psi}(\mathbf{x}, t): \mathbb{R}^D \times [0,1] \to \mathbb{R}^D$ that maps initial positions to their locations at time $t$. The relationship between these three fundamental objects is summarized schematically in Fig.~\ref{fig:flow}.

Flow matching reformulates the deep generative modeling problem as learning a velocity field $\boldsymbol{v}(\mathbf{x}, t)$ that transports samples from a simple prior distribution $p_0(\mathbf{x}) = \mathcal{N}(\mathbf{0}, \mathbf{I})$ at time $t = 0$ to the data distribution $p_1(\mathbf{x}) = p_{\mathrm{data}}(\mathbf{x})$ at time $t = 1$. Unlike score-based diffusion models that first corrupt data through a forward process and then learn to reverse it, flow matching directly parameterizes the velocity field of the CNF whose solution trajectories connect noise and data.

Just as score matching learns the gradient of log-density without requiring the normalization constant, flow matching learns the velocity field without computing intractable posteriors. By leveraging the fact that we can sample from the joint distribution $p(\mathbf{x}_t, \mathbf{x}_1)$ through the forward simulation $\mathbf{x}_t = (1 - t)\mathbf{x}_0 + t\mathbf{x}_1$ following~\cite{tong2024improving}, the velocity can be learned by minimizing:
\begin{equation}
	\mathcal{L}(\boldsymbol{\theta}) = \mathbb{E}_{t \sim \mathcal{U}[0,1], \mathbf{x} \sim p_t(\mathbf{x})}\left[\left\|\boldsymbol{v}(\mathbf{x}, t) - \boldsymbol{v}_{\boldsymbol{\theta}}(\mathbf{x}, t)\right\|_2^2\right], \label{eq:flow-loss}
\end{equation}
where $\boldsymbol{v}(\mathbf{x}, t) = (\mathbf{x}_1 - \mathbf{x}_0)$ is the true velocity along the straight-line path connecting $\mathbf{x}_0$ to $\mathbf{x}_1$, and $\boldsymbol{v}_{\boldsymbol{\theta}}(\mathbf{x}, t)$ denotes a neural network parameterization of the velocity with parameters $\boldsymbol{\theta}$.

The advantages of flow matching stem from its direct approach to learning the velocity field, offering substantial efficiency gains over traditional diffusion models. First, the training objective regresses on bounded velocity vectors rather than potentially unbounded score functions, leading to significantly more stable training dynamics and faster convergence. Second, the straight-line trajectories in the conditional paths result in smoother velocity fields that are fundamentally easier for neural networks to approximate, enabling more efficient sampling with fewer function evaluations. Third, the deterministic nature of the flow eliminates the need for stochastic sampling, directly reducing computational overhead while maintaining generation quality.

Flow matching naturally supports fewer-step generation through its continuous formulation, often achieving comparable quality with $10$ to $20\times$ fewer sampling steps than traditional diffusion models. This implies the empirical success of flow matching in constructing highly efficient diffusion models that balance generation quality with computational efficiency.

\begin{table}[t]
	\centering
	\caption{Representative computational profiling of efficient diffusion methods. Estimates are normalized for generating a single $512\times512$ image at 16-bit floating-point (FP16) precision on a standard NVIDIA RTX 3090 GPU with batch size 1. \emph{NFE} denotes the required number of forward passes through the backbone (\eg U-Net~\cite{ronneberger2015u}) during sampling; \emph{Latency (s)} represents the end-to-end generation time (including both codec and backbone) evaluated in PyTorch without hardware-specific compiler accelerations (\eg TensorRT~\cite{nvidia_tensorrt}); \emph{VRAM (GB)} measures peak GPU memory utilization, which is dominated by model weights and context overheads. $^\ddagger$Methods perform diffusion sampling in the latent space. Note that the reported results serve as reference averages that reflect the current experimental setup rather than definitive performance outcomes. \textbf{Bold} indicates the best result and \underline{underline} indicates the second best result for each metric.}
	\label{tab:efficiency}
	\renewcommand{\arraystretch}{1.5}
	\setlength{\tabcolsep}{5pt}
	\rowcolors{2}{white}{black!3!white}
	\begin{tabular}{l|r|c|c|c}
		\toprule
		\rowcolor{white}
		\textbf{Category} & \textbf{Selected Method} & \textbf{NFE} $\downarrow$ & \textbf{Latency} $\downarrow$ & \textbf{VRAM} $\downarrow$ \\
		\midrule\midrule
		Baseline        & DDPM~\cite{ho2020denoising}                      & $1000$ & $\sim$ $113.8$ & $\sim$ $9.8$ \\
		Dim.\ Reduc.    & LDM$^\ddagger$~\cite{rombach2022high}                & $20$   & $\sim$ $1.2$   & $\sim$ $6.5$ \\
		Know.\ Distill. & LCM$^\ddagger$~\cite{luo2023latent}              & $\mathbf{1}$    & $\sim$ $\mathbf{0.1}$   & $\sim$ $\underline{6.0}$ \\
		Struct.\ Prun.  & Diff-Pruning$^\ddagger$~\cite{fang2023structural} & $20$  & $\sim$ $0.9$   & $\sim$ $\mathbf{5.2}$ \\
		Cache Reuse     & DeepCache$^\ddagger$~\cite{ma2024deepcache}      & $20$   & $\sim$ $\underline{0.5}$   & $\sim$ $6.3$ \\
		Flow Match.     & Rectified Flow~\cite{liu2023flow}                & $\underline{10}$   & $\sim$ $1.1$   & $\sim$ $9.6$ \\
		\bottomrule
	\end{tabular}
\end{table}

\begin{remark}[Probability Paths as a Common Abstraction]
	\hspace{0.5em} Despite their distinct formulations, score-based diffusion models and flow matching models share a common mathematical abstraction: both construct a continuous probability path $\{p_t(\mathbf{x})\}_{t \in [0,1]}$ connecting a tractable prior $p_0$ to the data distribution $p_1$, and both learn a neural network that characterizes how samples move along this path. The fundamental difference lies in \emph{what} the network learns and \emph{how} the path is traversed:
	\begin{itemize}
		\item Score-based diffusion first defines a forward noising process (SDE) and then learns the score function $\nabla_{\mathbf{x}} \log p_t(\mathbf{x})$ to reverse it, where sampling follows either stochastic (reverse SDE) or deterministic (PF ODE) dynamics.
		\item Flow matching directly parameterizes the velocity field $\boldsymbol{v}_{\boldsymbol{\theta}}(\mathbf{x}, t)$ that transports samples along straight-line conditional paths, producing an ODE whose trajectories are inherently smoother and easier to integrate numerically.
	\end{itemize}
	
	\hspace{0.5em} This perspective can be further extended. The consistency models discussed above learn neither the score nor the velocity, but instead directly learn the mapping from any point on the ODE trajectory to its endpoint, effectively learning the \emph{analytical solution} of the PF ODE rather than its local dynamics. From this viewpoint, current diffusion models can be organized along three design axes: the geometry of the probability path (curved {\vs} straight), the dynamics governing transport (stochastic SDE {\vs} deterministic ODE), and the learning target of the neural network (score model $\boldsymbol{s}_{\boldsymbol{\theta}}$ or denoising network $\boldsymbol{\epsilon}_{\boldsymbol{\theta}}$ {\vs} velocity field $\boldsymbol{v}_{\boldsymbol{\theta}}$ {\vs} endpoint mapping $\boldsymbol{c}_{\boldsymbol{\theta}}$ or target sample $\mathbf{x}_0$)~\cite{li2025back}.
\end{remark}

To provide concrete deployment guidance, we summarize the computational characteristics of the five acceleration strategies in Table~\ref{tab:efficiency}. For a U-Net backbone~\cite{ronneberger2015u} operating in pixel space, the per-step complexity scales as $\mathcal{O}(C^2 HW)$, where $C$ denotes the channel dimension and $H, W$ denote the spatial resolution. The total inference cost is proportional to the number of function evaluations (NFE) $T$, yielding an overall complexity of $\mathcal{O}(T C^2 HW)$. Each of the surveyed strategies targets a different factor in this product: dimensionality reduction shrinks $HW$ via latent-space or wavelet-space operation; knowledge distillation reduces $T$ to as few as even one; structure pruning reduces $C$ by removing redundant channels; cache reuse skips redundant per-step evaluations by exploiting temporal feature stability; and flow matching straightens the sampling trajectory to require fewer numerical integration steps for equivalent discretization error. These profiling bounds equip researchers with quantifiable heuristics for selecting deployment strategies under specific hardware and latency constraints.

\subsection{Generalized Diffusion Models}
\label{sec:generalized}

\begin{table*}[t]
	\caption{Taxonomy of generalized diffusion models categorized by generalization targets. These three categories extend diffusion models to accommodate diverse modalities, domains, and tasks.}
	\label{tab:generalized-diffusion-models}
	\centering
	\renewcommand{\arraystretch}{1.5}
	\begin{tabularx}{\textwidth}{C{1.5cm}|C{0.8cm}|r|R|C{7cm}|C{1.5cm}}
		\toprule
		\textbf{Category} & \textbf{\#} & \textbf{Related Work} & \textbf{Venue} & \textbf{Illustration} & \textbf{Links} \\
		\midrule\midrule
		
		\multirow{5}{*}{\rotatebox{90}{\makecell{\textsc{Modality}\\\textsc{Expansion}}}}
		& \cellcolor{black!3!white} 1  & \cellcolor{black!3!white} MonoFormer~\cite{zhao2024monoformer} & \cellcolor{black!3!white} arXiv'24 & \multirow{5}{*}{\begin{minipage}[c][2.3cm][c]{\linewidth}\centering
				\includegraphics[height=2.5cm]{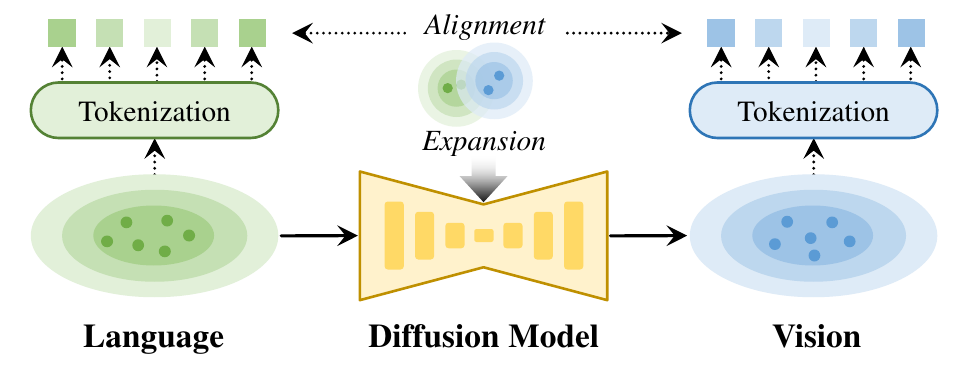}
		\end{minipage}}
		& \cellcolor{black!3!white} \href{https://github.com/MonoFormer/MonoFormer}{\github}\ \href{https://huggingface.co/MonoFormer}{\huggingface}\
		\href{https://monoformer.github.io}{\link} \\
		& 2  & Diffusion Forcing \cite{chen2024diffusion} & NeurIPS'24 & & \href{https://github.com/buoyancy99/diffusion-forcing}{\github}\
		\href{https://www.boyuan.space/diffusion-forcing}{\link} \\
		& \cellcolor{black!3!white} 3  & \cellcolor{black!3!white} Show-o \cite{xie2024show} & \cellcolor{black!3!white} ICLR'25 & & \cellcolor{black!3!white} \href{https://github.com/showlab/Show-o}{\github} \\
		& 4  & Transfusion \cite{zhou2024transfusion} & ICLR'25 & & \href{https://github.com/lucidrains/transfusion-pytorch}{\github} \\
		& \cellcolor{black!3!white} 5  & \cellcolor{black!3!white} UniDisc \cite{swerdlow2025unified} & \cellcolor{black!3!white} arXiv'25 & & \cellcolor{black!3!white} \href{https://github.com/alexanderswerdlow/unidisc}{\github}\
		\href{https://huggingface.co/aswerdlow/unidisc_interleaved}{\huggingface}\
		\href{https://unidisc.github.io/}{\link} \\
		\midrule
		
		\multirow{5}{*}{\rotatebox{90}{\makecell{\textsc{Domain}\\\textsc{Adaptation}}}}
		& \cellcolor{black!3!white} 1  & \cellcolor{black!3!white} DSB \cite{de2021diffusion} & \cellcolor{black!3!white} NeurIPS'21 & \multirow{5}{*}{\begin{minipage}[c][2.3cm][c]{\linewidth}\centering
				\includegraphics[height=2.5cm]{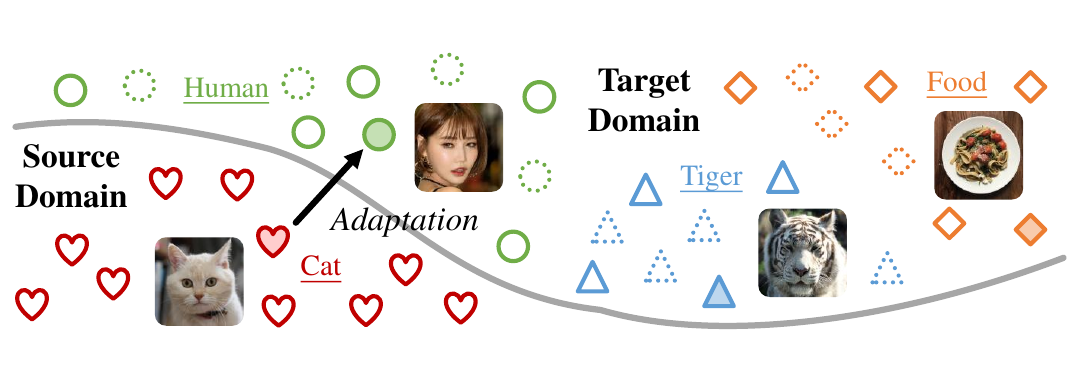}
		\end{minipage}}
		& \cellcolor{black!3!white} \href{https://github.com/JTT94/diffusion_schrodinger_bridge}{\github} \\
		& 2  & Composable Diffusion \cite{liu2022compositional} & ECCV'22 & & \href{https://github.com/energy-based-model/Compositional-Visual-Generation-with-Composable-Diffusion-Models-PyTorch}{\github}\
		\href{https://energy-based-model.github.io/Compositional-Visual-Generation-with-Composable-Diffusion-Models}{\link} \\
		& \cellcolor{black!3!white} 3  & \cellcolor{black!3!white} DreamBooth \cite{ruiz2023dreambooth} & \cellcolor{black!3!white} CVPR'23 & & \cellcolor{black!3!white} \href{https://github.com/google/dreambooth}{\github}\ \href{https://dreambooth.github.io}{\link} \\
		& 4  & I2SB \cite{liu2023i2sb} & ICML'23 & & \href{https://github.com/NVlabs/I2SB}{\github}\
		\href{https://i2sb.github.io/}{\link} \\
		& \cellcolor{black!3!white} 5  & \cellcolor{black!3!white} P2P-Bridge \cite{vogel2024p2p} & \cellcolor{black!3!white} ECCV'24 & & \cellcolor{black!3!white} \href{https://github.com/matvogel/P2P-Bridge}{\github}\
		\href{https://p2p-bridge.github.io}{\link} \\
		\midrule
		
		\multirow{5}{*}{\rotatebox{90}{\makecell{\textsc{Task}\\\textsc{Generalization}}}}
		& \cellcolor{black!3!white} 1  & \cellcolor{black!3!white} Diffuser \cite{janner2022planning} & \cellcolor{black!3!white} ICML'22 & \multirow{5}{*}{\begin{minipage}[c][2.3cm][c]{\linewidth}\centering
				\includegraphics[height=2.5cm]{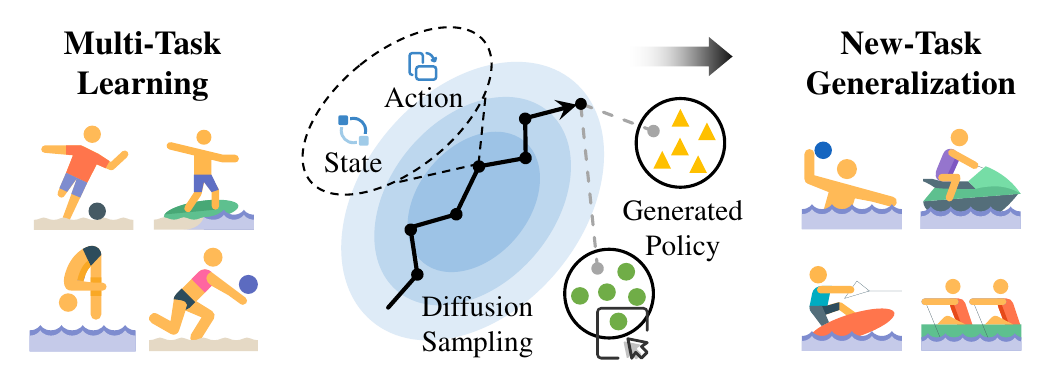}
		\end{minipage}} & \cellcolor{black!3!white}  \href{https://github.com/jannerm/diffuser}{\github}\
		\href{https://diffusion-planning.github.io/}{\link} \\
		& 2  & Diffusion Policy \cite{chi2023diffusion} & RSS'23 & &  \href{https://github.com/real-stanford/diffusion_policy}{\github}\
		\href{https://diffusion-policy.cs.columbia.edu}{\link} \\
		& \cellcolor{black!3!white} 3  & \cellcolor{black!3!white} DDPO \cite{black2024training} & \cellcolor{black!3!white} ICLR'24 & 
		& \cellcolor{black!3!white} \href{https://github.com/jannerm/ddpo}{\github}\
		\href{https://rl-diffusion.github.io}{\link} \\
		& 4  & C-LoRA \cite{smith2023continual} & TMLR'24 & & 
		\href{https://jamessealesmith.github.io/continual-diffusion}{\link} \\
		& \cellcolor{black!3!white} 5  & \cellcolor{black!3!white} B$^2$-DiffuRL \cite{hu2025towards} & \cellcolor{black!3!white} CVPR'25 & & \cellcolor{black!3!white} \href{https://github.com/hu-zijing/B2-DiffuRL}{\github} \\
		
		\bottomrule
	\end{tabularx}
\end{table*}

The remarkable success of diffusion models in generating high-quality samples has sparked significant interest in extending their capabilities beyond their original design scope. While initially conceived for specific data modalities and domains, recent advances have demonstrated that diffusion models possess inherent flexibility that enables adaptation across diverse modalities, domains, and tasks, which is precisely the requirement of task-specific multi-modal semantic communications.

The generalization of diffusion models represents a crucial frontier in generative modeling, addressing three fundamental dimensions: \emph{modality expansion} through hybrid architectures, \emph{domain adaptation} via optimal transport theory, and \emph{task generalization} through integration with decision-making frameworks. The taxonomy of generalized diffusion models and each category's representative works are illustrated in Table~\ref{tab:generalized-diffusion-models}.

\subsubsection{Modality Expansion}
Diffusion models excel at continuous high-dimensional data like images but face fundamental challenges with discrete sequential data like text or temporal signals like audio. This motivates hybrid approaches combining diffusion and autoregressive models to handle multiple modalities simultaneously.

\paragraph{Unified Multimodal Architectures}
Recent work integrates diffusion and autoregressive paradigms through innovative architectures. Transfusion~\cite{zhou2024transfusion} introduces a unified model that simultaneously trains on discrete language tokens and continuous image patches using a shared transformer backbone. The model combines autoregressive cross-entropy loss for language tokens with denoising loss for image patches, weighted by a balancing coefficient. This joint training enables coherent multimodal understanding and generation, bridging discrete and continuous domains within a single framework.

Diffusion Forcing~\cite{chen2024diffusion} introduces a training paradigm where a diffusion model denoises sequences with independent noise levels for each token. It trains a causal next-token predictor that generates one or several future tokens while keeping past tokens unchanged. This approach combines the adaptability of next-token prediction, which allows variable-length generation, with the guidance capability of full-sequence diffusion, which steers sampling toward preferred outcomes. It thereby unifies autoregressive prediction and diffusion-based reasoning, enabling smooth long-horizon generation and consistent planning across sequential domains.

\paragraph{Cross-modal Generation and Translation}
High-fidelity cross-modal generation hinges on the quality of the conditioning signal that bridges modalities. DALL-E 3~\cite{betker2023improving} demonstrates how caption quality directly impacts text-to-image generation. The key insight is that conventional web-crawled image-text pairs are noisy and lack descriptive detail, causing diffusion models to ignore fine-grained prompt semantics. To address this, a bespoke image captioner is first trained to produce highly descriptive synthetic captions for the training images, and the text-to-image diffusion model is then retrained on these enriched captions. By improving the textual conditioning signal at training time rather than modifying the generation architecture, this approach substantially narrows the semantic gap between user intent and visual synthesis, achieving markedly better prompt adherence.

Recent advances in unified multimodal models exemplified by Show-o~\cite{xie2024show} push this integration further by combining autoregressive and diffusion modeling within a single transformer architecture. Rather than applying uniform autoregressive generation to all modalities, Show-o strategically employs modality-specific modeling strategies: text tokens are processed autoregressively with causal attention for sequential reasoning, while image tokens are generated through discrete denoising diffusion with full attention for high-quality visual synthesis. This hybrid approach is unified through an omni-attention mechanism that seamlessly coordinates between the two modeling paradigms, enabling the model to generate and reason over interleaved sequences of images and text while maintaining the efficiency of language models and the generation quality of diffusion models.

\subsubsection{Domain Adaptation}
While diffusion models excel within their training distribution, real-world applications often require generation in substantially different domains. Domain adaptation enables efficient transfer of learned representations to new domains without extensive retraining. Optimal transport, particularly through \emph{Schr\"odinger bridges}, provides principled approaches for this adaptation.

\paragraph{Schr\"odinger Bridges for Domain Transfer}
The Schr\"odinger bridge problem offers an elegant framework for connecting two probability distributions through a stochastic process that minimizes transport path energy and deviation from Brownian motion. This enables construction of optimal transport maps between the source distribution $p_0$ and target distribution $p_1$. The optimization seeks a path measure $P$ satisfying boundary conditions $P_0 = p_0$ and $P_1 = p_1$, while minimizing:
\begin{equation}
	\min_{P: P_0 = p_0, P_1 = p_1} \mathbb{E}_P\left[\int_0^1 \frac{1}{2}\|\boldsymbol{f}(\mathbf{x}, t)\|^2_2 \mathrm{d}t\right] 
	+ D_{\mathrm{KL}}(P \parallel Q),
\end{equation}
where $\boldsymbol{f}(\cdot, t)$ is the drift guiding transport dynamics, $Q$ denotes reference Brownian motion, and the Kullback-Leibler (KL) divergence encourages solutions close to natural diffusion. Here the drift corresponds to the controlled component of the stochastic process described by $P$, and the first term penalizes the kinetic energy of this control signal. The combined objective seeks the stochastic process closest to uncontrolled Brownian motion while satisfying the prescribed boundary distributions.

Diffusion Schr\"odinger Bridge (DSB)~\cite{de2021diffusion} provides a practical algorithm for learning Schr\"odinger bridges between arbitrary distributions through Iterative Proportional Fitting (IPF). The method alternates between fitting forward and backward drift networks, progressively refining the transport map to satisfy both boundary conditions simultaneously. Building upon the forward and reverse SDEs introduced in Eq.~\eqref{eq:diffusion-forward-sde} and~\eqref{eq:diffusion-reverse-sde}, DSB learns modified drift functions that replace the standard drift in the forward process and the score-based drift in the reverse process. These learned drifts guide samples from source to target during the forward process (time $0 \to 1$), while the reverse process enables generation from the target domain. The coupling between forward and reverse processes ensures that the learned transport map remains consistent and invertible, enabling bidirectional translation between domains.

It is instructive to contrast Schr\"odinger bridges with the flow matching framework introduced in Section~\ref{sec:efficient}. Standard flow matching learns deterministic ODE paths from a fixed Gaussian prior to the data distribution, whereas Schr\"odinger bridges generalize this setting to transport between two \emph{arbitrary} distributions through stochastic paths that minimize a kinetic energy objective. When the source distribution is Gaussian, the Schr\"odinger bridge reduces to a diffusion model with an optimized noise schedule; when both endpoints are data distributions, it enables tasks such as unpaired domain translation that lie beyond the scope of conventional flow matching. The OT-CFM variant of conditional flow matching~\cite{tong2024improving} approximates this optimal transport coupling in a simulation-free manner, bridging the gap between the two paradigms.

\paragraph{Few-shot Domain Adaptation}
Domain adaptation often requires learning from limited target examples. DreamBooth~\cite{ruiz2023dreambooth} demonstrates effective few-shot adaptation by fine-tuning on small target sets while preserving prior knowledge. The training objective balances adaptation to target instances with maintaining general class knowledge through a two-term loss: the first optimizes denoising on target samples with specialized identifiers (\eg \emph{``a photo of \texttt{[what]} dog''}), while the second regularizes using prior preservation samples from general class prompts (\eg \emph{``a photo of a dog''}). These terms are combined with a weighting coefficient to prevent overfitting while maintaining diverse generation.

\paragraph{Compositional Domain Generalization and Interpolation}
Compositional approaches enable generalization to unseen domain combinations. Composable Diffusion~\cite{liu2022compositional} demonstrates how domain-specific models compose at inference through score function combination. Score functions from different models combine linearly to guide generation satisfying multiple domain constraints simultaneously. This strategy assumes approximate independence between constraints and works best for complementary domains. For example, separate diffusion models conditioned on \emph{``a painting of a forest''} and \emph{``a river at sunset''} can be composed through appropriate score weighting to generate images satisfying both descriptions simultaneously, enabling zero-shot generalization to novel concept combinations.

The Image-to-Image Schr\"odinger Bridge (I2SB) framework~\cite{liu2023i2sb} provides a tractable instantiation of Schr\"odinger bridges for paired image-to-image translation. By exploiting the availability of paired source and target samples, I2SB analytically marginalizes the boundary conditions and reduces the bridge learning problem to a conditional denoising objective compatible with standard DDPM-style training~\cite{ho2020denoising}. This simulation-free formulation avoids the iterative forward-backward procedures required by general Schr\"odinger bridge solvers, making it both scalable and straightforward to implement. During inference, the learned bridge transports degraded inputs toward clean reconstructions along a time-indexed stochastic path, where intermediate bridge states naturally blend characteristics of both domains and enable smooth transitions with controllable strength.

\subsubsection{Task Generalization}
Diffusion models extend beyond data generation to decision-making and sequential planning, traditionally dominated by reinforcement learning. This has led to diffusion policies that reformulate control and planning as conditional generation tasks, enabling powerful task generalization.

\paragraph{Diffusion Policies for Sequential Decision Making}
Planning with Diffusion (\aka Diffuser)~\cite{janner2022planning} demonstrates model-based reinforcement learning by generating entire trajectories. Rather than learning separate dynamics and policy models, it jointly learns trajectory distribution over sequences $\boldsymbol{\tau} = (\mathbf{s}_0, \mathbf{a}_0, \mathbf{s}_1, \mathbf{a}_1, \ldots, \mathbf{s}_T)$ from initial state $\mathbf{s}_0$ (then action $\mathbf{a}_0$) over a planning horizon $T$, capturing both environment dynamics and policy. High-reward trajectories are preferentially sampled through guidance during reverse diffusion, performing inference-time planning by biasing generation toward desirable outcomes without additional training.

Further on, Diffusion Policy~\cite{chi2023diffusion} reconceptualizes action selection as conditional generation, sampling actions from a diffusion model conditioned on state observations and goals. Rather than directly predicting actions, the policy generates actions by denoising from Gaussian noise $\mathbf{a}_T \sim \mathcal{N}(\mathbf{0}, \mathbf{I})$ to clean actions $\mathbf{a}_0$ through learned reverse diffusion. The reverse process refines actions iteratively with Gaussian transitions:
\begin{equation}
	p_{\boldsymbol{\theta}}(\mathbf{a}_{t-1}|\mathbf{a}_t, \mathbf{s}) = 
	\mathcal{N}(\mathbf{a}_{t-1}; \boldsymbol{\mu}_{\boldsymbol{\theta}}(\mathbf{a}_t, t, \mathbf{s}), \sigma_t^2 \mathbf{I}),
\end{equation}
where $\boldsymbol{\mu}_{\boldsymbol{\theta}}$ is the learned mean function and $\sigma_t^2 \mathbf{I}$ is the isotropic covariance, with action $\mathbf{a}$ and state observation $\mathbf{s}$ being vector-valued\footnote{Note that the subscript $t$ in $\mathbf{a}_t$ denotes the diffusion denoising step within the action generation process, which is distinct from the environment time step in the sequential decision-making problem.}. This naturally handles multi-modal action distributions crucial for tasks with multiple valid solutions, while iterative refinement provides implicit planning where early steps capture high-level strategy and later steps refine execution details.

Specifically, when deployed in multi-agent systems, each agent can run an independent diffusion policy conditioned on its local observation to produce provisional future trajectories. Semantic tokens received from neighboring agents then enter as additional conditioning vectors, blending local perception with shared intent during the denoising process. This mechanism serves as the computational engine for the ``Local Proposal'' step in multi-agent coordination protocols, as will be detailed in Section~\ref{sec:agent}.

\paragraph{Continual and Active Learning}
Diffusion models facilitate continual learning where models adapt to new tasks while retaining previous knowledge. C-LoRA~\cite{smith2023continual} tackles catastrophic forgetting in diffusion models through continually self-regularized low-rank adaptation. Specifically, C-LoRA applies low-rank updates to the key-value projection matrices of cross-attention layers in the U-Net. When learning a new concept, the accumulated magnitude of past LoRA~\cite{hu2022lora} weight deltas acts as an element-wise penalty (via Hadamard product) on new updates, discouraging large changes to parameters that encode previously learned concepts. This self-regularization mechanism balances plasticity and stability without storing or replaying past data. Combined with randomly initialized customization prompts that omit the concept-class name to avoid token collisions, C-LoRA enables continual sequential learning of fine-grained visual concepts.

Active learning leverages uncertainty quantification in probabilistic generation~\cite{gal2017deep}. Sampling multiple trajectories from different noise initializations reveals uncertainty through output variance. High variance indicates regions benefiting from additional training data, providing a natural acquisition function for selecting informative samples. Chan~\etal~\cite{chan2024estimating} formalize this intuition through HyperDM, which combines the conditional diffusion model with a Bayesian hyper-network~\cite{krueger2017bayesian} to disentangle two distinct sources of uncertainty from a single trained model. By sampling different network weights from the hyper-network and running the conditional diffusion process multiple times for each weight configuration, the framework decomposes the total predictive variance into an aleatoric component (the mean of per-configuration sample variances, reflecting irreducible data noise) and an epistemic component (the variance of per-configuration sample means, reflecting insufficient model knowledge). The epistemic uncertainty map directly identifies regions where the model lacks training coverage, providing a principled acquisition signal for active data selection without requiring ensembles of independently trained models.

\paragraph{Feedback-driven Optimization}
Recent advances integrate human feedback into diffusion training. DDPO (Denoising Diffusion Policy Optimization)~\cite{black2024training} recasts the iterative denoising procedure as a multi-step Markov decision process (MDP): each denoising transition $p_{\boldsymbol{\theta}}(\mathbf{x}_{t-1}|\mathbf{x}_t, \mathbf{y})$ is treated as a stochastic policy action conditioned on the current noisy state $\mathbf{x}_t$ and the task prompt $\mathbf{y}$, while a scalar reward $r(\mathbf{x}_0, \mathbf{y})$ is assigned only to the final sample $\mathbf{x}_0$. The objective maximizes the expected reward over complete denoising trajectories:
\begin{equation}
	\mathcal{J}(\boldsymbol{\theta}) = \mathbb{E}_{\mathbf{y},\, \mathbf{x}_{0:T} \sim p_{\boldsymbol{\theta}}} \left[ r(\mathbf{x}_0, \mathbf{y}) \right].
\end{equation}

Because each transition is an isotropic Gaussian with known mean and variance, its log-likelihood is available in closed form, enabling exact per-step policy gradient estimation. The resulting gradient decomposes across denoising steps as:
\begin{equation}
	\nabla_{\boldsymbol{\theta}} \mathcal{J}(\boldsymbol{\theta}) = \mathbb{E} \left[ r(\mathbf{x}_0, \mathbf{y}) \sum_{t=1}^{T} \nabla_{\boldsymbol{\theta}} \log p_{\boldsymbol{\theta}}(\mathbf{x}_{t-1}|\mathbf{x}_t, \mathbf{y}) \right],
\end{equation}
where $\nabla_{\boldsymbol{\theta}} \log p_{\boldsymbol{\theta}}(\mathbf{x}_{t-1}|\mathbf{x}_t, \mathbf{y})$ denotes the Fisher score\footnote{Fisher score is the gradient of log density with respect to \emph{model parameters}, unlike Stein score which differentiates with respect to \emph{random variables}.} at each step. This formulation enables fine-tuning diffusion models for diverse downstream objectives such as aesthetic quality, safety constraints, and prompt-image alignment via vision-language model feedback.

\paragraph{Online Learning and Adaptation}
Diffusion models integrate with online learning for real-time adaptation. Diffusion-ES~\cite{yang2024diffusiones} combines diffusion-based trajectory generation with gradient-free evolutionary search for black-box optimization. At each iteration, a population of trajectories is sampled from a pre-trained diffusion model and scored by a (potentially non-differentiable) reward function. High-scoring trajectories are then mutated through a truncated diffusion process~\cite{zheng2022truncated}: a small number of forward noising steps are applied to perturb the trajectory, followed by the same number of reverse denoising steps to project it back onto the data manifold. This truncated noise-denoise cycle serves as a structure-preserving mutation operator, enabling efficient exploration of the solution space without requiring gradients or model parameter updates. This approach enables inference-time trajectory optimization for tasks with black-box objectives, including autonomous driving scenarios with non-differentiable safety constraints and language-shaped reward functions generated through few-shot LLM prompting.

%% file: sec/4_inverse-problem.tex
\section{Generative Semantic Communications:\\An Inverse Problem Lens} \label{sec:inverse}

The convergence of semantic communications and generative models presents a paradigm shift toward generative semantic communications. This section explores a novel perspective that frames the semantic decoding process as an \emph{inverse problem}, where the receiver must recover the original semantic information from noisy, distorted, or incomplete channel measurements. By leveraging the power of deep generative models, particularly through conditional diffusion models, we can address the inherent ill-posedness of this channel-involved reconstruction challenge, even in scenarios where the forward channel model is completely unknown.

The inverse problem formulation naturally captures the essence of generative semantic communications: the receiver's task is not merely to decode symbols but to reconstruct the underlying semantic content that best explains the received measurements through generative models.

\subsection{Problem Statement}
\label{subsec:problem_statement}

Consider a semantic communication system where a source signal $\mathbf{x} \in \mathcal{X}$ containing semantic information needs to be transmitted through a noisy channel. The encoding process maps the source to a semantic latent representation $\mathbf{z} = \mathcal{E}_{\boldsymbol{\phi}}(\mathbf{x})$, where $\mathcal{E}_{\boldsymbol{\phi}}: \mathcal{X} \to \mathcal{Z}$ is a semantic encoder parameterized by $\boldsymbol{\phi}$, and $\mathcal{Z}$ represents the semantic latent space~\cite{qin2025neural}.

According to Eq.~\eqref{eq:forward-model}, the communication system introduces degradation through a forward operator $\mathcal{A} = \mathcal{H} \circ \mathcal{E}_{\boldsymbol{\phi}}$, where $\mathcal{A}:\mathcal{X} \to \mathcal{Y}$ accounts for both the encoding transformation $\mathcal{E}_{\boldsymbol{\phi}}$ parameterized by $\boldsymbol{\phi}$ (\eg nonlinear transform coding~\cite{balle-ntc}) and channel effects $\mathcal{H}:\mathcal{Z} \to \mathcal{Y}$ (\eg Rayleigh fading). This yields the received measurement $\mathbf{y}$:
\begin{equation}
	\mathbf{y} = \mathcal{A}(\mathbf{x}) + \mathbf{n} = \mathcal{H}(\mathcal{E}_{\boldsymbol{\phi}}(\mathbf{x})) + \mathbf{n}, \label{eq:forward-comm}
\end{equation}
where $\mathbf{n} \sim \mathcal{N}(\mathbf{0}, \sigma_{\mathbf{n}}^2\mathbf{I})$ represents additive Gaussian channel noise with variance $\sigma_{\mathbf{n}}^2$ for simplicity\footnote{Some signal-dependent noise, \eg Poisson, cannot be written as additive noise but can also be dealt with according to Eq.~\eqref{eq:forward-comm}.}, and $\mathcal{Y}$ denotes the measurement space.

It is worth emphasizing that the mathematical utility of this inverse problem formulation lies in its abstraction: the channel operator $\mathcal{H}$ serves as a unified container that can encapsulate arbitrarily complex degradation pipelines beyond the simplified AWGN model. For instance, in frequency-selective fading scenarios, $\mathcal{H}$ acts as a convolution with the multipath channel impulse response~\cite{yang2022ofdm, zhu2022ofdm, xiao2025frequency}; for time-varying channels, the operator becomes time-indexed $\mathcal{H}(\cdot, t)$ within each coherence block~\cite{huang2019deep, romano2025investigating}, as will be further discussed in Section~\ref{subsec:potential_solutions}; and for nonlinear radio frequency (RF) impairments such as power amplifier saturation~\cite{borel2021linearization} and low-resolution analog-to-digital converter (ADC) clipping~\cite{liu2019low}, these distortions are absorbed into $\mathcal{H}$ without altering the posterior sampling methodology, since the guidance gradient is computed via automatic differentiation~\cite{baydin2018automatic} regardless of operator linearity. This generality ensures that the formulation natively accommodates practical wireless impairments encountered in real-world deployments.

\begin{figure}[t]
	\centering
	\includegraphics[width=0.95\columnwidth]{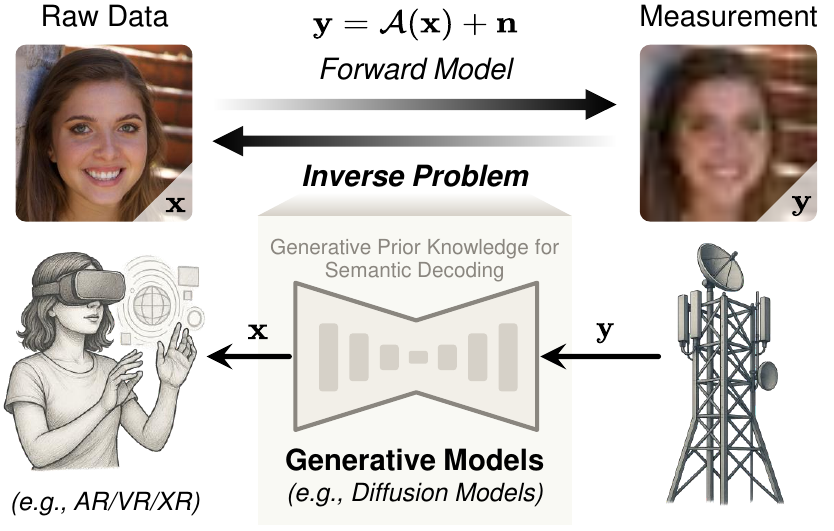}
	\caption{Generative model-enabled decoding process of semantic communications from the lens of inverse problems. \emph{Notations:} $\mathbf{x}$ denotes the raw data, $\mathcal{A}(\cdot)$ represents the forward operator, $\mathbf{y}$ denotes the measurement (\aka the channel received degraded signals), and $\mathbf{n}$ is the measurement noise.}
	\label{fig:inverse}
\end{figure}

In general, the generative model-enabled decoding process of semantic communications from the lens of inverse problems is illustrated in Fig.~\ref{fig:inverse}, where the channel received signals are meanwhile the measurement of generative semantic decoding for source recovery through solving the communication inverse problem given the forward model.

The semantic decoding task at the receiver can be formulated as the following inverse problem~\cite{daras2024survey}:
\begin{equation}
	\hat{\mathbf{x}} = \arg\min_{\mathbf{x} \in \mathcal{X}} \|\mathbf{y} - \mathcal{A}(\mathbf{x})\|^2_2 + \lambda \mathcal{R}(\mathbf{x}),
\end{equation}
where $\mathcal{R}(\mathbf{x})$ is a regularization term encoding \emph{prior knowledge} about the semantic structure of raw data, and $\lambda$ is a regularization parameter balancing data fidelity and prior constraints.

\subsubsection{Mathematical Ill-posedness}

The inverse problem in generative semantic communications exhibits fundamental challenges that render it ill-posed in the Hadamard sense:

\paragraph{Non-uniqueness} Multiple distinct source signals may map to similar semantic representations, particularly when the semantic encoder $\mathcal{E}_{\boldsymbol{\phi}}$ performs dimensionality reduction. Given a measurement $\mathbf{y}$, the set of feasible solutions $\mathcal{S}(\mathbf{y}) = \{\mathbf{x} \in \mathcal{X} : \|\mathbf{y} - \mathcal{A}(\mathbf{x})\|^2_2 \leqslant \varepsilon\}$ may contain multiple semantically equivalent but syntactically different elements, where $\varepsilon$ denotes a tolerance level determined by the measurement noise statistics.

\paragraph{Instability} Small perturbations in the received signal can lead to large variations in the reconstructed content. The condition number\footnote{Here, $\kappa(\mathcal{A})$ is the condition number of $\mathcal{A}$, defined as the ratio between its largest and smallest singular values. A large condition number indicates that the inverse problem is ill-conditioned, meaning small perturbations in the input can cause disproportionately large errors in the reconstruction.} of the forward operator $\mathcal{A}$ often exceeds practical bounds:
\begin{equation}
	\kappa(\mathcal{A}) = \|\mathcal{A}\| \cdot \|\mathcal{A}^{\dagger}\| \gg 1,
\end{equation}
where $(\cdot)^{\dagger}$ denotes the pseudo-inverse operator. This mathematical instability directly translates to extreme sensitivity to channel noise in wireless scenarios.

\paragraph{Non-existence} For severely degraded channels, there may be no source signal that exactly satisfies the measurement constraint, requiring relaxation to approximate solutions.

\subsubsection{Domain-specific Communication Challenges}

Based on the mathematical ill-posedness discussed above, we identify three critical challenges in semantic communications stemming from strong noise, highly nonlinear transformations, and time-varying channel states (Fig.~\ref{fig:challenges}):

\begin{figure}[t]
	\centering
	\includegraphics[width=\columnwidth]{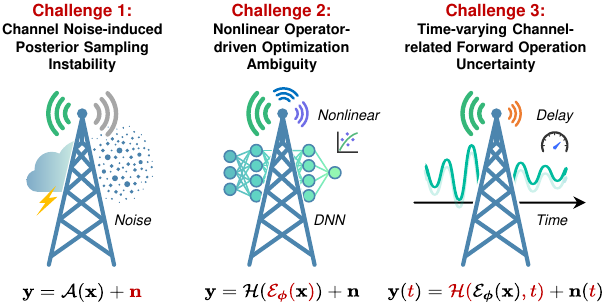}
	\caption{Three types of critical domain-specific challenges in generative semantic communications. \emph{Challenge 1:} Strong channel noise causes unstable posterior sampling and reconstruction variance. \emph{Challenge 2:} Highly nonlinear encoding creates multiple local minima leading to optimization ambiguity. \emph{Challenge 3:} Time-varying channels introduce unknown forward operations requiring adaptive inversion strategies.}
	\label{fig:challenges}
\end{figure}

\paragraph{Channel noise-induced posterior sampling instability} Channel noise is unavoidable in communication systems, particularly in wireless data transmission, where noise can significantly degrade the quality of raw data and confuse decoding. When the noise power is high relative to the signal power, the posterior distribution $p(\mathbf{x}|\mathbf{y})$ becomes highly dispersed, leading to unstable sampling. In wireless scenarios, this occurs under poor channel conditions where the signal-to-noise ratio (SNR) is extremely low. The uncertainty in received signals introduces greater randomness to the posterior sampling process of generative models, manifesting as increased variance and reduced consistency in reconstructed samples.
	
Fixed statistical channel models are typically employed to address noise-related issues, using channel coding and modulation techniques to mitigate the effects of noise on the raw data~\cite{haykin1988digital}. However, under extreme noise conditions, these models can fail to capture the complexity of noise, resulting in error accumulation during the decoding process.
	
Therefore, extreme noise can weaken or even invalidate posterior sampling of generative models at the receiver side, and addressing the instability caused by channel noise is one of the critical challenges for achieving high-fidelity and resilient transmission.
	
\paragraph{Nonlinear operator-driven optimization ambiguity} Modern semantic communication systems often employ DNNs for encoding, making $\mathcal{A}$ highly nonlinear~\cite{dai2022nonlinear}. This nonlinearity creates multiple local minima during optimization:
\begin{equation}
	\nabla_{\mathbf{x}} \|\mathbf{y} - \mathcal{A}(\mathbf{x})\|^2_2 = -2\mathbf{J}_{\mathcal{A}}^{\top}(\mathbf{x})(\mathbf{y} - \mathcal{A}(\mathbf{x})),
\end{equation}
where $\mathbf{J}_{\mathcal{A}}$ is the Jacobian of the forward operator $\mathcal{A}$. The complex structure of this Jacobian in neural network-based systems leads to gradient vanishing or explosion, causing the optimization of deep generative models to bias towards specific features, failing to maintain global semantic coherence.

In ultra-low bitrate scenarios, where data compression is more aggressive, the ambiguity is further amplified, exacerbating the uncertainty and risk of misjudgment in optimization objectives. In such cases, stochastic posterior sampling of diffusion models often proves to be ineffective.
	
\paragraph{Time-varying channel-related forward operation uncertainty} In real-world wireless scenarios, the forward operator $\mathcal{A}$ changes over time due to channel fading, mobility, and other environmental factors~\cite{yuan2024channel}. This means we face a blind inverse problem where $\mathcal{A}$ is unknown or time-varying:
\begin{equation}
	\mathbf{y}(t) = \mathcal{A}(\mathbf{x},t) + \mathbf{n}(t),
\end{equation}
where $\mathcal{A}(\cdot, t)$ represents the time-varying operator. This uncertainty makes it impossible to directly apply standard inverse problem solvers, requiring adaptive strategies that can handle unknown or changing forward operations.

When $\mathbf{y}$ becomes highly dynamic and $\mathcal{A}$ is unpredictable, posterior sampling of generative models may fail to accurately reconstruct the source at the receiver due to the uncertain or unknown forward operations, potentially generating samples that deviate from the true data manifold.

\subsubsection{Bayesian Formulation for Semantic Communications}

To handle the ill-posedness systematically, a Bayesian perspective is tailored for generative semantic communications. The posterior distribution of the source given measurement is:
\begin{equation}
	p(\mathbf{x}|\mathbf{y}) = \frac{p(\mathbf{y}|\mathbf{x})p(\mathbf{x})}{p(\mathbf{y})} \propto p(\mathbf{y}|\mathbf{x})p(\mathbf{x}),
\end{equation}
where the likelihood term, assuming Gaussian noise in the channel, takes the form:
\begin{equation}
	p(\mathbf{y}|\mathbf{x}) = \mathcal{N}(\mathcal{A}(\mathbf{x}), \sigma^2\mathbf{I}) \propto \exp\left(-\frac{\|\mathbf{y} - \mathcal{A}(\mathbf{x})\|^2_2}{2\sigma^2}\right).
\end{equation}

In practice, the likelihood $p(\mathbf{y}|\mathbf{x})$ is governed by a time-varying channel that is only partially observable via pilot estimation. Consequently, the mapping between $\mathbf{x}$ and $\mathbf{y}$ is neither explicit nor stationary as the above closed-form expression may suggest. Posterior sampling must therefore account for this reality by either propagating the channel estimation uncertainty into the guidance term or treating the channel parameters as latent variables within a blind inverse formulation, both of which will be detailed in Section~\ref{subsec:potential_solutions}.

The prior $p(\mathbf{x})$ encodes our knowledge about the semantic structure of valid source signals. In the context of diffusion models, this prior is learned from training data and captures the manifold of semantically meaningful content. Hence, the maximum a posteriori (MAP) estimate:
\begin{equation*}
	\hat{\mathbf{x}}_{\mathrm{MAP}} = \arg\max_{\mathbf{x}} p(\mathbf{x}|\mathbf{y}) = \arg\max_{\mathbf{x}} \log p(\mathbf{y}|\mathbf{x}) + \log p(\mathbf{x})
\end{equation*}
provides a principled approach to regularized reconstruction that balances measurement fidelity with semantic coherence.

Therefore, from an inverse problem perspective, generative semantic communications fundamentally shift the classical data transmission paradigm from MLE to a MAP formulation.

Specifically, the expressiveness of the learned generative prior directly governs how closely a diffusion-based semantic receiver can approach information-theoretic capacity limits. In the semantic decoding posterior $p(\mathbf{x}|\mathbf{y}) \propto p(\mathbf{y}|\mathbf{x})p(\mathbf{x})$, the true source prior $p(\mathbf{x})$ is approximated by a learned diffusion prior $p_{\boldsymbol{\theta}}(\mathbf{x})$. According to rate-distortion theory~\cite{cover06}, this prior mismatch incurs an excess semantic distortion bounded by the KL divergence:
\begin{equation}
	\Delta D \leqslant \frac{1}{\lambda^{\star}} D_{\mathrm{KL}}(p(\mathbf{x}) \| p_{\boldsymbol{\theta}}(\mathbf{x})),
\end{equation}
where $\Delta D$ denotes the excess semantic distortion and $\lambda^{\star}$ represents the slope of the semantic rate-distortion function at the operating point. Minimizing this divergence is therefore a fundamental prerequisite for approaching the optimal semantic compression rate~\cite{niu2024mathematical}. Recent theoretical analyses~\cite{debortoli2022convergence, chen2023sampling} establish that the discrepancy between the true data distribution $p(\mathbf{x})$ and the learned diffusion marginal $p_{\boldsymbol{\theta}}(\mathbf{x})$ is bounded by:
\begin{equation}
	\mathrm{TV}(p(\mathbf{x}), p_{\boldsymbol{\theta}}(\mathbf{x})) \leqslant \mathcal{O}(T^{-1} + \varepsilon_{\mathrm{score}}),
\end{equation}
where $T$ is the total number of denoising steps and $\varepsilon_{\mathrm{score}}$ denotes the score estimation error of the score-based diffusion model $\boldsymbol{s}_{\boldsymbol{\theta}}$. By Pinsker's inequality\footnote{Pinsker's inequality establishes a linkage between the maximum deviation in probability for any event (\eg TV distance) and the information-theoretic penalty of using an approximate prior (\eg KL divergence), formally stated as $\mathrm{TV}(p(\mathbf{x}), p_{\boldsymbol{\theta}}(\mathbf{x})) \leqslant \sqrt{\frac{1}{2} D_{\mathrm{KL}}(p(\mathbf{x}) \| p_{\boldsymbol{\theta}}(\mathbf{x}))}$.}, the TV distance and KL divergence are intrinsically coupled. This mutual dependence indicates that bounding the distributional discrepancy simultaneously minimizes the information-theoretic divergence. Consequently, reducing the score estimation error $\varepsilon_{\mathrm{score}}$ strictly guarantees a small KL divergence, which in turn tightly constrains the excess distortion $\Delta D$ (\ie $\varepsilon_{\mathrm{score}} \!\downarrow \implies \mathrm{TV\&KL} \!\downarrow \implies \Delta D \!\downarrow$). This theoretically justifies why high-capacity diffusion models with accurate score estimation are mathematically necessary to push generative semantic communications toward theoretical capacity limits.

In essence, interpreting diffusion-based semantic decoding from this Bayesian MAP formulation demonstrates that minimizing semantic distortion fundamentally depends on deploying various mechanisms to enhance the sampling capabilities of the generative diffusion prior.

\subsection{Potential Solutions}  \label{subsec:potential_solutions}

The ill-posed nature of the semantic communication inverse problem necessitates sophisticated solution strategies that can effectively leverage prior knowledge while maintaining computational tractability. Here, we present three potential solutions based on diffusion models.

\subsubsection{Training-time Conditional Diffusion Models for Semantic Reconstruction}

Diffusion models provide a powerful framework for solving inverse problems by learning to reverse a gradual noising process. In semantic communications, a training-time conditional diffusion model can be employed to directly learn the posterior distribution $p(\mathbf{x}|\mathbf{y})$, enabling effective semantic reconstruction at the receiver.

As discussed in Section~\ref{sec:conditional}, the key idea is to condition the score model $\boldsymbol{s}_{\boldsymbol{\theta}}(\mathbf{x}|\mathbf{y}, t)$ on the channel measurement $\mathbf{y}$. Specifically, the score model is trained to predict the conditional score given both the noisy state and the degraded measurement, thereby embedding channel knowledge into the generative process.

Representative works along this line include Gen-SC~\cite{wei2024language} and CDM-JSCC~\cite{yang2024rate}. The former pre-parses captions corresponding to the source images and transmits them as textual prompts carrying semantic information. At the receiver side, these prompts are fed into a fine-tuned LDM to generate samples aligned with the source semantics. The latter, in contrast, employs a conditional diffusion model in the pixel-level data space~\cite{yang2023lossy}, guided by images compressed through a rate-adaptive encoder, thereby indirectly conveying semantics.

However, the content generated by Gen-SC often exhibits significant deviations from the original source, while CDM-JSCC suffers from the difficulty of training conditional diffusion models and shows strong sensitivity to adverse wireless channel conditions. When the joint encoder-decoder fails, the reconstruction can collapse dramatically.

The core problem is that these methods rely on manually crafted semantic conditions to steer the generative process. However, such handcrafted conditions may fail to capture the complexity of real channel degradations, making it hard to train models that can robustly handle time-varying channels.

\subsubsection{Inference-time Conditional Diffusion Models for Semantic Inversion}

Compared to the training-time conditional diffusion model-based solution, the inference-time method based on estimator guidance provides a more flexible framework for solving inverse problems in generative semantic communications by combining learned priors with measurement consistency.

At the receiver, channel measurement $\mathbf{y}$ is employed as the external guidance for score-based generative sampling with measurement consistency. Consider different channel states with known or unknown forward operators, two cases are presented below.

\paragraph{Inversion for Known Forward Operators} When the forward operator $\mathcal{A}$ is known or can be accurately estimated, inversion strategies such as DPS~\cite{chung2022diffusion} guides the posterior sampling process using the measurement according to Eq.~\eqref{eq:dps}, where the measurement guidance term $\nabla_{\mathbf{x}} \mathcal{R}\left(\mathbf{y}, \mathcal{A}(\hat{\mathbf{x}}_{0|t})\right)$ ensures that the generated samples remain consistent with the channel received signal, while the diffusion prior guides the reconstruction towards semantically meaningful content. This balance is particularly crucial for human semantic communications where both fidelity to the recovered signal and semantic coherence are important.

Note that the reconstruction quality of the above inversion process is inherently sensitive to the guidance scale $\gamma$ in Eq.~\eqref{eq:dps}, which dictates the balance between the diffusion prior $p_{\boldsymbol{\theta}}(\mathbf{x})$ and the measurement likelihood $p(\mathbf{y}|\mathbf{x})$. In dynamic wireless environments where the channel SNR fluctuates rapidly, a fixed $\gamma$ is often suboptimal: an overly large $\gamma$ amplifies channel noise into the reverse sampling trajectory, degrading reconstruction quality at low SNR, whereas an insufficient $\gamma$ causes the unconditional prior to dominate and leads the diffusion model to hallucinate reconstructions that entirely deviate from the transmitted semantics. To address this sensitivity, three channel state information (CSI)-adaptive calibration strategies can be considered:
\begin{itemize}
	\item \emph{Noise-level scheduling}: Setting $\gamma \propto 1/\sigma_{\mathbf{n}}^2$, which directly maps the pilot-estimated channel noise variance $\sigma_{\mathbf{n}}^2$ to the guidance scale, ensuring that the guidance strength decreases gracefully as noise power grows~\cite{bao2021adjscc}.
	\item \emph{Step-dependent annealing}~\cite{bansal2023universal}: Applying weaker measurement guidance at early diffusion steps where the score prior dominates the coarse semantic structure, and stronger guidance at later refinement steps where measurement fidelity matters most.
	\item \emph{Pseudo-inverse guidance}: $\Pi$GDM~\cite{song2023pseudoinverseguided} computes an analytic likelihood gradient using the pseudo-inverse of the forward operator, eliminating heuristic $\gamma$ tuning entirely when the channel matrix is available.
\end{itemize}

These strategies provide principled mappings from instantaneous channel conditions to the guidance scale, enabling robust posterior sampling under non-stationary wireless links.

Beyond hyperparameter sensitivity, the DPS gradient approximation itself introduces a structural limitation. The core approximation $\nabla_{\mathbf{x}} \log p_t(\mathbf{y}|\mathbf{x}) \approx \nabla_{\mathbf{x}} \log p(\mathbf{y}|\hat{\mathbf{x}}_{0|t})$ relies on the posterior mean $\hat{\mathbf{x}}_{0|t}$ accurately capturing the measurement dependency, which holds under mild nonlinearities but degrades when the forward operator involves severe nonlinear distortions (\eg power amplifier saturation) or significant model mismatch~\cite{daras2024survey}. In such regimes, the approximation error accumulates across reverse sampling steps, leading to biased or unstable reconstructions. To mitigate this, Decoupled Annealed Posterior Sampling (DAPS)~\cite{zhang2025improving} decouples consecutive samples in the diffusion sampling trajectory and employs a noise annealing process (\eg Langevin dynamics), allowing each iterate to explore the solution space more freely while ensuring the time-marginal distributions converge to the true posterior. This decoupled structure prevents early-step approximation errors from propagating through the entire sampling chain, substantially improving stability for nonlinear inverse problems. For linear channel operators, $\Pi$GDM~\cite{song2023pseudoinverseguided} offers an alternative by computing the likelihood gradient analytically through the pseudo-inverse of the forward operator, bypassing the gradient approximation entirely and thus eliminating the associated bias.

\paragraph{Blind Inversion for Unknown Forward Operators} In practical scenarios where the channel state is unknown or time-varying, the vanilla inversion methods are extended to handle blind inverse problems. According to Section~\ref{sec:conditional}, the joint estimation of source and channel follows:
\begin{equation}
	(\hat{\mathbf{x}}, \hat{\mathcal{A}}) = \arg\min_{\mathbf{x}, \mathcal{A}} \|\mathbf{y} - \mathcal{A}(\mathbf{x})\|^2_2 + \lambda \mathcal{R}(\mathbf{x}) + \gamma \mathcal{R}(\mathcal{A}), \label{eq:blind-comm}
\end{equation}
where the hat notation $\hat{\mathbf{x}}$ and $\hat{\mathcal{A}}$ indicate estimated variables, $\mathcal{R}(\mathbf{x})$ and $\mathcal{R}(\mathcal{A})$ are regularization terms encoding prior knowledge on the source and forward operator, and $\lambda, \gamma$ are weighting coefficients controlling the strength of the corresponding regularizations.

This can be solved through alternating optimization based on Eq.~\eqref{eq:blind-dps}, where we iteratively update the source estimate using DPS with the current channel estimate, then update the channel estimate based on the current source reconstruction. 

It should be noted that the above blind parametric formulation rests on two key assumptions: \emph{(i)}~the channel can be tightly parameterized by a low-dimensional vector $\boldsymbol{\vartheta}$ (\eg a multipath Rayleigh fading channel $\mathcal{H}_{\boldsymbol{\vartheta}}$ with a small number of independent paths~\cite{yang2022ofdm}), and \emph{(ii)}~the channel remains quasi-static over the duration of the reverse diffusion inference, \ie the coherence time exceeds the total sampling latency. When either condition is violated, such as in rapidly time-varying or non-stationary environments where the channel statistics shift within a single inference window, the parametric blind assumption collapses, and the alternating optimization in Eq.~\eqref{eq:blind-dps} may diverge or converge to a physically inconsistent solution. In such regimes, fallback strategies are necessary, such as real-time pilot-based channel tracking~\cite{ye2017power} that continuously updates $\mathcal{H}$ during sampling, or hybrid approaches that initialize the diffusion posterior with a deterministic minimum mean square error (MMSE) warm-start before applying generative refinement~\cite{arvinte2022score}.

When the above conditions are satisfied, the blind formulation is computationally lightweight in practice. The joint estimation appends a gradient update $\nabla_{\boldsymbol{\vartheta}} \|\mathbf{y} - \mathcal{H}_{\boldsymbol{\vartheta}}(\mathcal{E}_{\boldsymbol{\phi}}(\hat{\mathbf{x}}_{0|t}))\|^2_2$ at each reverse step, introducing marginal per-step overhead proportional to the dimension of $\boldsymbol{\vartheta}$. Crucially, the core diffusion prior $\boldsymbol{s}_{\boldsymbol{\theta}}$ is pre-trained offline on the source data distribution and requires no retraining when channel conditions change; only the low-dimensional channel parameters $\boldsymbol{\vartheta}$ are adapted online during inference. This separation enables effective channel estimation-free semantic decoding with minimal additional latency beyond the standard reverse sampling process~\cite{wang2024diffcom}. These boundary conditions and feasibility considerations together delineate the current applicability of blind diffusion-based semantic decoding and highlight the need for adaptive inference mechanisms in highly dynamic wireless environments.

Moreover, a common and normal assumption underlying both the known and blind inversion strategies above is that the channel noise $\mathbf{n}$ follows a Gaussian distribution, which justifies the squared $\ell_2$ measurement residual $\|\mathbf{y} - \mathcal{A}(\hat{\mathbf{x}}_{0|t})\|^2_2$ as the data-fidelity term~\cite{chung2022diffusion, chung2023parallel}. However, practical wireless environments frequently exhibit non-Gaussian disturbances such as impulsive interference (\eg co-channel collisions in dense networks), burst errors, and heavy-tailed residuals induced by imperfect equalization. Under these conditions, the $\ell_2$ gradient can be dominated by a small number of outlier samples, destabilizing the reverse sampling trajectory. Two robust alternatives can mitigate this fragility: \emph{(i)}~replacing the $\ell_2$ norm with a Huber loss~\cite{huber1964robust}, which behaves quadratically for small residuals but transitions to linear penalization beyond a threshold, naturally down-weighting burst-error outliers; and \emph{(ii)}~adopting a Student-$t$ log-likelihood~\cite{jylanki2011robust}, whose heavier tails accommodate impulsive residuals without inflating the guidance gradient. Developing principled blind posterior sampling algorithms that jointly handle non-stationary, non-Gaussian interference and unknown forward operators remains an open research challenge, as further discussed in Section~\ref{sec:open}.

\paragraph{Robust Inversion Under Weak Diffusion Priors}
The inversion strategies discussed above implicitly assume that the receiver's diffusion prior is well-matched to the transmitted content: that is, the generative model has been trained on data drawn from the same or a similar distribution as the source data. In practice, however, this assumption is frequently violated. The receiver may only have access to a generic pre-trained model (\eg trained on diverse natural images) while the transmitter sends domain-specific content (\eg medical imagery, satellite photos, or industrial inspection data). Moreover, computational constraints at the edge receiver may force heavy truncation of the diffusion sampling process to just a few denoising steps, severely degrading the generative fidelity of the prior itself. These scenarios give rise to what Jia~\etal~\cite{jia2026weak} term \emph{weak diffusion priors}: models that are either domain-mismatched or low-fidelity due to inference-time truncation.
	
A key question for the practical deployment of diffusion-based semantic decoding is therefore: under what conditions can weak priors still yield reliable reconstructions? The answer, as established in~\cite{jia2026weak}, depends critically on the \emph{informativeness} of the channel measurement $\mathbf{y}$. The central theoretical insight is rooted in Bayesian posterior consistency: when the measurement provides sufficient information about the source, the likelihood term $p(\mathbf{y}|\mathbf{x})$ dominates the posterior distribution, effectively ``washing out'' the influence of the prior. This phenomenon can be formalized through the notion of \emph{identifiability}: if candidate signals that differ from the ground truth $\mathbf{x}^\star$ can be reliably distinguished through the forward model (\ie their projected measurements $\mathcal{A}(\mathbf{x})$ and $\mathcal{A}(\mathbf{x}^\star)$ differ substantially in many observed dimensions), then even a severely mismatched prior cannot mislead the posterior away from $\mathbf{x}^\star$. Concretely, for linear forward models $\mathbf{y} = \mathbf{A}\mathbf{x} + \mathbf{n}$, the posterior concentration strengthens exponentially with the effective observed dimensionality $m$ of the measurement: incorrect candidate signals typically disagree with the observation across many dimensions simultaneously, and the accumulated evidence overwhelms whatever bias the prior might introduce. As a result, very different priors can produce nearly identical reconstructions when the observed information is rich.
	
This theoretical framework maps naturally onto the generative semantic communication setting. In typical scenarios where the semantic encoder transmits a moderately compressed latent representation, the channel measurement retains substantial information about the source, corresponding to the data-informative regime where weak priors provably suffice. For example, consider a semantic encoder with compression ratio $r = m/D$ (where $D$ is the source dimension and $m$ is the measurement dimension): when $r$ is not too aggressive (\eg $r > 0.3$ for random linear measurements), the effective observed dimensionality is large enough to ensure posterior concentration regardless of prior quality. Conversely, under extreme compression (\eg $r < 0.1$) or when the forward operator masks contiguous semantic regions (\eg block-structured information loss), the measurement becomes insufficiently informative, and reconstruction quality becomes unavoidably prior-dominated. In this regime, the diffusion prior must provide strong semantic generation capability to fill in missing content, and domain mismatch leads to visible artifacts.
	
These findings carry two practical implications for semantic communication system design. First, they provide a principled justification for deploying generic pre-trained diffusion models as semantic decoders without domain-specific fine-tuning: as long as the semantic encoding rate is above a task-dependent threshold, the measurement consistency mechanism in Eq.~\eqref{eq:dps} is sufficient to steer even an out-of-domain prior toward accurate reconstruction. Second, they delineate a clear boundary between the measurement-dominated regime (where the encoder design and channel quality determine reconstruction accuracy) and the prior-dominated regime (where the generative model's domain coverage becomes the bottleneck), offering guidance for jointly optimizing semantic compression ratios and diffusion prior selection. These design considerations are further revisited in the context of human and machine semantic communication applications in Section~\ref{sec:human} and~\ref{sec:machine}.

\subsubsection{Semantic-aware Regularization for Diffusion Posterior Sampling}

The measurement regularization $\nabla_{\mathbf{x}} \|\mathbf{y} - \mathcal{A}(\hat{\mathbf{x}}_{0|t})\|^2_2$ employed in the preceding inversion strategies corresponds to the maximum-likelihood estimator under the Gaussian noise assumption, enforcing that the reconstruction remains consistent with the received signal. However, measurement consistency alone is insufficient to guarantee semantic correctness: at low SNR, where the measurement $\mathbf{y}$ itself is heavily corrupted, an observation-consistent reconstruction may faithfully reproduce the noise pattern while drifting to an entirely wrong semantic class. In other words, the $\ell_2$ data-fidelity term provides a \emph{necessary} condition for signal-level accuracy, but it does not provide a \emph{sufficient} condition for preserving task-relevant meaning. This gap motivates the introduction of an explicit semantic regularizer that anchors the sampling trajectory to the correct semantic manifold regardless of channel quality.

The semantic nature of the generative semantic communication systems allows for specialized regularization that preserves meaning rather than pixel-level fidelity:
\begin{equation}
	\mathcal{R}_{\mathrm{sem}}(\mathbf{x}) = \|\mathcal{E}_{\mathrm{sem}}(\mathbf{x}) - \mathcal{E}_{\mathrm{sem}}(\mathbf{x}_{\mathrm{ref}})\|^2_2,
\end{equation}
where $\mathcal{E}_{\mathrm{sem}}$ is a semantic feature extractor (\eg CLIP~\cite{radford2021learning} and DINO~\cite{caron2021emerging} encoders), and $\mathbf{x}_{\mathrm{ref}}$ is a reference with similar semantic content. This regularization can be integrated into the DPS framework through an additional guidance term $\lambda \nabla_{\mathbf{x}}\mathcal{R}_{\mathrm{sem}}(\hat{\mathbf{x}}_{0|t})$, where $\hat{\mathbf{x}}_{0|t}$ denotes the posterior mean given $t$, and $\lambda$ is a hyperparameter controlling the strength of semantic guidance. This additional gradient can be seamlessly combined with the score-based update during the posterior sampling of diffusion models to bias the generation toward semantically coherent reconstructions, potentially unifying understanding, reconstruction, and generation~\cite{li-imagefolder, yao-rec&gen}.

Combining measurement and semantic regularization, the full posterior score for diffusion-based generative semantic decoding takes the form:
\begin{equation*}
	\boldsymbol{s}(\mathbf{x}|\mathbf{y}, t) \approx \underbrace{\boldsymbol{s}_{\boldsymbol{\theta}}(\mathbf{x}, t)}_{\text{prior}} + \gamma \underbrace{\nabla_{\mathbf{x}} \mathcal{R}(\mathbf{y}, \mathcal{A}(\hat{\mathbf{x}}_{0|t}))}_{\text{measurement reg.}} + \lambda \underbrace{\nabla_{\mathbf{x}} \mathcal{R}_{\mathrm{sem}}(\hat{\mathbf{x}}_{0|t})}_{\text{semantic reg.}},
\end{equation*}
where $\gamma$ and $\lambda$ weight the measurement consistency and semantic guidance respectively. An interesting ``gradient contention'' arises between the latter two terms at low SNR: when $\mathbf{y}$ is heavily corrupted, the measurement gradient pulls the sampling trajectory toward noisy, potentially semantically incorrect reconstructions, while the semantic gradient steers it toward the correct semantic class regardless of noise. These two gradients can become antiparallel when the measurement error is large, causing the sampling trajectory to oscillate or converge slowly. When $\gamma$ is large relative to $\lambda$, the reconstruction overfits the degraded channel measurement at the expense of semantic integrity; when $\lambda$ dominates $\gamma$, semantic consistency is preserved but signal-level fidelity may suffer. Resolving this conflict requires either careful scheduling of the relative weights across diffusion steps, or introducing a robust anchor that inherently reconciles both objectives, a concrete instantiation of which is the confirming constraint mechanism adopted by DiffCom~\cite{wang2024diffcom} in Section~\ref{sec:human}.

%% file: sec/5_application.tex
\section{Diffusion Models for\\Generative Semantic Communications} \label{sec:app}
Traditional communication systems treat the receiver as a passive component that simply inverts the encoder's operations. Generative semantic communications fundamentally change this paradigm by making the receiver an active participant in the reconstruction process~\cite{grassucci2024generative}. Instead of deterministic inversion, the receiver now synthesizes outputs that best match the channel received signal while conforming to the semantic constraints of the application domain. This generative approach connects naturally to inverse problem theory. When signals traverse an encoder and a noisy channel, the receiver faces an ill-posed inverse problem~\cite{zhang2024semantic, chen2024commin}: multiple plausible reconstructions may explain the same received measurements.

Generative models address this ambiguity by incorporating learned priors that capture the statistical structure of valid outputs, thereby guiding the reconstruction toward semantically meaningful solutions. Among various generative modeling techniques, diffusion models have become increasingly prominent in semantic communications due to several practical advantages. They learn to reverse a gradual noising process, which allows them to generate high-quality samples without the training instabilities often encountered in adversarial approaches. More importantly for semantic communications, diffusion models naturally handle conditional generation: they can incorporate various side information such as text descriptions, compressed features, or noisy channel outputs to guide the generation process~\cite{zhao2025lrgd}. Additionally, their iterative refinement mechanism provides inherent robustness to perturbations from both compression and channel noise~\cite{pei2025latent}.

It is worth clarifying the specific role of diffusion models within the semantic communication pipeline: they serve as the \emph{generative semantic decoder} at the receiver, responsible for reconstructing source content from degraded channel measurements via conditional posterior sampling. The communication-theoretic grounding of this role is established through the forward operator decomposition $\mathcal{A}(\mathbf{x}) = \mathcal{H}(\mathcal{E}_{\boldsymbol{\phi}}(\mathbf{x}))$ formulated in Section~\ref{sec:inverse}, where $\mathcal{E}_{\boldsymbol{\phi}}$ denotes the neural semantic encoder and $\mathcal{H}$ models the physical wireless channel. This decomposition explicitly couples the generative modeling process to link-layer operations: the encoder $\mathcal{E}_{\boldsymbol{\phi}}$ implements neural source-channel coding~\cite{qin2025neural} that maps source data into channel-input representations, while the diffusion decoder inverts the composite degradation by leveraging learned data priors under measurement guidance from the received signal $\mathbf{y}$.

Diffusion-based generative semantic communications serve three distinct use cases (see Fig.~\ref{fig:comm}), each with its own design priorities and performance metrics:
\begin{figure*}[t]
	\centering
	\includegraphics[width=\textwidth]{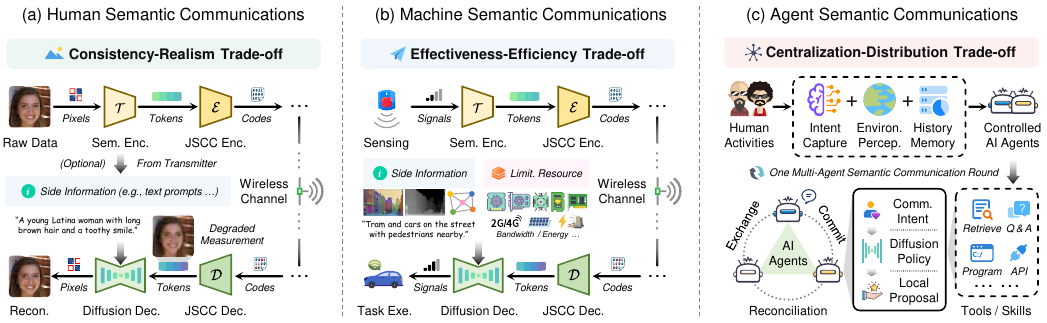}
	\caption{Three typical scenarios of diffusion-based semantic communications. (a) Fidelity-oriented human semantic communications aiming to optimize the consistency-realism trade-off. (b) Task-specific machine semantic communications aiming to optimize the effectiveness-efficiency trade-off. (c) Intent-driven agent semantic communications aiming to optimize the centralization-distribution trade-off. \emph{Abbreviations:} ``Enc.'' and ``Dec.'' stand for ``Encoder'' and ``Decoder'', respectively. In line with the standardized coding workflow proposed in~\cite{qin2025neural}, Joint source-channel coding (JSCC) is adopted for compressive coding in both human and machine semantic communication scenarios, where ``tokens'' and ``codes'' refer to ``semantic tokens'' and ``latent codes'', respectively. Additionally, ``Exe.'' stands for ``Execution'' and ``Percep.'' for ``Perception''. All abbreviations are introduced for brevity.}
	\label{fig:comm}
\end{figure*}

\begin{itemize}
	\item \textbf{Fidelity-oriented Human Semantic Communications}: When communicating for human consumption, such as images or video, the system must balance two competing objectives. On one hand, the reconstructed content should accurately preserve the original information. On the other hand, it should appear natural and realistic to human perception. We refer to this as the consistency-realism trade-off. Diffusion models excel here because they can generate perceptually realistic outputs while being conditioned on features that preserve semantic content. 
	
	\item \textbf{Task-specific Machine Semantic Communications}: When the receiver is a machine performing a specific task, communication efficiency becomes paramount. The question becomes: what is the minimal information needed to achieve the task objective? For instance, an autonomous vehicle may only need to transmit information relevant for obstacle detection rather than full scene reconstruction. Diffusion models help extract and refine task-relevant features from compressed representations, enabling effective task execution under strict bandwidth and energy constraints.
	
	\item \textbf{Intent-driven Agent Semantic Communications}: In multi-agent systems, coordination can be achieved through centralized or distributed architectures. Centralized approaches use powerful hubs that maintain global state but may create bottlenecks. Distributed approaches employ specialized agents that operate autonomously but must align their understanding through communication. Diffusion models facilitate this alignment by enabling agents to share and refine probabilistic representations of their intents and plans. This allows scalable coordination even when agents have limited views of the environment.
\end{itemize}

We begin with human-centric semantic communications in Section~\ref{sec:human}, first examining how diffusion models enable generative compression without channel effects. This establishes the core mechanisms. We then incorporate wireless channel impairments to develop complete end-to-end systems. Next, we turn to machine-centric semantic communications in Section~\ref{sec:machine}, showing how diffusion priors can be specialized for task-specific objectives under resource constraints. Finally, we explore multi-agent systems in Section~\ref{sec:agent}, demonstrating how diffusion-based intent representations and alignment enable effective coordination in distributed settings.

\subsection{Fidelity-oriented Human Semantic Communications} \label{sec:human}
Human-oriented semantic communications are fundamentally fidelity-driven: the system design must not only minimize bitrate to conserve transmission bandwidth, but also ensure that human perceptual experience at the receiver is fully satisfied. This requirement becomes especially stringent in high-resolution immersive applications such as virtual reality (VR), telepresence, and real-time streaming. Against this backdrop, two major paradigm shifts define the move from traditional discriminative designs to diffusion-based solutions.
\begin{itemize}
	\item \textbf{Paradigm Shift in Evaluation}: The quality criteria for human-centric data transmission have evolved from distortion measures toward perceptual metrics. These measures emphasize alignment with the natural data distribution rather than pixel-level similarity. Correspondingly, the optimization objective has shifted from the classical rate-distortion (RD) to the rate-distortion-perception (RDP) trade-off~\cite{blau2018perception}, where perceptual realism becomes an explicit dimension in system design.
	
	\item \textbf{Paradigm Shift in Reconstruction}: At the receiver, the decoding process has transitioned from discriminative regression to generative sampling. Traditional autoencoder-based decoders aimed to approximate the source deterministically, whereas diffusion-based generators leverage data-driven priors to sample reconstructions. Because the diffusion model has learned the underlying data distribution, sampling trajectories can balance generation probability and complexity, ultimately producing reconstructions that are both consistent with the transmitted semantics and realistic to human perception~\cite{theis2024makes}.
\end{itemize}

In the following, we expand on how diffusion models concretely enable fidelity-oriented human semantic communications. We first highlight the intrinsic link between generative compression and semantic transmission, showing that compression forms the necessary foundation for transmission.

\subsubsection{Consistency-Realism Trade-off}
As we all know, data compression serves as a cornerstone in end-to-end wireless transmission, fundamentally underpinning source coding. Its primary goal is to reduce the number of bits needed to represent useful information. Shannon's information theory~\cite{shannon1948mathematical} addresses both lossless compression, through the source coding theorem, and lossy compression, through the \emph{rate-distortion theorem}, with the latter being particularly crucial for visual transmission applications. A key challenge in lossy compression, specifically when compressing a single continuous source sample, lies in \emph{quantization}, which is the process of mapping continuous infinite values to a discrete finite set. This process, while analogous to analog-to-digital conversion, introduces quantization errors and potentially impedes gradient backpropagation.

In the context of lossy compression, the discretized representation (rate) and the error arising from the quantization (distortion) should be carefully traded off. Contemporary approaches have largely shifted from manual design to \emph{neural compression}, leveraging neural networks, particularly autoencoders. In a typical autoencoder framework, the encoder $\mathcal{E}$ transforms an input image $\mathbf{x}$ into a quantized latent code $\hat{\mathbf{z}} = \left \lfloor \mathcal{E}(\mathbf{x}) \right \rceil$ through the quantization operator $\left \lfloor \cdot \right \rceil$, while the decoder $\mathcal{D}$ generates an approximated reconstruction $\hat{\mathbf{x}} = \mathcal{D}(\hat{\mathbf{z}})$. Drawing from Shannon's rate-distortion theory, neural compression optimizes the rate-distortion (RD) trade-off through end-to-end neural network training, minimizing the mean squared error (MSE) according to:
\begin{equation}
	\mathcal{L} = R(\hat{\mathbf{z}}) + \lambda  D(\mathbf{x}, \hat{\mathbf{x}}),
\end{equation}
where $R(\hat{\mathbf{z}}) = -\log_2 P(\hat{\mathbf{z}})$ is an estimate of the bitrate, with $P$ representing the probability model of $\hat{\mathbf{z}}$. The distortion measure $D(\mathbf{x}, \hat{\mathbf{x}})$ commonly employs metrics such as MSE or peak signal-to-noise ratio (PSNR), while $\lambda$ serves as a rate-controlling hyperparameter balancing the RD trade-off.

Beyond traditional discriminative neural networks like auto-encoders, deep generative models have emerged as preferred solutions, owing to their superior capacity for capturing complex data distributions and generating perceptually pleasing outputs. However, conventional distortion measures $D(\mathbf{x}, \hat{\mathbf{x}})$, typically based on pixel-wise differences, often lead to unnatural artifacts and oversmoothing in reconstructions, particularly at low bitrates. As Blau and Michaeli \cite{blau2018perception} observe, this can result in mode averaging behavior manifesting as blurriness, failing to ensure that reconstructed images faithfully reflect natural image distributions. This limitation has motivated a shift toward perceptual quality metrics $d(p_{\mathbf{x}}, p_{\hat{\mathbf{x}}})$, where lower values indicate better quality, utilizing various distribution divergence measures such as Kullback-Leibler (KL), Jensen-Shannon (JS), or Wasserstein distances.

Incorporating perceptual losses, rate-distortion theory has evolved into \emph{rate-distortion-perception theory} \cite{blau2019rethinking}. The rate-distortion-perception (RDP) trade-off can be formally optimized through the following formulation:
\begin{align}
	R(\delta, \varphi) := &\min_{p_{\hat{\mathbf{x}} | \mathbf{x}}} \, I(\mathbf{x}; \hat{\mathbf{x}}), \\
	& \  \st \ \mathbb{E}\left[ D(\mathbf{x}, \hat{\mathbf{x}}) \right] \leqslant \delta, \enspace d(p_{\mathbf{x}}, p_{\hat{\mathbf{x}}}) \leqslant \varphi,\nonumber
\end{align}
where $I(\mathbf{x}; \hat{\mathbf{x}})$ represents mutual information~\cite{cover06}, $\delta$ bounds the expected distortion of decoded signals, and $\varphi$ constrains perceptual quality. Notably, perceptual loss and distortion differ from each other, thus minimizing one does not necessarily reduce the other.

Here, the RDP trade-off is instantiated as a \emph{consistency-realism trade-off}, where consistency (\ie faithfulness) measures the content consistency between the original and recovered data at receiver side, while realism (\ie naturalness) refers to the perceptual quality of recovered data.

\subsubsection{Diffusion-based Generative Compression}
In this context, implementing neural compression through deep generative models yields \emph{generative compression}, offering two significant advantages:
\begin{itemize}
	\item \textbf{Distribution Fitting}: The sophisticated distribution fitting capabilities of generative models improve entropy coding, enabling effective data compression at low bitrates.
	
	\item \textbf{Perception Alignment}: The inherent perception-oriented properties of deep generative models contribute to improved fidelity of compressed images, resulting in outputs that better align with human visual perception.
\end{itemize}

In terms of evaluation metrics for visual quality assessment (VQA) in fidelity-driven generative compression, three categories of VQA metrics have been developed in~\cite{zhai-vqa}, including full-reference, reduced-reference, and no-reference metrics, among which the full-reference (\eg LPIPS~\cite{zhang2018unreasonable}, DISTS~\cite{ding2022IQA}) and no-reference metrics (\eg NIQE~\cite{mittal2012niqe}, FID~\cite{heusel2017gans}) are usually employed in practice, effectively optimizing the consistency-realism trade-off.

With appropriate evaluation metrics, the objective of generative compression is to design a generator capable of producing reconstructions ranging from low MSE (prioritizing consistency) to high perceptual quality (prioritizing realism), or anything in between, all derived from the compressed representations~\cite{agustsson2023multi}. Score-based diffusion models are well-suited for robust generative compression due to their inherently stochastic generative process, which use stochastic posterior sampling to create multiple reconstructions from a corrupted input, enhancing the resilience against various perturbations. CDC~\cite{yang2023lossy}, a pioneering diffusion model-based generative compression method, replaces deterministic decoders with more expressive conditional diffusion models. It introduces discrete ``content'' latent variables from raw images as conditioning factors, while detail-related ``texture'' variables are synthesized during the reverse sampling process of conditional diffusion models.

Generative compression's advantages are most evident in the exceptional perceptual quality of compressed samples, even at extremely low bitrates. This property well satisfies the requirements of human semantic communications due to the joint consideration of ultra-low bitrates and perfect perceptual quality. Here, two representative approaches implemented using conditional diffusion models are introduced for reference: PerCo~\cite{careil2023towards} and IPIC~\cite{xu2024idempotence}.

PerCo enhances image quality and eliminates bitrate dependency through conditional latent diffusion models. This approach addresses the persistent artifacts in deterministic generative compression methods (\eg NTC~\cite{balle-ntc}, HiFiC~\cite{mentzer2020high}), which remain even when training with perceptual or adversarial losses at low bitrates. The LDM employed by PerCo conditions the decoding process on both vector-quantized latent representations and image captions (from BLIP-2~\cite{li2023blip}), providing rich contextual information. Through extensive experiments on the Kodak dataset\footnote{Kodak dataset (\url{https://r0k.us/graphics/kodak}): $24$ images for evaluation, each $768\times512$ or $512\times768$ in size.}, PerCo demonstrates superiority over state-of-the-art codecs at bitrates from $0.1$ to $0.003$ bits per pixel (bpp), achieving compression using less than $153$ bytes per image on average. As mentioned in Section~\ref{sec:conditional}, we can find that PerCo actually employs the training-time conditioning in Eq.~\eqref{eq:training-canonical} to finetune the conditional latent diffusion model, with global textual conditions as the class label $\mathbf{y}$ and local latent codes as the denoised sample $\mathbf{x}$.

Similarly, one can also employ the inference-time conditioning to construct conditional diffusion models for generative compression, which leads to the introduction of IPIC. It innovatively applies the concept of ``idempotence'' to lossy image compression at ultra-low bitrates. Idempotence refers to the stability of the image codec under re-compression, which is actually a conceptual interpretation of estimator guidance ability of inference-time conditional diffusion models to solve inverse problems.

Technically, IPIC demonstrates that conditional generative model-based codecs satisfy idempotence, and that an unconditional generative model with an idempotence constraint is equivalent to a conditional generative codec. Based on this insight, IPIC proposes constructing a generative codec by integrating idempotence constraints into an unconditional diffusion model. From the lens of inverse problems, idempotence implicitly ensures that the forward model is either invertible or admits a pseudo-inverse, allowing the generative compression to be framed as an inverse problem with a closed-form solution, which can be effectively addressed using inversion methods. For further theoretical details and experimental results, please refer to~\cite{xu2024idempotence}.

A potential challenge for diffusion model-based generative compression lies in the fact that the guidance provided by side information may be insufficient to enable the conditional diffusion model to generate samples that match the original image's texture details, even though the perceptual quality may be satisfactory. Content consistency may deteriorate, especially in ultra-low bitrate compression scenarios. A feasible solution is to refine side information to strengthen its guiding role, such as incorporating multi-level conditions~\cite{xu2025picd}. Another approach is to optimize the conditional diffusion model, improving the model’s controllable sampling, as discussed in Section~\ref{sec:conditional}, for instance by considering improved variants of DPS~\cite{zhang2025improving}. Additionally, the time cost of recursive sampling remains an inherent challenge for diffusion-based generative compression. This issue has been further discussed in Section~\ref{sec:efficient}.

In conclusion, conditional diffusion models show great promise in enabling efficient generative data compression, which produces perceptually pleasing samples with improved fidelity at the receiver side, thus paving the way for fidelity-oriented human-centric generative transmission.

\subsubsection{Human-centric Generative Transmission}
Combining aforementioned domain-specific challenges in Section~\ref{sec:inverse} with generative compression methods, we present a detailed introduction to a novel end-to-end human-centric generative visual semantic communication paradigm that utilizes off-the-shelf score-based diffusion priors with stochastic posterior sampling, as proposed in \cite{wang2024diffcom} with the framework DiffCom. The core insight of this work is that \emph{channel received signal is a natural fine-grained condition to guide diffusion posterior sampling}, which improves the perceptual quality of reconstructed images and preserves fidelity without heavily relying on bandwidth costs or the received signals quality.

Channel noise is an inherent issue in wireless transmission, especially when the channel is in poor condition. Noise not only disrupts the transmitted signal but also leads to severe distortion in the received signal. Traditional communication systems rely on fixed channel coding and modulation schemes, but as noise intensifies, the effectiveness of these methods diminishes, particularly when it comes to maintaining high perceptual quality in reconstructed signals.

In the context of end-to-end DeepJSCC framework~\cite{bourtsoulatze2019deep}, DNNs are widely introduced to realize powerful neural encoders. However, this introduces a problem: such highly nonlinear encoders are hard to invert, making the reconstruction task highly ill-posed\cite{erdemir2023generative}. As a result, optimization for posterior sampling becomes too ambiguous to return reconstructions that truly align with the source.

Therefore, in practical communication setups, one faces the challenge of working with highly noisy channel-received signals as the measurement, while performing inversion through a complex and highly nonlinear forward model. These factors significantly elevate the difficulty of keeping stable and effective DPS, which in turn impacts the quality of reconstructions.

To tackle these issues, DiffCom introduces a key insight: traditional inverse solvers often overfit the forward operator $\mathcal{A}$. In other words, while the log-likelihood gradient $\nabla_{\mathbf{x}} \log p_t(\mathbf{y} | \mathbf{x})$ approximation in DPS may lead to a small-enough measurement distance, inspecting $\hat{\mathbf{x}}_{0|t}$ easily reveals that it is not well-aligned with the received signal $\mathbf{y}$. In light of the instability and ambiguity caused by high-power noise and the nonlinear nature of neural encoders in communication setups, DiffCom emphasizes the need for a critical balance: exploring $\mathbf{x}_t$ in proximity to the generative prior for the underlying semantic structure, while exploiting the received latent signal to enhance measurement consistency.

Specifically, DiffCom introduces a confirming constraint and an adaptive ancestral sampling strategy to counteract noise and nonlinear interference, thereby improving both the speed and effectiveness of posterior sampling.

The core idea behind the confirming constraint is to introduce an additional constraint during posterior sampling to ensure that the samples generated by the diffusion model not only satisfy the received signal conditions but also better align with the true data distribution. This approach effectively addresses the common optimization instability and ambiguity problem, ensuring that the generated images are consistent in both local details and global structures. The extended objective function for posterior sampling in DiffCom is $\boldsymbol{s}(\mathbf{x}|\mathbf{y}, \mathbf{x}_{0|\mathbf{y}}, t) \approx \boldsymbol{s}_{\boldsymbol{\theta}}(\mathbf{x}, t) + \nabla_{\mathbf{x}} \log p_t(\mathbf{y}, \mathbf{x}_{0|\mathbf{y}} | \mathbf{x})$, and can be further approximated as follows (supposing the received latent signal $\mathbf{y}$ as the measurement):
\begin{equation*}
	\underbrace{\boldsymbol{s}_{\boldsymbol{\theta}}(\mathbf{x}, t)}_{\text{prior (pixel)}} + \gamma \nabla_{\mathbf{x}} \underbrace{\| \mathbf{y} - \mathcal{H}(\mathcal{E}_{\boldsymbol{\phi}}(\hat{\mathbf{x}}_{0|t})) \|_2^2}_{\text{measure. reg. (latent)}}  + \lambda \nabla_{\mathbf{x}} \underbrace{\| \mathbf{x}_{0|\mathbf{y}} - \Tilde{\mathbf{x}}_{0|t} \|_2^2}_{\text{confirm. reg. (pixel)}},
\end{equation*}
where $\gamma$ is the step size of measurement regularization, and $\lambda$ controls the strength of confirming constraint in the pixel-level data space. Notably, $\mathcal{E}_{\boldsymbol{\phi}}(\hat{\mathbf{x}}_{0|t}) = \hat{\mathbf{z}}_{0|t}$ represents the latent posterior mean produced by the pre-trained neural JSCC encoder $\mathcal{E}_{\boldsymbol{\phi}}(\cdot)$ parameterized by $\boldsymbol{\phi}$, and $\mathbf{x}_{0|\mathbf{y}} = \mathcal{D}_{\boldsymbol{\varphi}}(\mathbf{y})$ is denoted as a deterministic MSE-optimized reference reconstructed with the paired neural JSCC decoder $\mathcal{D}_{\boldsymbol{\varphi}}(\cdot)$ parameterized by $\boldsymbol{\varphi}$. Besides, we denote $\Tilde{\mathbf{x}}_{0|t} = \mathcal{D}_{\boldsymbol{\varphi}}(\mathcal{H}^{\dagger}(\mathcal{H}(\mathcal{E}_{\boldsymbol{\phi}}(\hat{\mathbf{x}}_{0|t}))))$, which explicitly considers the noisy wireless channel-related forward operator $\mathcal{H}$ and nonlinear JSCC codec.

\begin{figure}[t]
	\centering
	\includegraphics[width=\columnwidth]{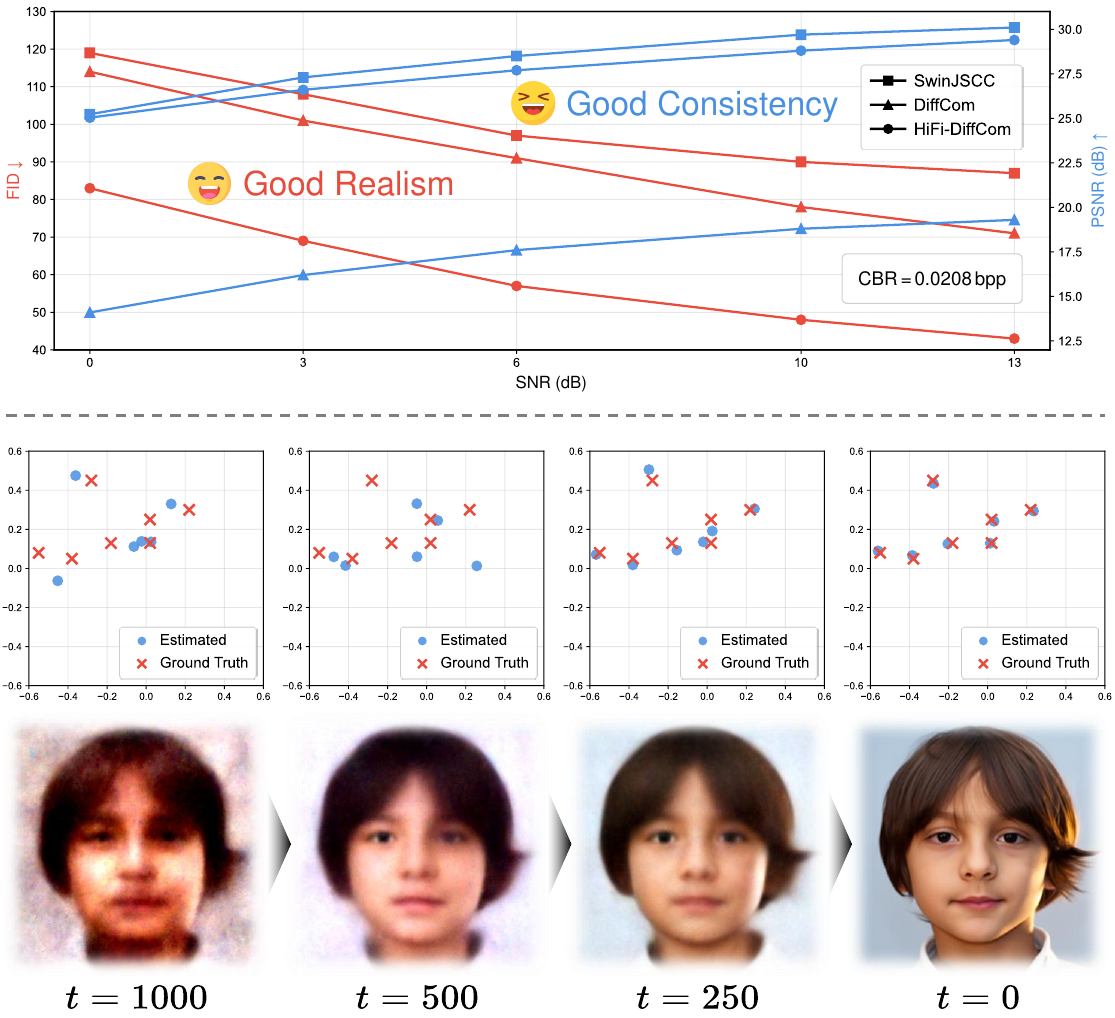}
	\caption{Primary quantitative and qualitative results reproduced from \cite{wang2024diffcom} over an AWGN channel. \emph{Notations:} Channel bandwidth ratio (CBR) is the ratio between transmitted and original signal dimensions. \emph{Setups:} CBR is $0.0208$ bpp. Metrics FID (realism, lower $\downarrow$ better) and PSNR (consistency, higher $\uparrow$ better) are evaluated for SwinJSCC~\cite{yang2024swinjscc}, vanilla DiffCom, and HiFi-DiffCom on FFHQ dataset~\cite{karras2019style}. Below the dashed line: qualitative results where channel response $\redcross$ is estimated by $\bluedot$. First row shows estimated $\hat{\mathbf{h}}_{0|t}$ at $t=1000, 500, 250, 0$; second row shows posterior mean $\hat{\mathbf{x}}_{0|t}$ evolution. \emph{Conclusions:} DiffCom series balances consistency and realism better than the baseline, with stable blind inversion demonstrating effectiveness for channel estimation-free transmission.}
	\label{fig:case-result}
\end{figure}

By introducing the confirming constraint $\| \mathbf{x}_{0|\mathbf{y}} - \Tilde{\mathbf{x}}_{0|t} \|_2^2$, the model not only optimizes perceptual quality (ensuring samples lie on the data manifold) but also aligns the reconstruction with the deterministic baseline $\mathbf{x}_{0|\mathbf{y}}$, ensuring the generated sample does not deviate from the global characteristics of raw data.

To speed up posterior sampling and enhance robustness, DiffCom introduces a simple yet effective strategy to adaptively adjust the starting point of the sampling process to counteract noise interference. Instead of starting from a isotropic Gaussian distribution (pure noise), the model begins sampling from a lightly noisy signal, \ie the reference sample $\mathbf{x}_{0|\mathbf{y}}$ obtained from the channel received signal. In the meanwhile, DiffCom determines an optimal intermediate time step aligned with wireless channel quality referred to \cite{saidutta2021joint}. By running the ancestral sampling via Eq.~\eqref{eq:diffusion-reverse-sde} from $\mathbf{x}_{0|\mathbf{y}}$, DiffCom can generate high-fidelity reconstruction results with much fewer sampling steps. By the way, to distinguish from the vanilla DiffCom, \cite{wang2024diffcom} refers to this improved method, which incorporates the confirming constraint and adaptive sampling, as HiFi-DiffCom.

The time-varying channel environment exacerbates the uncertainty in channel responses, leading to inconsistencies in the observed forward operation between the transmitter and receiver, which may result in decoding errors during real-time communications. In other words, the actual channel response $\mathbf{h}^*$ in wireless operations $\mathcal{H}^*$ is nearly unknown, and should be approximated by $\mathbf{h}$, \ie estimating the unknown forward operator $\mathcal{H}_{\boldsymbol{\vartheta}}$ parameterized by $\boldsymbol{\vartheta}$, which is often derived using channel estimation (CE) algorithms~\cite{haykin1988digital}. 

To tackle extreme scenarios where CE is unavailable or highly inaccurate, which are modeled as blind inverse problems caused by time-varying channels, \cite{wang2024diffcom} introduces Blind-DiffCom incorporating an extra score-based diffusion model that samples the channel response based on its distribution \cite{chung2023parallel}. Blind-DiffCom further considers the OFDM transmission configurations adopted in \cite{yang2022ofdm}, modeling a multipath Rayleigh fading channel with $L$ independent paths, and $\mathbf{h} \sim \mathcal{C} \mathcal{N}(\mathbf{0}, \sigma_{\mathbf{h}}^2 \mathbf{I}_{L})$. For detailed setups, please refer to \cite{yang2022ofdm}. Similar in Eq.~\eqref{eq:blind-comm}, the source and channel vector can be jointly estimated. A natural derivation of posterior score can be acquired from Eq.~\eqref{eq:blind-dps} as follows:
\begin{equation}
	\boldsymbol{s}(\mathbf{x}, \mathbf{h} | \mathbf{y}, t) \approx \boldsymbol{s}_{\boldsymbol{\theta}}(\mathbf{x}, t) + \gamma \nabla_{\mathbf{x}} \mathcal{R}\left(\mathbf{y}, \mathcal{H}_{\hat{\boldsymbol{\vartheta}}}(\mathcal{E}_{\boldsymbol{\phi}}(\hat{\mathbf{x}}_{0|t}))\right) , \label{eq:blind-diffcom}
\end{equation}
where $\mathcal{H}_{\hat{\boldsymbol{\vartheta}}}$ denotes a wireless channel-related forward operator parameterized by the estimated channel posterior mean $\hat{\mathbf{h}}_{0|t}$, which is computed as Eq.~\eqref{eq:posterior-mean}. By recursively running the joint estimation process in Eq.~\eqref{eq:blind-diffcom}, Blind-DiffCom incurs only minimal additional computational costs, but achieves significantly effective DPS for the blind semantic decoding inverse problems, thus extending the generalization boundary of end-to-end generative visual semantic communication based on diffusion models to in-the-wild scenarios.

The primary qualitative and quantitative results of DiffCom simulated over an additive white Gaussian noise (AWGN) channel are reproduced and illustrated in Fig.~\ref{fig:case-result}. For more experimental setups and results, please refer to \cite{wang2024diffcom}.

In a word, DiffCom introduces a novel end-to-end generative visual semantic communication framework toward humans by incorporating score-based diffusion priors, effectively addressing the three key challenges in wireless image semantic transmission. Whether it is through HiFi-DiffCom to resolve instability and optimization ambiguity in posterior sampling, or through Blind-DiffCom to handle forward operation uncertainty in time-varying channels, DiffCom demonstrates strong adaptability to dynamic channel environments and complex noise conditions. By leveraging the channel-received signal as a fine-grained condition, DiffCom achieves high-fidelity image reconstruction while reducing reliance on bandwidth costs and signal quality, providing a more efficient and resilient solution for human-centric generative semantic communications.

\subsection{Task-specific Machine Semantic Communications} \label{sec:machine}
Task-specific machine semantic communications address scenarios in which the receiver is a \emph{machine} (\eg a perception or decision module in autonomous driving, uncrewed aerial vehicle (UAV) swarms, or industrial inspection)~\cite{shi2023task}. In this regime, the goal is not to reproduce pixel-level content for human consumption but to deliver just enough information for the downstream task to succeed under strict resource budgets (\eg bandwidth, energy, memory, and latency)~\cite{letaief2019roadmap}. This reframes ``what to transmit'' and ``what to reconstruct'' from signal fidelity to \emph{utility for the task}. In particular, perception tasks (\eg detection, segmentation, tracking) and decision tasks (\eg control, navigation, manipulation) can operate directly on compact semantic representations~\cite{qin2025neural}, while deep generative models at the receiver can denoise, complete, or reconcile partial semantics to recover task-critical structure under channel impairments. Diffusion models are especially attractive here~\cite{liu2025two}: their conditional sampling naturally fuses the received semantics with strong learned priors, improving robustness and utility when measurements are noisy, sparse, or quantized.

\subsubsection{Effectiveness-Efficiency Trade-off}
A defining characteristic of machine-centric settings is the \emph{effectiveness-efficiency} trade-off: maximize downstream task effectiveness (\eg mean Average Precision (mAP) and Intersection over Union (IoU) for perception, success rate and cumulative return for decision) subject to tight efficiency constraints (\eg spectral efficiency, device memory and floating-point operations (FLOPs), inference latency, and energy). Practically, this trade-off is tuned by choosing what ``semantic cues'' to extract and transmit (\eg object masks, keypoints, bird’s eye view (BEV) grids, and high-level tokens), how to compress them (\eg entropy models, vector quantization, and pruning), and how to exploit them at the receiver (\eg direct task heads, and conditional generation with measurement consistency). Diffusion models contribute on both sides: \textit{(i)} they \emph{improve effectiveness} by hallucinating task-relevant detail consistent with the received semantics; and \textit{(ii)} they \emph{improve efficiency} by enabling aggressive source compression (and even lossy or partial semantics) while maintaining task performance through conditional generation at the receiver.

\subsubsection{Diffusion-enabled Machine Perception}
A representative line of work demonstrates that transmitting compact semantic maps or features and regenerating task-sufficient content with diffusion at the receiver yields resilience to channel noise and strong task utility. Grassucci \etal \cite{grassucci2023generative} advocate replacing bit-accurate recovery with semantic regeneration: the transmitter conveys compressed semantic descriptions (\eg segmentation maps), and a diffusion model at the receiver synthesizes consistent images or intermediate representations that preserve the semantics needed by perception models. The key insight is to let the generative prior ``fill in'' high-frequency details and resolve ambiguities induced by the channel, improving downstream perception compared with purely discriminative reconstructions at similar rates. This paradigm (called GESCO) reports robustness across channel conditions and demonstrates that conditional diffusion models can serve as a semantics-to-signal bridge for perception tasks, rather than targeting strict pixel fidelity. 

Concretely, a machine-perception pipeline proceeds as follows. The transmitter extracts task-centric representations (\eg class-agnostic masks, panoptic labels, sparse keypoints, or learned tokens) and entropy-codes them under a rate budget. The receiver runs a conditional diffusion model whose inputs include the received semantics (and possibly channel side information). The diffusion sampler produces either \textit{(i)} a \emph{proxy signal} (\eg a regenerated view) on which legacy detectors/segmenters operate, or \textit{(ii)} \emph{task-focused embeddings} by decoding into the feature space consumed by a lightweight task head. In both cases, the model leverages strong priors to compensate for missing details and channel artifacts, improving detection/segmentation utility at low rates relative to discriminative decoders. The same mechanism extends naturally to multi-view or BEV settings by conditioning on fused semantics (\eg map priors or cross-sensor tokens) before sampling.

\subsubsection{Resource-constrained Machine Decision}
While diffusion priors enhance task utility, their memory and compute demands can exceed what edge devices tolerate~\cite{yang2022semantic}. Pignata \etal \cite{pignata2024lightweight} introduce Q-GESCO, a \emph{lightweight} semantic diffusion model tailored for resource-constrained semantic communication. The core idea is post-training quantization of the diffusion model to reduce both model size and FLOPs while retaining generative capability for semantic regeneration. Q-GESCO reports up to $75\%$ memory savings and $79\%$ FLOPs reduction with performance comparable to full-precision counterparts and robustness across channel noises, enabling embedded receivers to harness diffusion priors within tight budgets. This provides a practical recipe: train a conditional diffusion model for task utility, then apply post-training quantization (and compatible sampling schedules) to deploy on-device without retraining the communication stack.

At a more economical level, diffusion model-based semantic communications shift energy expenditure from the transmitter to the receiver, and the system achieves a net energy benefit if and only if the transmission energy saved by sending fewer bits offsets the neural inference cost at the receiver:
\begin{equation}
	\Delta E_{\mathrm{tx}} := E_{\mathrm{tx}}^{\mathrm{conv}} - E_{\mathrm{tx}}^{\mathrm{sem}} \geqslant E_{\mathrm{inf}},
\end{equation}
where $E_{\mathrm{tx}}^{\mathrm{conv}}$ and $E_{\mathrm{tx}}^{\mathrm{sem}}$ denote the per-sample transmission energy of conventional and semantic codecs respectively, and $E_{\mathrm{inf}}$ represents the receiver-side neural inference energy. This break-even inequality formalizes the fundamental mechanism of trading communication energy for computation energy~\cite{mao2017survey, shao2021communication}: the left-hand side captures the bandwidth savings enabled by aggressive semantic compression, while the right-hand side reflects the computational burden of reconstructing the missing details via iterative diffusion sampling. In practice, $E_{\mathrm{inf}}$ remains substantial on edge devices. For instance, even with optimized implementations, generating a single $512\times512$ image via a 20-step Stable Diffusion v1.4 pipeline takes approximately $12$ seconds on flagship mobile GPUs (\eg Adreno 740 on Samsung S23 Ultra and A16 on iPhone 14 Pro Max)~\cite{chen2023speed}; extrapolating to a full-step DDPM~\cite{ho2020denoising} receiver with $1000$ function evaluations, $E_{\mathrm{inf}}$ would dominate the system energy budget, and the break-even condition is satisfied only under severely constrained link budgets (\eg ultra-low SNR or ultra-narrow bandwidth) where the transmission savings are correspondingly large. The practical path to improving this balance is to compress $E_{\mathrm{inf}}$ through the acceleration and quantization techniques surveyed in Section~\ref{sec:efficient}: Consistency Models~\cite{song2023consistency} collapse the sampling trajectory to as few as sub-10 steps or even $1$ step~\cite{luo2023latent} (see Table~\ref{tab:efficiency}), while training-free 8-bit (INT8) post-training quantization~\cite{migacz20178bit, shang2023post} further reduces per-step energy. These complementary optimizations can push the break-even point into a regime that is viable for edge deployment~\cite{liu2025two}.

The computational profiling summarized in Table~\ref{tab:efficiency} can be mapped to concrete 5G or 6G service categories to guide deployment decisions. Full-step diffusion models with NFE $\geqslant50$ incur multi-second latencies and are suited primarily to offline or delay-tolerant broadband scenarios where reconstruction quality is prioritized over real-time response~\cite{guo2025enhancing, ntontin2025vision}. Moderately accelerated models operating at NFE $\approx10$ to $20$ (\eg latent diffusion with fast solvers) bring the end-to-end latency into the sub-second range, making them compatible with enhanced mobile broadband (eMBB) applications such as high-resolution image and video streaming~\cite{yin2025generative, yan2025semantic, xie2025wireless, li2026goal}. Single-step generators (\eg Consistency Models~\cite{song2023consistency, luo2023latent} with NFE $=1$), when combined with INT8 quantization~\cite{shang2023post}, can push generative reception toward the sub-100\,ms latency regime relevant to selected edge inference scenarios. However, meeting the stringent sub-1\,ms latency requirements of ultra-reliable low-latency communications (URLLC) remains beyond the reach of current diffusion architectures, as discussed in Section~\ref{sec:open}.

From a system-design lens, the resource-aware diffusion-based machine semantic communication framework unlocks several complementary operating modes along the effectiveness-efficiency frontier, each corresponding to a different allocation of communication bandwidth and receiver-side computation:
\begin{itemize}
	\item \textbf{Semantic-first, Generation-optional}: The transmitter conveys minimal semantic cues (\eg task masks or compact tokens) via aggressive source-channel coding. The receiver first attempts direct task inference on the received semantics without invoking the diffusion model; conditional diffusion sampling is activated only when the channel quality degrades below a threshold or when the task head's prediction confidence falls short, thereby amortizing the computational cost of iterative generation across time.
	\item \textbf{Hybrid Task-utility Paths}: A lightweight task head produces a provisional decision from the received semantics immediately upon arrival. In parallel, a budget-limited diffusion process (\eg a small number of denoising steps or a single-step consistency model~\cite{song2023consistency}) refines the semantic features or regenerates a proxy view for the same input. The two outputs are then fused, for example through confidence-weighted ensembling or late feature concatenation, yielding a final decision that combines the low-latency advantage of the discriminative path with the distributional robustness of the generative path~\cite{nguyen2025contemporary}.
	\item \textbf{Semantic Latent Space Sampling}: Rather than decoding to the high-dimensional pixel space, the diffusion model operates entirely within a compact semantic latent space~\cite{qin2025neural} that is pre-aligned with the downstream task backbone (\eg the feature space of a detection or segmentation network). This eliminates the cost of full-resolution image synthesis and avoids a subsequent re-encoding step, reducing the per-sample denoising cost while preserving task utility. The resulting latent samples can be directly consumed by the task head, closing the loop between generative decoding and task execution in a single representation space.
\end{itemize}

\subsection{Intent-driven Agent Semantic Communications} \label{sec:agent}

Modern multi-agent systems, from embodied robots and aerial swarms to networked industrial cells, increasingly communicate \emph{intent} rather than raw data. By intent we mean compact, structured summaries of goals, constraints, and anticipated actions that let peers coordinate without reconstructing full sensory streams~\cite{zhang2024intellicise}. Generative semantic communications operationalize this idea: agents exchange low-rate messages that serve as conditions for stochastic generators at the receiver, which then synthesize the beliefs, trajectories, or task plans most compatible with local context. Diffusion models are particularly attractive here. Their iterative denoising embodies a principled way to fuse priors with partial evidence; their conditioning interfaces admit heterogeneous side information (\eg role descriptors, task prompts, peer hypotheses, confidence estimates); and their stochasticity scales gracefully from single-agent inference to population-level coordination where uncertainty and diversity are features, not bugs.

\subsubsection{Centralization-Distribution Trade-off}
A central design axis is the balance between \emph{centralized} and \emph{distributed} intent processing. Centralized hubs (a ``semantic control plane'') offer global consistency, strong cross-agent credit assignment, and amortized compute for large diffusion priors, but risk bottlenecks, single points of failure, and reduced agility. Fully distributed schemes maximize flexibility and robustness, each agent carries a lightweight prior and communicates only when local surprises arise, but may suffer from partial observability, redundant computation, and non-convergent negotiations. In practice, intent-driven systems benefit from hybrid hierarchies~\cite{dev2025advanced}: edge agents perform fast, on-board intent forecasting and event-triggered messaging; regional aggregators reconcile conflicting intents via short consensus rounds; and optional cloud controllers refine long-horizon plans at low frequency. Diffusion models fit naturally at all three layers~\cite{yang2024agent}: small distilled priors for reflexive edge behavior, mid-size conditional priors for neighborhood fusion, and rich world-model priors for strategic replanning. The operative balance is a moving target shaped by scene complexity, fleet size, link reliability, and latency-energy budgets; the goal is enough centralization to align global semantics, and enough distribution to keep the system resilient and economical.

\subsubsection{Representing and Communicating Intent}
Intent messages should be \textit{(i)} compact, \textit{(ii)} compositional, and \textit{(iii)} uncertainty-aware. A practical recipe is to tokenize intent into a small set of latent factors (\eg goal, affordance mask, temporal horizon, safety margin, confidence)~\cite{jiang2023motiondiffuser}, and let a conditional diffusion prior at the receiver sample trajectories, occupancy forecasts, or action proposals conditioned on those factors and on local measurements. Compared to exchanging pixel maps or dense features, these messages encode what matters for joint behavior, while the receiver’s generator reconstructs how to realize it in situ. Text-like prompts (\eg mission directives, role descriptions) and graph descriptors (\eg who-acts-on-what) can serve as additional conditions. Multi-modality is natural: an agent may attach a handful of visual or geometric tokens only when they change the posterior materially, yielding sparsity by design. Crucially, uncertainty is first-class~\cite{demiris2007prediction}: intents are interpreted as distributions (\eg a set of plausible routes with calibrated likelihoods), enabling peers to reason about collisions, complementarity, and contingency plans.

To quantify the compactness advantage, consider an intent vector encoding $K$ temporal waypoints over a planning horizon, where each waypoint comprises spatial coordinates and a scalar confidence value. This requires only $\mathcal{O}(K)$ floating-point values, typically on the order of hundreds of bits after entropy coding, which is orders of magnitude smaller than transmitting raw sensor data (\eg LiDAR point clouds or camera frames). The communication payload thus scales logarithmically with the perception dimensionality, making intent-based exchange fundamentally more bandwidth-efficient than data-level or feature-level sharing.

\subsubsection{Coordination Protocols with Diffusion Priors}

\begin{figure}[t]
	\centering
	\includegraphics[width=\columnwidth]{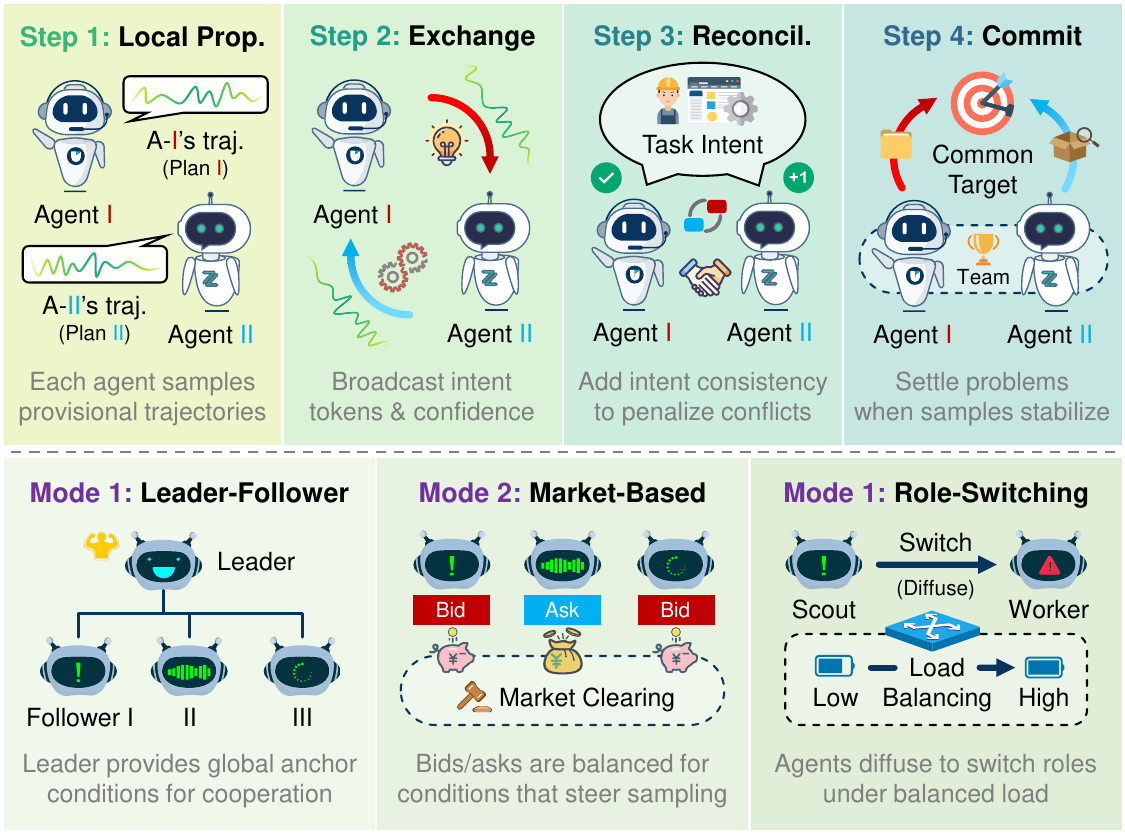}
	\caption{Coordination protocols with diffusion models in intent-driven agent semantic communications. \emph{Top:} Four-step minimal round for multi-agent coordination to achieve task-specific targets under intent constraints. \emph{Bottom:} Three typical variant modes for flexible coordination strategies.}
	\label{fig:agent}
\end{figure}

Intent alignment can be cast as iterative denoising across agents, see Fig.~\ref{fig:agent}. A minimal round proceeds as:

\begin{enumerate}
	\item \textbf{Local Proposal}: Each agent generates provisional future trajectories by sampling from a local diffusion prior conditioned on its current intent and observation. Concretely, this step is driven by the diffusion policy $p_{\boldsymbol{\theta}}(\mathbf{a}_{t-1}|\mathbf{a}_t, \mathbf{s})$ introduced in Section~\ref{sec:generalized}, where each agent runs an independent reverse sampling process conditioned on its local observation $\mathbf{s}$ to produce candidate action trajectories. When semantic tokens from neighboring agents are available from prior exchange rounds, they enter as additional conditioning vectors that blend local perception with shared intent. Drawing on Diffuser~\cite{janner2022planning}, each proposal is not a single action but a full trajectory rollout $\boldsymbol{\tau} = (\mathbf{s}_0, \mathbf{a}_0, \mathbf{s}_1, \mathbf{a}_1, \ldots, \mathbf{s}_T)$ over a planning horizon $T$, so that peers can reason jointly about temporal dependencies rather than myopic next-step actions.
	
	\item \textbf{Exchange}: Each agent broadcasts a compact intent summary to its neighbors, consisting of the compressed trajectory proposal (\eg a small set of temporal waypoints), an associated confidence score, and optionally a role or task identifier~\cite{demiris2007prediction}. Following the bandwidth-efficient intent representation defined above, this payload remains on the order of $\mathcal{O}(K)$ floating-point values after entropy coding, ensuring that the communication overhead scales with the planning granularity rather than the sensory dimensionality. Event-triggered transmission can further reduce overhead: an agent transmits only when its local posterior shifts beyond a surprise threshold (\eg upon encountering a new obstacle or receiving a role reassignment), and remains silent otherwise~\cite{qin2025generative}.
	
	\item \textbf{Reconciliation}: Upon receiving neighboring intent tokens, each agent augments its denoising update with a guidance term that enforces inter-agent consistency. Concretely, this guidance takes the form of a differentiable cost that penalizes spatiotemporal collisions and resource conflicts (\eg overlapping coverage regions or bandwidth contention) while rewarding complementarities (\eg formation diversity and task coverage). From the perspective of inference-time conditioning, this cost functions as a measurement-consistency gradient analogous to estimator guidance: it steers the local diffusion trajectory toward proposals that are individually plausible \emph{and} collectively compatible, without requiring a centralized coordinator. When multiple feasible coordination strategies coexist, the stochastic nature of diffusion sampling naturally explores this multi-modal solution space~\cite{janner2022planning}, enabling agents to discover diverse yet compatible plans rather than collapsing to a single consensus point.
	
	\item \textbf{Commit}: Agents iterate the Exchange-Reconciliation loop until successive intent samples converge, \ie the Wasserstein distance or KL divergence between consecutive proposal distributions falls below a predefined tolerance. At convergence, each agent commits to its current trajectory and begins execution. The commit criterion can be relaxed in time-critical scenarios~\cite{yang2024agent}: if the latency budget expires before full convergence, agents commit to their best current proposal, relying on the diffusion prior's manifold constraint to ensure that even partially reconciled plans remain physically feasible. Subsequent rounds can then refine the plan in a receding-horizon fashion, analogous to model predictive control with a generative world model~\cite{ha2018world, hafner2019dream, alonso2024diffusion}.
\end{enumerate}

A natural concern is the robustness of this protocol when intent tokens are corrupted or lost due to channel impairments during the Exchange step. In such cases, the ``Local Proposal'' mechanism provides an inherent safety buffer: an agent experiencing token loss temporarily falls back to its local diffusion prior conditioned solely on its own observation, producing conservative trajectory proposals that remain physically plausible but are unaware of neighbors' latest intents. The consequence is a transient coordination lag, manifesting as a temporarily suboptimal formation or redundant task coverage, rather than a catastrophic failure such as a collision. This is because the diffusion prior still constrains the generated trajectories to lie on the manifold of feasible motions. Once the communication link is restored in subsequent exchange rounds, the Reconciliation step re-integrates the recovered intent tokens, and coordinated behavior resumes. This graceful degradation property distinguishes diffusion-based coordination from deterministic consensus protocols, where a missing message can stall the entire decision pipeline.

Variants of the four-step protocol adapt to different organizational structures, including \emph{(i)} \emph{leader-follower modes} (leaders provide anchor conditions), \emph{(ii)} \emph{market-based modes} (bids/asks become conditions that steer sampling), and \emph{(iii)} \emph{role-switching modes} (an agent can ``diffuse'' between roles under load). In leader-follower mode, a designated leader agent first generates an anchor trajectory proposal using a richer diffusion prior (or more denoising steps), and broadcasts it as a strong conditioning signal; follower agents then run their local diffusion policies conditioned on this anchor, effectively reducing the reconciliation burden to aligning with a single reference plan. In market-based mode, agents attach utility bids or resource asks to their intent tokens; the reconciliation step incorporates these bids as additional guidance gradients that steer sampling toward allocatively efficient outcomes, analogous to reward-guided trajectory optimization in Diffuser~\cite{janner2022planning}. In role-switching mode, an agent can transition between roles (\eg from scout to worker or relay) by conditioning its diffusion policy on a different role descriptor token, allowing the same generative backbone to produce qualitatively different behavior without retraining.

Because diffusion sampling is incremental, communication across all modes can be \emph{event-triggered}: an agent transmits only when its local posterior shifts meaningfully (\eg upon detecting a new obstacle or receiving a role reassignment), and remains silent otherwise. This yields graceful bandwidth scaling with scene complexity rather than with fleet size.

Beyond multi-agent planning coordination, the diffusion policy strategy~\cite{chi2023diffusion} mentioned in Section \ref{sec:generalized} generalizes naturally to multi-user network resource allocation in a plug-and-play~\cite{graikos2022diffusion} or graph-based manner~\cite{sun2023difusco}. In this setting, the abstract action vector $\mathbf{a}$ encodes transmit power levels, bandwidth partitions, and scheduling assignments across the network. The policy iteratively denoises a random initialization of action $\mathbf{a}$ conditioned on the instantaneous network state (\eg CSI and queue backlogs) toward an allocation that maximizes aggregate system utility subject to quality-of-service (QoS) constraints. Compared to conventional optimization solvers~\cite{sun2018learning, he2019model} that typically converge to a single fixed point, stochastic generative sampling offers a structural advantage: when multiple feasible allocation strategies coexist, the diffusion policy characterizes the full solution manifold rather than collapsing to a single local optimum~\cite{janner2022planning}. This multi-modal nature is especially valuable for non-convex resource allocation problems that admit multiple near-optimal configurations~\cite{du2024enhancing}, enabling the network controller to select or blend strategies based on real-time operational priorities.

%% file: sec/6_open-issue.tex
\section{Open Issues} \label{sec:open}

\begin{figure*}[t]
	\centering
	\includegraphics[width=0.95\textwidth]{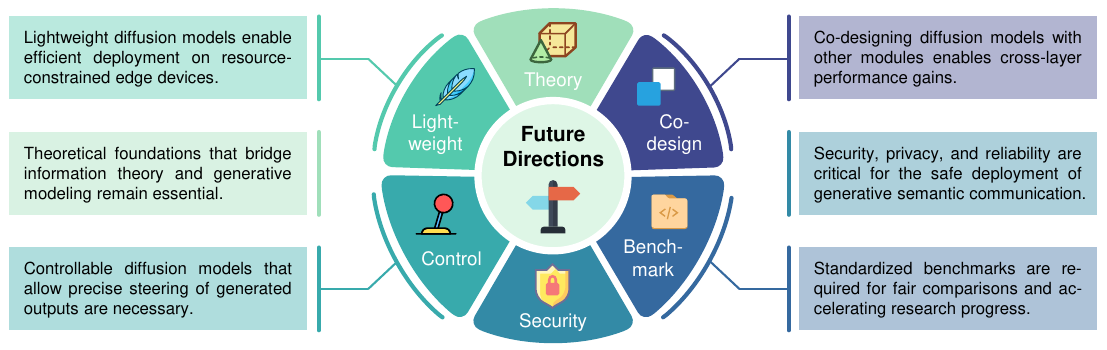}
	\caption{Promising future research directions for diffusion model-based generative semantic communications.}
	\label{fig:future-directions}
\end{figure*}

While diffusion models have demonstrated remarkable potential in advancing generative semantic communications, the field remains in its infancy. Theoretical gaps, technical barriers, and unexplored research directions still need to be addressed to fully realize their transformative potential. This section discusses key open issues across three dimensions: theoretical limitations, technical challenges, and future directions.

\subsection{Theoretical Limitations}

Despite impressive empirical success, a unified theoretical foundation for diffusion-based semantic communications is lacking. First, existing analyses of diffusion models primarily focus on their convergence properties in image generation, often assuming clean conditions and unlimited compute. Extending these analyses to communication settings, where priors, measurements, and channels interact, remains an open problem. Specifically, there is no rigorous characterization of how conditional diffusion sampling behaves under different channel noise distributions, unknown or time-varying forward operators, and semantic-aware regularizations. 

Second, performance guarantees are not yet well understood. Traditional communication theory is grounded in Shannon capacity and rate-distortion functions, while semantic communication introduces new metrics such as semantic fidelity, perceptual realism, and task effectiveness~\cite{guo2024survey}. The absence of a clear information-theoretic framework that unifies diffusion priors, semantic encoding, and generative decoding prevents systematic analysis of achievable trade-offs. It remains unclear how to define semantic capacity regions, or how generative models influence rate-fidelity-complexity relationships.

Third, the stochastic nature of diffusion sampling raises questions about reproducibility and controllability~\cite{yang2023diffusion}. While diversity is beneficial for generative tasks, it can be problematic in communications where reliability is paramount. Establishing theoretical tools to control and quantify stochastic variability, such as posterior concentration bounds~\cite{polson2018posterior} or semantic Wasserstein distances~\cite{paty2019subspace} under noisy channels, is still an open research frontier. A concrete step toward quantifying this reliability is to define a \emph{semantic outage probability}:
\begin{equation}
	P_{\mathrm{out}}^{\mathrm{sem}} = \Pr\!\left(d_{\mathrm{sem}}(\hat{\mathbf{x}}, \mathbf{x}) \geqslant \epsilon\right),
\end{equation}
where $d_{\mathrm{sem}}$ denotes a task-relevant semantic distance metric (\eg cosine distance in a CLIP~\cite{radford2021learning} embedding space), $\hat{\mathbf{x}}$ is the diffusion-decoded output, $\mathbf{x}$ is the original source, and $\epsilon$ represents an application-specific error tolerance. This metric naturally convolves two distinct sources of uncertainty: the extrinsic randomness of channel fading and the intrinsic stochasticity of the diffusion reverse sampling process. Bounding $P_{\mathrm{out}}^{\mathrm{sem}}$ under realistic channel distributions would provide the communication-theoretic reliability guarantees that the field currently lacks. A promising tool for this purpose is conformal prediction~\cite{angelopoulos2023conformal}, which constructs distribution-free prediction sets satisfying $\Pr(d_{\mathrm{sem}}(\hat{\mathbf{x}}, \mathbf{x}) \leqslant \epsilon) \geqslant 1-\alpha$ for a user-specified confidence level $1-\alpha$, without requiring Gaussian noise assumptions or knowledge of the true posterior. These finite-sample coverage guarantees are inherently compatible with the stochastic nature of diffusion sampling and the non-stationarity of wireless channels, offering a principled path toward verifiable semantic reliability.

\subsection{Technical Challenges}

On the technical side, several practical issues hinder the deployment of diffusion models in real-world semantic communication systems. 

\subsubsection{Computational Overhead} Diffusion models involve iterative denoising procedures with hundreds of steps, which can be prohibitively expensive for latency-sensitive applications such as immersive media streaming, autonomous driving, and multi-agent coordination. Although recent advances discussed in Section~\ref{sec:efficient} have reduced sampling steps, further acceleration, particularly under strict resource budgets, remains crucial. As profiled in Table~\ref{tab:efficiency}, even aggressively accelerated single-step generators combined with model quantization achieve end-to-end latencies on the order of hundreds of milliseconds, which suffices for eMBB applications but falls well short of the sub-1\,ms requirement imposed by URLLC~\cite{wang2025semantic, lu2025generative}. Bridging this gap likely requires a paradigm shift beyond conventional iterative denoising, such as amortized single-pass generators that distill diffusion priors into feed-forward networks~\cite{yin2024one}, or hardware-software co-design with dedicated neural processing units (NPUs) that parallelize the score evaluation pipeline. Until such breakthroughs materialize, diffusion-based semantic receivers should be regarded as complementary to, rather than replacements for, conventional low-latency decoding in safety-critical communication links.

\subsubsection{Robustness to Channel Conditions} Realistic wireless environments involve time-varying fading, interference, and unpredictable noise. Robust sampling on degraded channel measurements is still challenging. Current approaches either assume accurate channel knowledge or rely on handcrafted regularizations, both of which are brittle in practice. A particularly underexplored dimension is robustness to non-Gaussian noise. As discussed in Section~\ref{sec:inverse}, the standard $\ell_2$ measurement guidance implicitly assumes Gaussian disturbances and can destabilize under impulsive interference or heavy-tailed channel residuals. While robust loss substitutions (\eg Huber loss, Student-$t$ likelihood) offer pointwise remedies, a systematic framework for diffusion posterior sampling under unknown and possibly non-stationary noise distributions is still lacking. Promising directions include learning the noise model jointly with the source prior via score matching on the residual distribution, and incorporating distributionally robust optimization~\cite{rahimian2019distributionally} to guarantee worst-case performance over a family of plausible noise models.

\subsubsection{Semantic Alignment and Control} Generative semantic communications rely on conditioning mechanisms to align generated content with intended meaning. However, robustly controlling the generation process, especially under diverse tasks, multi-modal conditions, or conflicting agent intents, requires more principled conditioning strategies. Current solutions remain heuristic, often leading to semantic drift or misalignment.

\subsubsection{System Integration} Integrating diffusion models into full-stack communication systems, including semantic encoders, physical-layer codecs, and network protocols, poses nontrivial engineering challenges. Diffusion models are often trained in isolation from encoders and channels, leading to mismatches between training and deployment. Joint end-to-end optimization remains technically demanding.

\subsection{Future Directions}

Looking ahead, several promising research directions may overcome these limitations and bring generative semantic communications closer to practical deployment, see Fig.~\ref{fig:future-directions}.

\subsubsection{Unified Theoretical Frameworks} A pressing need is to establish rigorous foundations bridging information theory, inverse problems, and generative modeling. This includes defining semantic capacity measures, analyzing achievable regions under diffusion priors, and deriving generalization bounds for semantic-aware posterior sampling. Extending score-based formulations to explicitly account for channel uncertainty and semantic constraints could lead to principled performance guarantees.

\subsubsection{Lightweight and Adaptive Models} Advances in model compression, adaptive noise schedules, and conditional consistency distillation could enable real-time inference on edge devices. Dynamic computation graphs~\cite{cao2024diffusione} that adjust denoising depth based on channel quality or task importance represent a promising direction for resource-aware deployment.

\subsubsection{Semantic Control Mechanisms} Developing controllable diffusion models that allow precise steering of generated outputs is essential. Techniques such as classifier-free guidance, semantic token conditioning, or learned control fields can be adapted to communication tasks, enabling flexible trade-offs between fidelity, realism, and task effectiveness.

\subsubsection{Cross-layer Semantic-Physical Co-design} Diffusion models should not operate as isolated modules but be co-designed with semantic encoders, channel estimators, and network schedulers. For example, diffusion priors can guide adaptive modulation, and network protocols can allocate resources based on semantic importance~\cite{du2024enhancing}. Joint optimization across layers may unlock significant performance gains. This co-design imperative is particularly acute in emerging 6G application scenarios that impose stringent and heterogeneous constraints. In aerial-aided edge networks, UAV-mounted semantic communication nodes must jointly optimize trajectory planning, transmission scheduling, and on-board diffusion inference under tight energy and latency budgets~\cite{emami2025diffusion, zhang2024multi}; cross-modal attention mechanisms can further enable resource-constrained aerial platforms to fuse multi-source semantics via lightweight diffusion models~\cite{liu2024cross}. In vehicular semantic communication networks, cooperative perception among vehicles demands ultra-low-latency exchange of compressed semantic representations (\eg BEV features) over lossy vehicle-to-everything (V2X) links, where conditional diffusion models have shown promise in reconstructing task-critical scene understanding from heavily compressed transmissions~\cite{mao2025diffcp}. More broadly, the integration of generative AI techniques into unmanned vehicle swarms~\cite{liu2024generative} and semantic edge computing architectures~\cite{zhang2025semantic} calls for cross-layer frameworks that dynamically adapt the semantic codec, the diffusion sampling depth, and the physical-layer resource allocation to the mobility pattern, link reliability, and mission criticality of each node.

\subsubsection{Benchmarks and Standardization} The field would benefit from standardized datasets, metrics, and evaluation protocols that reflect realistic communication scenarios, including perception tasks, multi-agent settings, and low-SNR regimes. Establishing shared benchmarks will accelerate progress and ensure fair comparisons between methods.

\subsubsection{Safety, Security, and Ethics} Finally, as generative semantic communications move closer to deployment, issues of reliability, privacy, and robustness must be addressed. Stochastic generative systems may introduce unpredictable behaviors, and malicious conditions could be exploited for adversarial attacks. Formal verification, adversarial robustness, and trustworthy intent exchange are emerging research frontiers.


%% file: sec/7_conclusion.tex
\section{Conclusion} \label{sec:conclusion}
This article has provided a systematic guideline of diffusion models for generative semantic communications, establishing theoretical foundations while charting practical pathways for implementation. Through comprehensive coverage of score-based fundamentals, conditioning mechanisms, efficiency optimizations, and generalization techniques, we have equipped researchers with essential tools for designing next-generation semantic communication systems. The inverse problem framework offers a unified mathematical perspective that naturally connects semantic reconstruction to established signal inversion methodologies, opening new avenues for algorithmic innovation. Analysis across human-centric, machine-centric, and agent-centric scenarios demonstrates the versatility of diffusion models in addressing diverse communication requirements. Moreover, open challenges and promising future research directions have been discussed, highlighting opportunities for advancing diffusion-based semantic communications.